\RequirePackage{fix-cm}
\documentclass[natbib,smallextended]{svjour3}       
\smartqed  
\usepackage{graphicx}
\usepackage[colorlinks=true,citecolor=blue,urlcolor=blue,breaklinks]{hyperref}
\def\simlt{\mathrel{\rlap{\lower 3pt\hbox{$\sim$}}
        \raise 2.0pt\hbox{$<$}}}
\def\simgt{\mathrel{\rlap{\lower 3pt\hbox{$\sim$}}
        \raise 2.0pt\hbox{$>$}}}
\journalname{The Astronomy and Astrophysics Review}
\begin{document}

\title{Hosts and environments: a (large-scale) radio history of  AGN and star-forming galaxies}


\titlerunning{Hosts and environments of radio-active AGN and star-forming galaxies}        

\author{Manuela Magliocchetti}

\institute{M. Magliocchetti\at
              INAF-IAPS, Via Fosso del Cavaliere 100, 11033, Rome, Italy \\
              \email{manuela.magliocchetti@inaf.it}           
}

\date{Received: date / Accepted: date}

\maketitle

\begin{abstract}
Despite their relative sparseness, during the recent years it has become more and more clear that extragalactic radio sources (both AGN and star-forming galaxies) constitute an extremely interesting mix of populations, not only because of their intrinsic value, but also for their fundamental role in shaping our Universe the way we see it today. Indeed, radio-active AGN are now thought to be the main players involved in the evolution of massive galaxies and clusters. At the same time, thanks to the possibility of being observed up to very high redshifts, radio galaxies can also provide crucial information on both the star-formation history of our Universe and on its Large-Scale Structure properties and their evolution. In the light of present and forthcoming facilities such as LOFAR, MeerKAT and SKA that will probe the radio sky to unprecedented depths and widths, this review aims at providing the current state of the art on our knowledge of extragalactic radio sources in connection with their hosts, large-scale environments and cosmological context. 
\end{abstract}

\setcounter{tocdepth}{3} 
\tableofcontents

\section{Introduction}
\label{intro}
During the past years, radio galaxies have quickly moved from being an interesting class of extragalactic objects to become one of the fundamental bricks of the wall of our comprehension of the physical processes that shape galaxy formation and evolution. Indeed, radio-emitting AGN have been invoked as the major players in action to halt star-formation in massive galaxies, with an effect that would also prevent their masses to reach unobserved values and produce the local `read and dead' population of ellipticals both in terms of physical properties and number density (the so-called `radio-mode' feedback, e.g., \citealt{croton, bower, fanidakis, weinberger}). Furthermore, a very strong connection between radio-AGN and galaxy clusters is now well established, with radio-AGN emission being responsible for a large part of their thermal state (see \citealt*{McNamara, brunetti} for reviews on this topic) as well as for  enhancing or suppressing star-formation within its member galaxies (e.g. \citealt{Gilli}; \citealt{salome}).

Extragalactic radio sources come in two flavours: AGN and star-forming galaxies. Radio-emitting star-forming galaxies, at least in the local universe, generally coincide with normal spiral and irregular galaxies, typically present low ($L_{1.4\, \rm GHz} \simlt 10^{22}$ W Hz$^{-1}$ sr$^{-1}$) luminosities and dominate the radio source counts below the mJy level (see \citealt{padovani} for a review). 
On the other hand, the population of radio-emitting AGN, much brighter and extending to much higher redshifts as observed by current facilities (indeed these were the first galaxies to be found above redshifts one, two, three and four -- cf.\ \citealt*{stern2} and references therein), is way more variegated. 
Historically, radio-AGN have been sub-divided in two different ways. \citet*{fanaroff} classified sources according to their radio morphologies as type I (FRI), in which the peak of radio emission is located near the core (edge-darkened) and type II (FRII), in which the peak of surface brightness is at the edge of the radio lobes far from the centre of emission (edge-brightened). A further classification scheme is instead based on the presence and relative intensity of high- and low-excitation lines in their
optical spectra (e.g., \citealt*{hine}) and divides them into high-excitation radio galaxies (HERGs) and low-excitation radio galaxies (LERGs). HERGs and LERGs are
believed to represent intrinsically different types of objects (see \citealt*{heckman} for a review), with LERGs accreting at low rates (luminosities below 1\% of their Eddington limit) and HERGs accreting at high rates (between 1\% and 10\%). 
 
As a matter of fact, the realm of radio-emitting AGN is even more complex than this. On top of compact/point-like (or FR0 cf.\ \citealt{Baldi1}) galaxies, radio-quiet AGN which are thought to emit at radio wavelengths mainly thanks to star-forming processes taking place within their hosts (cf.\ \citealt{padovani}) and quasars, blazars and BL Lacs believed to be the very same AGN-powered radio galaxies only seen at smaller viewing angles \citep{urry}, in the recent years new populations have been discovered, such as the Compact Symmetric Objects (CSO), the Compact Steep Spectrum (CSS) galaxies and the Gigahartz-Peaked galaxies (GPS). These are all powerful sources ($L_{1.4\, \rm GHz} > 10^{24}$ W Hz$^{-1}$ sr$^{-1}$), with GPS presenting very compact ($\simlt 1$ Kpc) radio emission with a peak in the GHz range (see \citealt*{odea}  for a review), and CSS being less compact ($\simlt 15$ Kpc) and with a peak around 100 MHz (see \citealt*{miley1} for a review). CSOs are instead the smallest sources, with sizes of less than a few hundred pc (see \citealt*{hardcastle4} for a review).
We further note that radio-AGN with resolved morphologies can also present themselves as Giant Radio Galaxies (GRG) if their extension goes beyond $\sim 0.7$ Mpc (e.g., \citealt*{willis}) or as Wide Angle/Narrow Angle Tail (WATs or NATs) galaxies when their jets do not lie along a common direction but show various degrees of bending (e.g \citealt*{owen}).


Many excellent reviews tackling in a very comprehensive way the AGN phenomenon and more in general radio emission are available in the literature (e.g., \citealt*{miley1}; \citealt{dezotti}; \citealt*{heckman, tadhunter1}; \citealt{padovani, padovani1, hardcastle4}). Our purpose is to extend the content of these reviews by presenting radio galaxies as cosmological objects in connection with the more general context of galaxy and large-scale structure formation and evolution. To this aim, we will present the current state of the art on our knowledge of extragalactic radio sources (both AGN and star-forming galaxies) from the point of view of their galaxy hosts, environments and cosmological behaviour, so to provide a link between radio activity on small (nuclear and sub-galactic) scales and environmental (Kpc to Mpc) properties.

Within this framework, we have decided to concentrate on a number of topics which are mostly still controversial and/or poorly understood. For what concerns the hosts of radio galaxies we will then discuss the role of their mass on radio emission, the FRI/FRII dichotomy, the cosmological evolution of radio-AGN in terms of star-formation activity within their hosts and, in the case of star-forming galaxies, the properties of the relationship between infrared and radio luminosities (the so-called IR-radio correlation) as obtained at the different radio frequencies and redshifts. Environments will instead be investigated by following three different methods. The choice is dictated by the fact that in general these different methods return different and complementary answers. For example, statistical techniques such as clustering analyses are the only ones capable of providing information on the dark matter content of the structures that host radio-AGN and star-forming galaxies. Furthermore, they can also offer a snapshot of the large-scale structure of our Universe as traced by these sources and on its evolution up to the earliest epochs. The other two methods, which consist in a) searching for structures around pre-selected radio-AGN and b) pinpointing radio-AGN within known structures might seem to be very similar to each other. As a matter of fact though, these two are also complementary since they return results which in some cases differ even by large factors. For instance, the first method finds that most radio-AGN reside within overdense structures, while according to the second method this is true only for a much smaller fraction of sources. This discrepancy reverberates e.g., onto the estimates of the life-time of the radio-active AGN phase and on a number of other issues, so that a direct comparison between the results obtained in these two different ways is not always possible. Furthermore, whenever structures such as groups or clusters of galaxies with available X-ray information are involved, the second method can also return fundamental information on the interaction between radio-AGN activity and the intracluster medium (ICM). 

The layout of this review is then the following: Sect.~\ref{sec:1} will summarize the main methods adopted in the literature to distinguish between radio-emitting AGN and star-forming galaxies. Section \ref{sec:2} will be devoted to the hosts. In more detail, Sect.~\ref{sec:2.1} will investigate the role of galaxy mass on radio (mainly AGN) activity, while Sect.~\ref{sec:2.2} will discuss the morphological dichotomy between FRI and FRII galaxies from the point of view of their hosts and nuclear/accretion properties (LERG vs HERG distinction). Section \ref{sec:2.3} will present a detailed analysis of the cosmological evolution of star-forming activity within radio-AGN hosts beyond $z \simgt 1$, while Sect.~\ref{sec:2.4} will review our current knowledge on the controversial issue of the IR-radio relation in star-forming galaxies, its dependencies and eventual redshift evolution. Section \ref{sec:3} will instead be dedicated to the environments of extragalactic radio sources. As previously discussed, these will be presented by following the method adopted to investigate their properties. So, after a brief introduction on the theory behind statistical methods such as the correlation function and cross-correlation function -- necessary to understand some of the works reviewed in this sub-section -- Sect.~\ref{sec:3.1} will present the results obtained so far by making use of the aforementioned techniques. Section \ref{sec:3.2} will instead show our current knowledge on the environmental properties of radio galaxies as obtained by pinpointing them within pre-defined over-densities such as groups and clusters of galaxies, while Sect.~\ref{sec:3.3} will do the same but in the case of structures identified around chosen radio sources. Our conclusions will then be summarized in Sect.~\ref{sec:4}.

Because of its contents, this review is thought to reach an audience which extends beyond radio astronomers. To this aim, it will present the various topics in a somehow didactic way. Each section will first introduce the general problem and then present a historical excursus of the most relevant results obtained throughout the years up to the present days with the aim to understand what has been done so far, what we think is understood, what is still missing and where (if any) contradictions stand. Since sample sizes, completeness issues and depths of the observations are all crucial quantities that can largely affect the outcome of statistical analyses such a those presented in the following sections, 
we will add these pieces of information when discussing most of the works.  

Before moving on though, an important point has to be made. Many names can be found in the literature for the radio-active AGN population. From the historically adopted radio-loud AGN to jetted AGN (\citealt{padovani}). We personally believe that none of these can correctly describe the considered sources, since the term radio-loud made more sense when radio surveys only sampled bright `monsters', while it is not anymore the case for current facilities that reach the sub-mJy/$\mu$Jy regime. On top of that, as it will become more clear in Sect.~\ref{sec:1}, radio-loudness does not have a univocal definition (e.g., \citealt{jarvis3}) since radio-emitting AGN can be selected in many different ways. At the same time, not all radio-active AGN present jetted structures, so even this terminology only provides a partial description of the radio AGN phenomenon. Our preference goes to the more general terms `radio-active AGN', `radio-emitting AGN' or alternatively `radio-AGN' to indicate all AGN emitting at radio wavelengths because of their nuclear activity. This is the wording that will be used throughout this review. 
We also stress that throughout the text we will use the word ``protocluster'' to indicate a non-virialized, overdense structure formed by galaxies and/or AGN, regardless of whether it will subsequently evolve into a cluster. Lastly, whenever mentioned, $h$ is the reduced Hubble constant.
Note that most (if not all) of the works presented in this review and published after 2000 adopt the standard cosmological model $\Omega_0\simeq 0.3$, $\Omega_\Lambda\simeq 0.7$, $h\simeq 0.7$ (e.g. \citealt{Aghanim}), with $\Omega_0$ and $\Omega_\Lambda$ respectively present density of (total) matter and dark energy. However, earlier results still consider a universe with  $\Omega_0 = 1$ and $\Omega_\Lambda = 0$.

\section{The difficult art of discerning between radio-active AGN and star-forming galaxies from continuum radio surveys}
\label{sec:1}
Extragalactic radio sources are a mixed bag of different astrophysical populations. And while until the 1990s the radio sky was dominated by bright enough objects to ensure they were \textit{bona fide} radio-active AGN, indeed with the advent of deep enough radio surveys it became clear that the counts below $\sim 1$ mJy start to be dominated by different objects such as star-forming galaxies (e.g., \citealt{windhorst, maglio3, prandoni} and references therein) and radio-quiet AGN, whereby -- despite the presence of an AGN -- radio emission mainly originates from star-forming processes within their hosts rather than from accretion onto a black-hole (e.g., \citealt{Kimball, condon2, bonzini, ceraj} -- see also \citealt{panessa} for a review on these objects and \citealt{padovani} for a review on the faint radio sky).
It follows that, unless the radio-AGN present morphological structures (such as lobes and jets) that uniquely identify them as such, distinguishing between these three different populations in continuum radio surveys is currently an impossible task. Information (both photometric and/or spectroscopic) at wavelengths different than radio is required, but even in this case the answers are not always univocal. 

As we will see later in Sect.~\ref{sec:2.1}, in the local universe radio-AGN are almost ubiquitously associated with massive elliptical galaxies, with little or no ongoing star-forming activity. This prompted early studies to introduce a value for the radio-to-optical (R band) ratio (defined as $q_R = F_{1.4\, \rm GHz} \cdot 10^{(R-12.5)/2.5}$, with $F$ expressed in mJy) larger than $\sim 30$ (e.g., \citealt{Kellermann, urry}) as an indicator for a radio-emitting AGN. At the same time, still in the local universe, star-forming galaxies (and radio-quiet AGN) present a very tight correlation between their far-infrared (FIR) and radio fluxes (e.g., \citealt{condon}, cf.\ Sect.~\ref{sec:2.4}). So, distinguishing between radio-AGN and star-forming galaxies at $z\simlt 1$ is rather doable.

The situation becomes more complicated once we move to redshifts higher than $z\sim 1-1.5$. The explanation is quite straightforward: $z \sim 1.5$ in fact marks the transition from a relatively quiet universe, to one dominated by AGN and cosmic star-formation activities (e.g., \citealt*{merloni}). In such a universe, many star-forming galaxies are found to host an AGN and the other way round (cf.\ Sect.~\ref{sec:2.3}), so a clear distinction between these two classes of sources based on their radio-to-infrared colours is extremely difficult. At the same time, galaxies at $z\sim 2$ are generally younger than those found in the local universe, therefore also an optical search for radio galaxies as red-and-dead sources is proven to be ineffective. 
This implies that new techniques have to be applied to try to distinguish between radio-emitting AGN and the other populations. \citet{padovani} (see also \citealt{hardcastle3} and references therein) provides an excellent summary of the different selection methods. Here we will just briefly list the more commonly used ones in view of their strengths and weaknesses:\\
a) \underline{Optical/NIR spectroscopy}: This method (e.g., \citealt*{Baldwin}; \citealt*{Veilleux}) provides the cleanest distinction between AGN and star-forming galaxies and is based on the absence or presence (and relative strength) of emission lines indicating AGN and/or star-forming activity in the optical/NIR spectra of radio galaxies. Despite its success, being spectroscopy a very time-consuming task, a wide application is however currently limited only to relatively local sources, or alternatively to sparse samples of bright objects in the more distant universe. Also, even if line emission in the spectra guarantees the presence of an AGN, it does not provide any information on whether radio emission is due to nuclear activity or to star-forming processes within the host galaxy, as is the case for radio-quiet AGN.\\
b) \underline{SED fitting}: This method (e.g., \citealt{smolcic5}) uses photometry at a number of wavelengths (the more the better) to produce a spectral energy distribution (SED) for the sources in exam. This can be thought as a lower-resolution version of the spectroscopy method, which only considers continuum emission instead of lines. Observations then get fitted with synthetic models for stellar evolution (e.g., \citealt*{bruzual}) to infer a number of galaxy properties, including presence and relevance of an AGN. Less precise but also less expensive than spectroscopy, it can be applied to much higher redshifts. As in the spectroscopy case though, it cannot discern whether in presence of an AGN radio emission originates from the central black hole or from surrounding star-forming regions.\\
c) \underline{Radio/FIR excess}: This method (e.g., \citealt{delvecchio}) selects as radio-AGN all sources that present a radio excess and lie below the IR-radio relation (cf.\ Sect.~\ref{sec:2.4}), and as star-forming galaxies those that either lie on the relation or show a FIR excess. As already discussed, it works very efficiently in the local universe. However, it has to rely on FIR observations which are currently unavailable for many radio sources. Also, it is proven to loose its ability at redshifts $z \simgt 1-1.5$ due to the cosmological evolution of the galaxies, most of them co-hosting AGN and star-forming processes at the same time.\\
d) \underline{Radio continuum slope}: This method (e.g., \citealt{alberts}) relies on radio observations taken at different frequencies. It works well at identifying flat-spectrum or inverted-spectrum AGN, but it is ineffective at distinguishing between steep-spectrum radio-AGN and star-forming galaxies as they generally present very similar values for the radio spectral index, $\alpha \sim 0.7$ (e.g., \citealt{dezotti}).\\
e) \underline{X-ray emission}: Most of the AGN emit in the X-ray, therefore another way to distinguish between a radio-AGN and a radio-active star-forming galaxy is by looking at the X-ray emission produced by a radio source. However, this method only (partially) works for very bright objects, since the majority of radio-AGN have X-ray luminosities lower than $\sim 10^{42}$ erg sec$^{-1}$, value below which X-ray emission might just as well be due to star-forming processes within the host (e.g., \citealt{fragos}). As a matter of fact, the fraction of radio-AGN also detected in the X-ray bands is quite low ($\sim 5$\%, e.g., \citealt{hickox}, value that can however increase according to selection criteria and flux densities, e.g., \citealt{maglio18}).\\
f) \underline{MIR colours}: Another possibility to distinguish between radio-AGN and star-forming galaxies is provided by the investigation of their mid-infrared (MIR) colours (e.g., \citealt{stern1}). However, this method is also shown to work well only for very luminous and radiatively efficient AGN, while it misses the majority of the radio-active AGN population since it is biased against lower MIR luminosities and higher redshifts (e.g., \citealt{donley1, mingo}), both typical of most radio-AGN.\\
g) \underline{Radio Luminosity}:  The last method presented here (\citealt{maglio13}) -- which we discuss in greater detail since it is not mentioned in the reviews of \citet{padovani} or \citet{hardcastle3} -- relies on the radio luminosity of the sources. Indeed, thanks to the steepness of the radio-luminosity function of star-forming galaxies as opposed to the flatter one exhibited by radio-AGN (e.g., \citealt{maglio4, mauch}), selecting sources brighter than the radio luminosity value at which the two distributions cross, virtually ensures that all \emph{all} of them will be radio-AGN. This holds in the local universe. At higher redshifts, the cosmological evolution of these populations, together with the existence of extreme starburst galaxies, implies that sources with star-forming rates above $\sim 10^3\, M_\odot$ yr$^{-1}$ will present radio luminosities $\simgt 10^{24}$ W Hz$^{-1}$ sr$^{-1}$ and therefore be mistakenly identified as radio-AGN (e.g., \citealt{hardcastle3}). However, such sources are relatively rare and furthermore \citet{maglio13} take the evolutionary issue into account by introducing a criterion based on the radio-luminosity function as provided up to redshifts $\sim 2.5$ by  \citet*{mcalpine}, who indeed find that star-forming galaxies evolve in a much stronger way, $\propto (1+z)^{2.5}$, than radio-selected AGN ($\propto (1+z)^{1.2}$). By using this result, \citet{maglio13} adopt an evolving luminosity threshold that scales with redshift as ${\rm Log_{10}}[L_{\rm cross}(z)] = {\rm Log_{10}}[L_{\rm 0, cross}]+z$ (with $L_{\rm 0, cross}=10^{21.7}$ W Hz$^{-1}$ sr$^{-1}$, roughly corresponding to the break of the local radio luminosity function of star-forming galaxies) up to $z=1.8$ and $L_{\rm cross}(z) =10^{23.5}$ W Hz$^{-1}$ sr$^{-1}$ for $z>1.8$, above which to select radio-AGN. Since the radio-AGN samples selected in this way are contaminated by very few interlopers (\citealt{maglio13, maglio14, maglio15}), the present method has the important advantage of being rather clean while only needing redshift determinations instead of expensive spectroscopy or multi-wavelength information for the sources in exam. On the other hand, it cannot identify star-forming galaxies and misses faint radio-AGN which end up in the mixed bag which also contains radio-quiet AGN.

The main points to be taken from the above discussion are then three:

\begin{enumerate}
\item Despite the enormous wealth of multi-wavelength information that is getting more and more accessible to the astronomical community, discerning between radio-active AGN and galaxies dominated by star forming activity is still not a straightforward task.
\item None of the methods highlighted above provides the `ultimate' selection, as they all suffer from various (and generally different) limitations.
\item Different techniques as a matter of fact select slightly different populations, which generally coincide only in the case of sources with very bright (or alternatively very faint) radio luminosities. This is an issue that always needs to be kept in mind when comparing results from different works adopting different selection methods.
\end{enumerate}

\section{Hosts of radio-active AGN and star-forming galaxies}
\label{sec:2}
This section presents results on the connection between radio-activity and host galaxy properties. It is divided into four different parts that will respectively tackle the connection between host (stellar) mass and radio emission (Sect.~\ref{sec:2.1}), the possible causes for the FRI-FRII dichotomy (Sect.~\ref{sec:2.2}), the cosmological evolution of radio-AGN in terms of star-formation activity within their hosts (Sect.~\ref{sec:2.3}) and -- in the case of star-forming galaxies --  the IR-radio relation, its eventual dependences and redshift evolution (Sect.~\ref{sec:2.4}).

\subsection{The role of mass on radio activity}
\label{sec:2.1}
It has been clear since many years now (\citealt{matthews}) that the majority of powerful radio-AGN is hosted within the most massive galaxies (stellar masses $M_* \sim 10^{11}\, M_\odot$ and above) known in the Universe. 
This result has been confirmed at least up to $z\sim 1$ (and in some cases beyond) both by direct estimates (e.g., {\citealt{heeschen, ekers, Auriemma, jenkins, best4, mauch, smolcic1, maglio14, sabater1, capetti2}) and also by the extremely tight correlation in the $M_K-z$ plane exhibited by the hosts of radio-AGN (e.g., \citealt{lilly1, eales, jarvis, debreuck, willott}) that can be explained by assuming that the observed $K$-band light is dominated by emission from the old stellar population and that radio-AGN have very similar (large) host masses all across half of the age of the Universe. More recent results based on rest-frame 1.6 $\mu$m and 4.5 $\mu$m photometry also report very high masses for the hosts of radio-AGN, at least up to $z\sim 3$ if not beyond (\citealt{seymour, debreuck1}; \citealt*{gurkan}; \citealt{drouart1}).

\begin{figure}
\centering
 \includegraphics[scale=0.18]{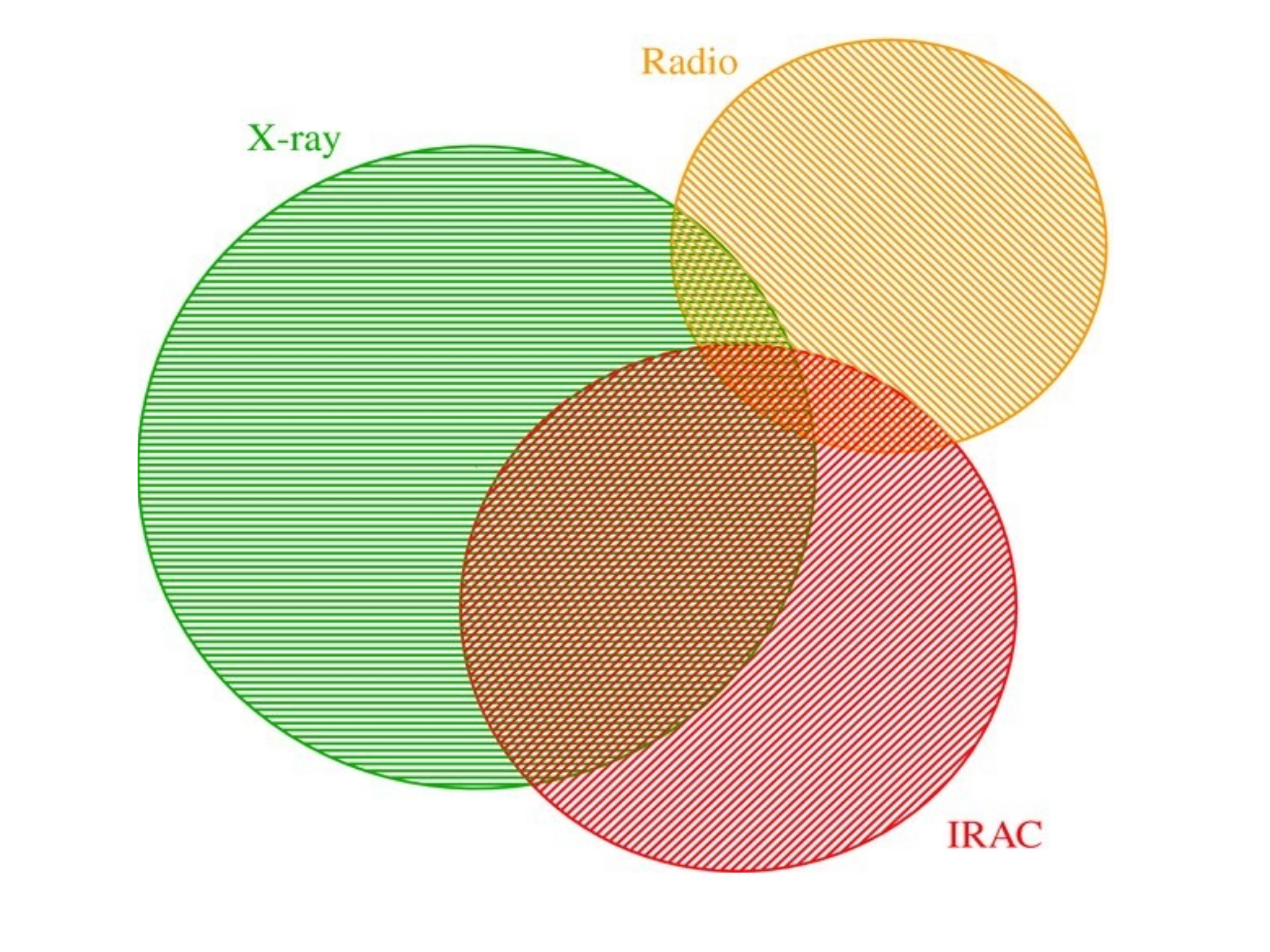} 
 \includegraphics[scale=0.30]{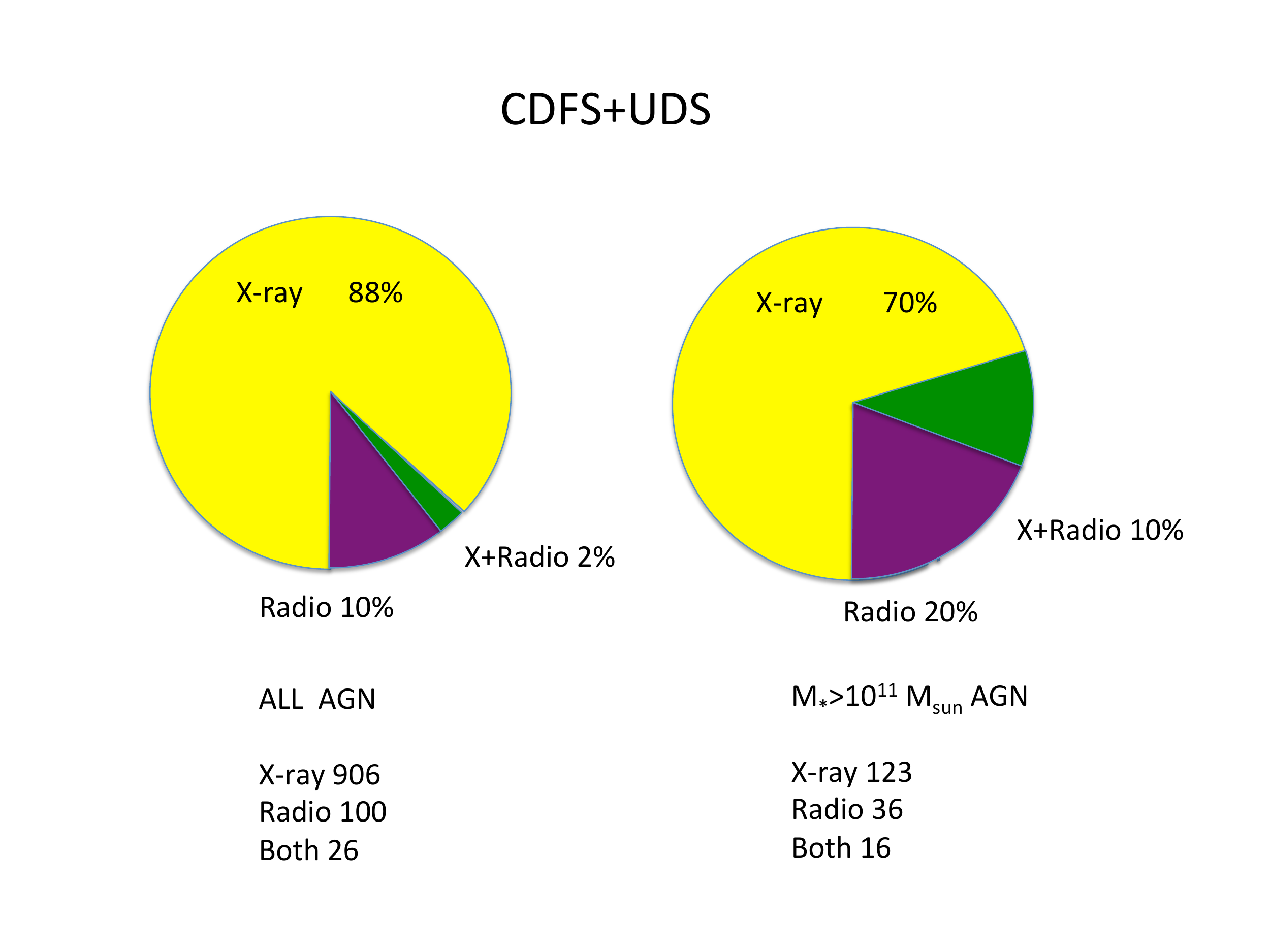}
 \caption{Top panel: Venn diagram showing the relative number of AGN with spectroscopic redshifts in the range 
 $0.25 <z< 0.8$, selected in the radio, X-ray, and near-infrared (NIR). Out of 122, $L_{1.4\, \rm GHz} >10^{23.8}$ W Hz$^{-1}$, radio-AGN, only 6 are also detected in the X-ray (down to $L_X = 10^{42}$ ergs s$^{-1}$) and IRAC (down to $L_{4.5 \rm \mu m} = 10^{43}$ ergs s$^{-1}$) bands. Similarly, out of 296 (199) X-ray-selected (IRAC-selected) AGN, only 6 (6) are also detected at radio wavelengths. Figure from \citet{hickox}. Bottom panel: Pie chart illustrating the contribution of the different emitters to the total (radio+X-ray) AGN population obtained by
combining together the UDS and CDFS fields. The yellow areas correspond to X-ray emitters, the purple to radio emitters while the green regions indicate the contribution from those AGN which simultaneously emit in the radio and X-ray bands. The left-hand pie shows the case for AGN associated with galaxies of all stellar masses, while that on the right corresponds to AGN residing within hosts with $M_* \ge 10^{11}$ $M_\odot$. Figure from \citet{maglio18}.}
\label{fig:hickox}       
\end{figure}

However, these results do not fully answer the question that naturally arises when considering radio-AGN: what causes their radio emission?
In a seminal paper by \citet{hickox}, it was found that the overlapping between families of AGN selected with different methods (X-ray, \textit{Spitzer}-IRAC bands and radio) was extremely small, especially when it came to radio emission from X-ray- and NIR-selected AGN ($\sim 5$\%). Between a third and a half of X-ray selected AGN were instead observed to have a counterpart in the NIR, and the other way around (cf.\ top panel of Fig.~\ref{fig:hickox}). Although the relative depths of the different samples can provide at least a partial answer to this effect (e.g., \citealt{ho}), it is also very possible that it might instead be mainly determined by more physical factors.  

On the other hand, there are instead cases where radio-AGN emission is almost ubiquitous. \citet{brown} used 1.4 GHz, NRAO VLA Sky Survey (NVSS -- \citealt{condon1}) observations of 396 local early-type galaxies selected from the 2MASS Extended Source Catalogue (\citealt{jarrett}), finding that virtually all 120 galaxies with absolute magnitudes $M_K< -25$ (corresponding to stellar masses $\simgt 10^{11}\,M_\odot$) were radio emitters with flux densities larger than $\sim 1$ mJy. This was in line with earlier findings (e.g., \citet{sadler1} and references therein).
Much more recently, \citet{capetti2} have come to the same conclusion for their complete sample of 188 local giant ellipticals with $M_K<-25$ taken from the 2MASS Redshift Survey (\citealt{huchra}) and with radio information from the LOFAR Two-Metre Sky Survey Data Release 2  (LoTSS DR2 -- \citealt{shimwell1}). Indeed, they show that about 78\% of these sources are detected by LOFAR above $L_{\rm 150 MHz} \simgt 10^{21}$ W Hz$^{-1}$, and that \emph{all} the brightest, $M_K< -25.8$, giant ellipticals are associated with radio-emission, even if for one case this might be due to star-forming processes.  The same result has also been obtained by \citet{grossova} who extend the work of \citet{dunn} by considering 42, very local (distance $< 100$ Mpc), optically and X-ray brightest early-type galaxies observed with the highest resolution VLA A configuration in the frequency range 1--2 GHz. Out of these 42 sources, 41 present detectable radio emission.

The discrepancy between the frequency of radio-AGN emission as obtained for local massive galaxies by the aforementioned works and for higher-redshift AGN as obtained by \citet{hickox} is really striking ($\sim$ 5\% vs $\sim$ 100\%). A number of factors could contribute to this disagreement, such as galaxy type, nuclear/accretion differences and/or redshift range. However, in the following we will mainly investigate the effects of the host mass.
Indeed, already 50 years ago, a number of works had envisaged a strong correlation between the probability for a galaxy to be the host of a radio-active AGN and its optical luminosity $L_{\rm opt}$ (e.g., \citealt{colla, Auriemma, hummel}).
\citet{best4} brought these early works to a much more solid statistical grounding by presenting an analysis performed for 420, $z<0.1$ radio-AGN brighter than 5 mJy (corresponding to a radio luminosity limit of $L_{1.4\, \rm GHz} \simgt 10^{23}$ W Hz$^{-1}$ at $z=0.1$), selected from the Faint Images of the Radio Sky at Twenty centimeters (FIRST -- \citealt{becker}) and NVSS surveys, and with spectroscopic information from the second data release of the Sloan Digital Sky Survey (SDSS -- \citealt{york}). Their results clearly showed that there is a steady increase of the fraction of galaxies that host a radio-AGN with increasing stellar mass $M_*$. The functional form is of the kind $F_{\rm radio-AGN}\propto M_*^{2.5}$ (which holds at at all radio luminosities), until a plateau of about 30\% is reached for $M_*\simgt 10^{11.7}\,M_\odot$. 
It is interesting to note that \citet{best4} also observe a correlation between $F_{\rm radio-AGN}$ and central black hole mass, which is however shallower than what found in the case of the host mass ($\propto M_{\rm BH}^{1.6}$). 

\begin{figure*}
\centering
\includegraphics[scale=0.4]{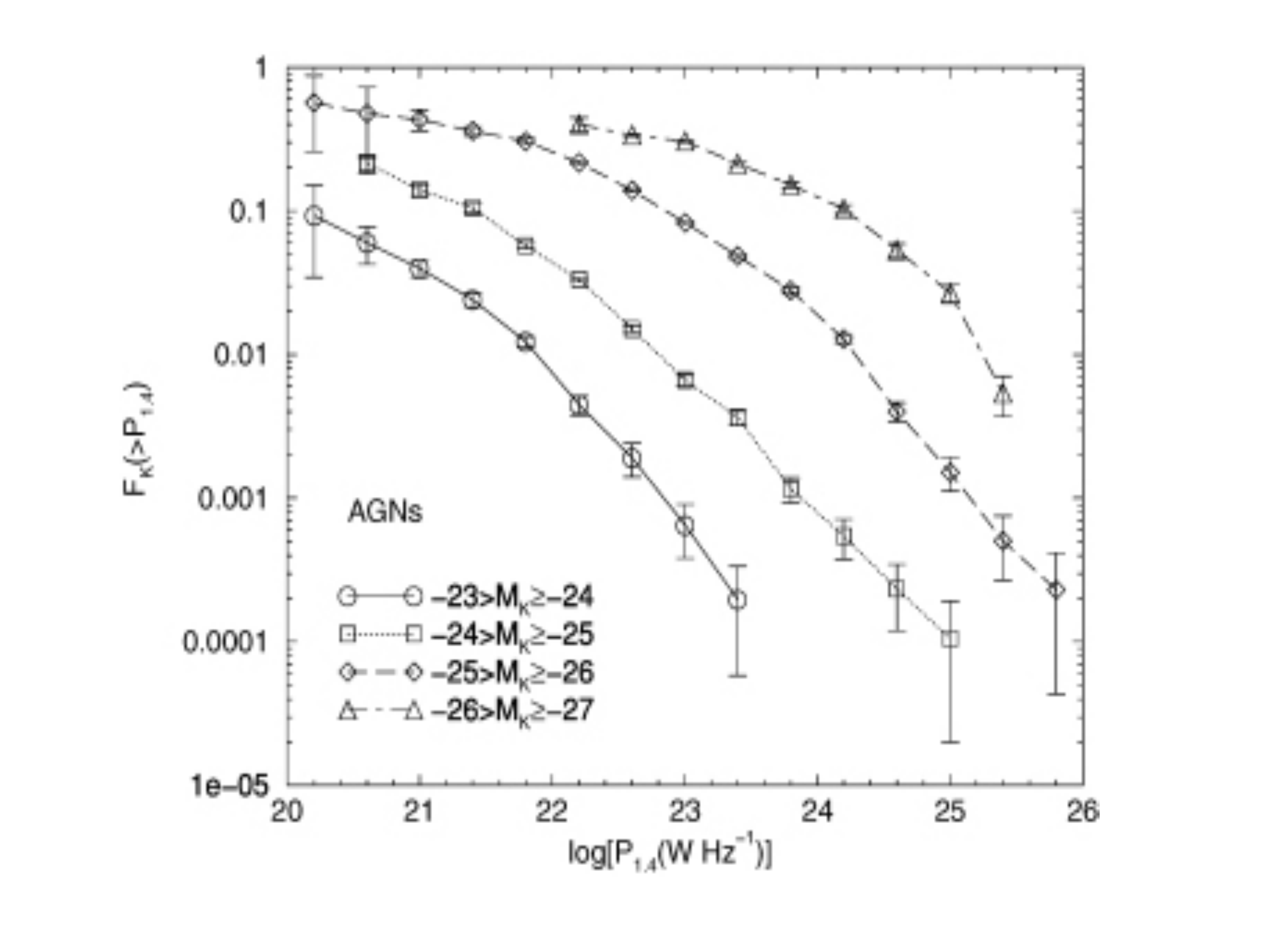}
\caption{Fraction of galaxies $F_K(>P_{1.4})$ which host a radio-AGN as a function of radio luminosity $P_{1.4}$ calculated in four bins of absolute $K$ magnitude ($M_K$) as labelled in the figure. Figure from \citet{mauch}.}
\label{fig:mauch}       
\end{figure*}

\citet{mauch} use a $z\simlt 0.15$ sample of 2661 radio-AGN selected from the NVSS survey with radio fluxes above 2.8 mJy, magnitudes $K>12.75$ and spectroscopic redshifts from the 6 degree Field Galaxy Redshift Survey (6dFGRS -- \citealt{jones}). These authors also find that the probability for a galaxy to host a radio-AGN has a strong dependence on the $K$-band luminosity (proxy for the stellar mass) of the host galaxy with a functional form $F_{\rm radio-AGN}\equiv F_{K}(>P_{1.4})\propto L_K^{2.1}$ (cf.\ Fig.~\ref{fig:mauch}). At least $\sim 60$\% ($\sim 100$\% if one includes upper errors) of their NIR brightest, $-26<M_K<-25$, sources are found to have a radio-AGN counterpart down to $\sim 10^{20}$ W Hz$^{-1}$. On the other hand, no correlation is observed between radio luminosities and $K$-band magnitudes, since radio-AGN span a wide range in $L_{1.4\, \rm GHz}$ but are almost all found in the most luminous NIR galaxies.
 
These results have been recently confirmed by the work of \citet{sabater1} who use the LoTTS Data Release 1 (DR1 -- \citealt{shimwell}) cross-matched with the seventh data release of the SDSS (DR7 -- \citealt{abazajian}) to identify with various methods 2021, $z<0.3$, radio-AGN. Their study also shows a strong dependence of the fraction of galaxies hosting a radio-AGN on their stellar mass and radio luminosity. In the most extreme case of $L_{\rm 150 MHz} \ge 10^{21}$ W Hz$^{-1}$ and stellar masses $M_*\ge 10^{11}\,M_\odot$, the analysis of \citet{sabater1} finds that such a fraction reaches the value of 100\% i.e., --  in agreement with the works of e.g., \citet{brown} and  \citet{capetti2} -- one has that \emph{all} local massive galaxies are switched on to a radio-emitting AGN. As in \citet{best4}, \citet{sabater1} also conclude that the probability for a galaxy to host a radio-active AGN has a stronger dependence on its host mass rather than on the mass of the powering black hole, suggesting a tighter physical link between radio-activity and gas fuelling with respect to nuclear properties. On a side, we further note that the same dataset used by \citet{sabater1} indicates that the morphology of the host galaxy is also related to radio-AGN activity, with sources with $L_{\rm 150 MHz} > 10^{23}$  W Hz$^{-1}$ more likely to be found within massive, round galaxies, with respect to fainter radio-AGN hosts which instead exhibit more elongated shapes with a distribution which is indistinguishable from that of inactive galaxies with the same stellar mass content (\citealt{zheng}). This confirms the very early results of e.g., \citet{heeschen} and \citet*{hummel}. 

\citet{smolcic1} repropose the \citet{best4} analysis but at intermediate ($0.3<z<0.7$) and high ($0.7<z<1.3$) redshifts for a sample of 610,  $L_{1.4\, \rm GHz} \ge 10^{23.6}$ W Hz$^{-1}$ radio-AGN selected from the VLA-COSMOS survey (\citealt{schinnerer}).  No evolution from the local results is found in the intermediate-redshift range, while in the high-redshift range the fraction of galaxies host of a radio-AGN is always higher than that derived up to $z\sim 0.7$ for all galaxy masses. This increase is however differential in the sense that low-mass galaxies at $z\sim 1$  present a stronger increment  with respect to higher-mass ones. 
Very similar results are obtained by \citet{williams} who present a study of the evolution of the fraction of radio-active ($L_{1.4\, \rm GHz}\ge 10^{24}$ W Hz$^{-1}$) AGN as a function of their host stellar mass, spanning from $z \sim 0$ to $z\sim 2$. The comparison between local and high-redshift observations is achieved by considering at $z<0.3$ the catalogue of \citet*{best6}, while in the range $0.5<z<2$ a catalogue of $\sim 1500$ radio-AGN resulting from combining the VLA-COSMOS survey (\citealt{schinnerer}) with $Ks<23.4$ sources selected from the COSMOS/UltraVISTA survey (\citealt{muzzin}). Also in this case, it is observed that the fraction of galaxies host of a radio-AGN shows a clear increase with host mass and redshift, with a functional form which goes as $\propto M^{2.7}$ in the local universe and as $\propto M$ at $1.5 <z< 2$ (cf.\ Fig.~\ref{fig:williams}).  This implies that radio-AGN activity amongst galaxies of different mass increases differently towards higher redshifts, with a more marked increment (more than an order of magnitude) observed amongst galaxies of lower ($M_*\simlt 10^{10.75}\,M_\odot$) mass. The radio-AGN fraction within higher mass galaxies ($M_* > 10^{11.5}\,M_\odot$) instead stays roughly constant in time. In agreement with the conclusions of \citet{smolcic1}, these trends are interpreted as being due to a rising contribution of AGN accreting in the radiative mode (see later in this Section) at those earlier epochs thanks to larger reservoirs of gas in lower-mass galaxies.

\begin{figure*}
\centering
\includegraphics[scale=0.4]{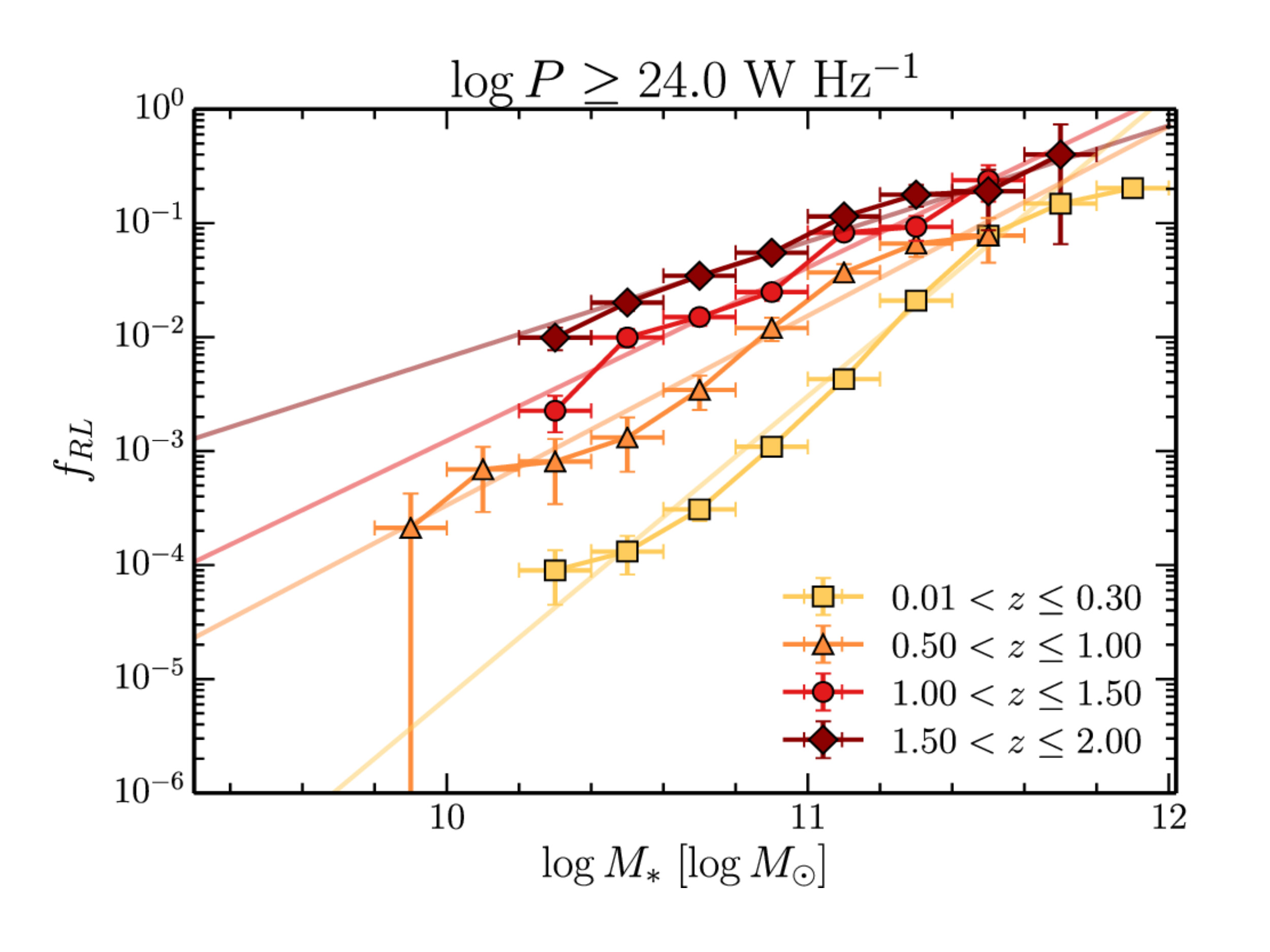}
 \caption{Fraction of galaxies hosting a radio-AGN with $L_{1.4\, \rm GHz} > 10^{24}$ W Hz$^{-1}$ ($f_{\rm RL}$) as a function of their stellar mass  in the four redshift bins highlighted in the plot. The coloured lines show linear fits over the stellar mass range $10^{10} < (M_*/M_\odot) < 10^{11.5}$. Figure from \citet{williams}.}
\label{fig:williams}       
\end{figure*}

In terms of redshift evolution, it is also worth noticing that both \citet{lin3} and \citet{mo2} report a strong increment of the fraction of radio-AGN within galaxies residing in the proximity of cluster centers. In more detail, \citet{lin3} analyse the recurrence of $L_{\rm 1.4 GHz}\ge 10^{24.7}$ W Hz$^{-1}$ radio-AGN detected with FIRST in galaxies belonging to the 400, $0.3<z<1$, brightest clusters from the Subaru Hypersuprime-Cam Survey \citep{aihara}, finding that this is a steep function of their stellar mass. \citet{mo2} obtain a similar result but based on 194, $L_{\rm 1.4 GHz}\ge 10^{25}$ W Hz$^{-1}$, FIRST-detected radio-AGN found within 500 Kpc from the centres of clusters selected at $0.7<z<1.5$ by the Massive and Distant Clusters of WISE Survey (MaDCoWS -- \citealt{gonzalez}). They further report 
that the fraction of radio-AGN within cluster galaxies is factor $\sim 3$ higher than within galaxies of the same mass but in the field. \citet{lin3} instead observe that the increased fraction of radio-AGN within large stellar mass galaxies at high, $z\simgt 0.8$, redshifts is due to blue massive galaxies, at variance with the predominance of red galaxies inhabited by a radio-AGN at lower-$z$. In agreement with e.g., \citet{smolcic1} and \citet{williams}, the authors conclude that this finding hints to a switch in the dominant accretion mode powering cluster radio galaxies, from cold/efficient at high-$z$ to hot/inefficient at low-$z$ (see below). 

Recently, \citet{maglio18} presented an analysis of 100 radio-emitting and 907 X-ray emitting AGN from the two cosmological UKIDSS Ultra Deep Field (UDS) and Extended Chandra Deep Field South (CDFS) regions, which showed that the stellar mass of AGN hosts is a fundamental quantity which determines their level of activity. Optical/NIR counterparts, together with precise redshift estimates (both spectroscopic and photometric) for these sources were taken from the VANDELS survey (\citealt{mclure5, pentericci3}). The depths of both radio (\citealt{simpson, miller1}) and optical/NIR observations ensure completeness up to at least $z\sim 3$. By using this data it was then found not only that all radio-AGN hosts are very massive, $M_* \simgt 10^{10.5}\,M_\odot$ galaxies, but that the probability for an X-ray-selected AGN to also emit at radio wavelengths is a very strong function of its host mass. Indeed, this increases from $\sim 1$\% to $\sim 13$\% when the stellar mass goes from $M_* \sim 10^{10.5}\,M_\odot$ to $M_* \sim 10^{11.5}\,M_\odot$. Within the same mass range, the chances for a radio-active AGN to also emit in the X-ray go from $\sim 15$\% $\sim 45$\%, value which reaches $\sim 78$\% in the deepest CDFS field. These findings, summarized in the bottom panel of Fig.~\ref{fig:hickox} in order to show the variations from the \citet{hickox} diagram for AGN of different masses,  led the authors to conclude that the mass of a galaxy host of an AGN plays a crucial role in determining the AGN level of activity, not only in the case of radio emission alone, but even more so for simultaneous emission in the various bands of the electromagnetic spectrum. 
No dependence was instead found on the cosmological evolution of these sources (i.e., the probability that an AGN will be simultaneously active at both X-ray and radio wavelengths is the same at all epochs) or on their luminosities (radio and/or X-ray), accreting level of the black hole responsible for the AGN signal or even on the star-forming activity of their hosts. 

All the results presented so far clearly indicate that radio emission from an AGN is strongly favoured within galaxies of large stellar mass. In the most extreme case of local and very massive ellipticals there is a general consensus on the fact that almost all them present radio emission at some level, mostly of AGN origin. On the other hand, studies performed at high redshifts indicate that galaxies, especially those with a lower stellar mass content, were more prone to host a radio-AGN than those in the local universe. The reason for such a differential trend in host mass is not totally understood, but could be explained by invoking different accretion properties (and different cosmological evolutions) for radio-AGN within galaxies of different mass. In this light, a possible connection could be envisaged with the two populations of HERGs and LERGs. As already briefly presented in the Introduction, radio-AGN can be divided into high-excitation radio galaxies (HERGs) and low-excitation radio galaxies (LERGs) according to the absence or presence and relative importance of emission lines in their optical spectra  (\citealt*{hine}).  
HERGs (which also include quasars) are believed to possess high accretion rates onto their super-massive black holes (producing a total luminosity $>0.01\; L_{\rm Edd}$, where $L_{\rm Edd}$ is the Eddington luminosity), as opposed to LERGs for which $\simlt 0.01\; L_{\rm Edd}$ (see \citealt*{heckman} for a review). It is further argued (e.g., \citealt*{hardcastle}) that the difference between HERGs and LERGs is dictated by different sources of fuel. In the first case black holes would accrete cold gas via an accretion disc which produces radiation in an efficient manner, while LERGs would be fuelled by the accretion of hot gas from the haloes of their host galaxies and surrounding environments through advection-dominated flows (ADAFs, e.g., \citealt*{narayan}) and release the accretion energy in the form of jets or winds (e.g., \citealt*{merloni1}).

Given the differences between the two populations of HERGs and LERGs, it is then worth investigating whether these also reverberate onto their host galaxies. Indeed, in the recent years, more and more works have considered the host properties of these two populations separately.
\citet*{heckman} (see also \citealt{tadhunter1}) present a detailed review of the LERGs vs HERGs distinction, so we refer the reader to their work. Here we will just briefly summarize the most important findings and concentrate on the more recent results which are mainly obtained at high redshifts. It is observed that in the local universe HERGs appear within systematically smaller, lower-mass  and less concentrated galaxies than LERGs and also exhibit higher levels of star-forming activity (e.g., \citealt{smolcic, smolcic4, best6}). Furthermore (\citealt{jannsen}), HERGs are preferentially found within green (i.e., in the process of being quenched) and blue galaxies, and -- at all radio luminosities -- the probability for a galaxy to host a LERG exhibits a steeper dependence on the mass ($\propto M_*^{2.5}$) with respect to a HERG  ($\propto M_*^{1.5}$). These results are also confirmed by IR observations which clearly show that the vast majority of LERGs is associated with galaxies with WISE (\citealt{wright}) colours that identify them as `early-type', as opposed to HERGs whose hosts are preferentially `late-type' (\citealt{sadler}), and also by the fact that HERGs are typically about four times more luminous than LERGs in the FIR, 250 $\mu$m band (\citealt{hardcastle2}).

On top of these differences, LERGs and HERGs are also found to present different cosmological evolutions at least up to $z\sim 1$ (e.g., \citealt{best7, pracy, butler2}), with LERGs substantially not evolving (or even showing negative evolution, at least for low-luminosity sources, \citealt{best7}) and HERGs presenting a strong degree of positive evolution (e.g., $\propto (1+z)^7$ for a pure luminosity evolution model in the work of \citealt{pracy}). It is to be noted though that there is some inconsistency between the different degrees of evolution as found for the two classes of sources by the various works. Also, cf.\ \citet{williams2} who find in the range $0.5<z<2$ no evolution for HERGs and a negative evolution at all radio luminosities for LERGs. 

From the above discussion it then appears that in the relatively local, $z<1$,  universe the hosts of LERGs and HERGs are quite different.
Things however seem to change at higher redshifts. \citet{fernandes} use \textit{Spitzer} information combined with a SED-fitting approach on a sample of 27, $z\sim 1$ radio-AGN spanning the range in radio luminosity $10^{25}<L_{\rm 151 MHz}/[{\rm W Hz^{-1}}]<10^{28}$ to find that HERG and LERG hosts do not differ from each other, since they all present large  ($10^{10.7}\le M_*/{\rm\,M_\odot}\le 10^{12}$) stellar masses. The only dichotomy shown by the \citet{fernandes} sample is in accretion modes, whereby HERGs accrete at a much higher rate ($\lambda/\lambda_{\rm Edd}\simgt 0.1$), while virtually all LERGs cluster around the value $\lambda/\lambda_{\rm Edd}\sim 0.01$ (see also \citealt{gurkan} for similar results on accretion rates based on WISE observations).

\citet{delvecchio} consider radio sources from the VLA-COSMOS 3 GHz Large Project (\citealt{smolcic3}) down to a 1$\sigma$ sensitivity of $2.3$ $\mu$Jy. Available excellent quality multi-wavelength information up to $z\sim 6$ allows to sub-divide the sample into 1604 high radiative luminosity AGN (HLAGN, close relatives of HERGs) and 1333 low radiative luminosity AGN (MLAGN, close relatives of LERGs) which are then investigated at the different cosmological epochs. Although a number of differences are observed between these two classes of sources also as a function of redshift, when only radio-active HLAGN (i.e., the 30\% of the HLAGN population as selected by \citealt{delvecchio} which emit at radio wavelengths because of the presence of a radio-active AGN and not ongoing star-formation within the hosts) are considered, the \citet{delvecchio} analysis shows that MLAGN and HLAGN host properties are largely overlapping, with basically no difference in either star-forming activity or (large) mass distributions at all cosmic epochs.  Also, no difference between the distributions of mechanical powers (obtained from  AGN-related radio emission) for these two classes of sources is observed.

\citet{butler} apply similar methods to those described in \citet{delvecchio} to select at 2.1 GHz and up to $z\sim 4$ 1729 LERGs, 1455 HERGs (out of which 296 are defined as radio-quiet by the authors, i.e., with radio emission powered by star-formation) and 558 star-forming galaxies (SFGs) from ATCA observations of the XMM extragalactic survey south field (XXL-S -- \citealt{pierre, butler1}). 76\% of the original radio sample is endowed with multi-wavelength information. According to this study, LERGs and radio-active HERGs tend to reside within hosts with substantially the same stellar mass content, gathering around a median value of $\langle M_*\rangle\sim 10^{11}\,M_\odot$, except for a more pronounced tail exhibited by the distribution of HERGs at lower stellar masses. Larger differences are instead observed in the star-forming properties of these sources, whereby LERGs are mainly found amongst lower star-formation (about a factor 5 less) and redder galaxies than HERGs. However, since the \citet{butler} analysis has been carried out without a distinction between more local and more distant sources, it is not possible to discern whether their results are due to intrinsic properties of the two HERG and LERG populations or whether these originate from some cosmological evolution. Indeed, LERGs in the \citet{butler} sample dominate below $z\sim1.2$, while they almost totally disappear at higher redshifts, when only HERGs are found. 

\citet{williams2} instead consider 624 radio-AGN with $L_{\rm 151 MHz}\ge 10^{25}$ W Hz$^{-1}$ in the redshift range $0.5\le z\le 2$ as observed with LOFAR on the Bo\"otes field (\citealt{williams1}). Multi-wavelength information for these sources is used to classify 297 HERGs, 138 LERGs and 199 SFGs. The relative relevance 
of these populations varies according to the considered redshift range (more HERGs and SFGs  beyond $z=1$). At all redshifts, LERGs appear in massive, $M_*\simgt 10^{11}\,M_\odot$, galaxies. Also HERGs are mainly hosted within massive galaxies, but their distribution presents a tail towards lower, $M_*\simlt 10^{10.5}\,M_\odot$ masses, especially beyond $z\sim 1$. Furthermore, while below $z\sim 1$ the $u-r$ colours of the galaxies host of LERGs and HERGs show marked differences, with LERGs being systematically redder than HERGs, at higher redshifts there is a larger overlap since HERGs are found within galaxies of all SFRs and LERGs shift to more star-forming systems. On the other hand, while the fraction of LERGs within the general galaxy population 
show no cosmological evolution -- result that suggests the same fuelling mechanism for this class of sources at all $z\simlt 2$ (\citealt{best7}) -- that of HERGs instead increases with redshift by a factor $\sim 3$ at all stellar masses. Also the functional form of the dependence of this fraction on the host mass is different, since \citet{williams2} find $F_{\rm radio-AGN}\propto M_*^{2-2.5}$ for LERGs, while in the case of HERGs the slope is much flatter, going from the $z\simlt 1$ value of $\sim 0.5$ to $\sim 1.2$ at $z\sim 2$. Such a differential evolution implies that beyond $z\sim 1$ the fraction of HERGs will always be higher than that of LERGs, except for the most massive galaxies which will be as likely to host a LERG as a HERG.

All the works considered so far converge at indicating that locally LERGs and HERGs do not only differ in terms of their accretion properties but also for what concerns their hosts, which are generally less massive, more star-forming and bluer for HERGs with respect to LERGs. The situation is however less clear at redshifts $z\simgt 1$, since there seems to be no agreement between conclusions originating from different studies. Indeed, in some cases it is found that HERGs are generally associated to more star-forming galaxies than LERGs (result which once again would hint to the need of cold gas in order to trigger efficient accretion), while in others it is instead reported a substantial overlap between the colours (and therefore star-forming properties) of these two populations. Furthermore, there does not seem to be much of a difference for what concerns the masses of their hosts, possibly except for a tail in the distribution of HERGs below $M_*\sim 10^{10.5}\,M_\odot$. These results which generally point to a less marked distinction between LERGs and HERGs with respect to what observed in the more local universe, likely stem from the cosmological evolution of the general galaxy population. Indeed, not only all galaxies become more star-forming as cosmic noon is approached, but also star-formation activity preferentially shifts to more massive systems (the so-called \emph{cosmic downsizing}, e.g., \citealt{cowie}). Another issue that should also be taken into account when comparing local with high-redshift (and also high-redshift with high-redshift) results is that, while in the local universe the LERG-HERG classification is based on the spectral properties of these sources, for $z\simgt 0.5$  it mainly has to rely on less precise multi-wavelength photometric information and on various selection methods. This gives rise to samples of LERGs and HERGs that might not only differ from each other, but that are also more contaminated than local ones.



\subsection{What causes the FRI-FRII dichotomy?}
\label{sec:2.2}
From the radio morphological point of view, FRI and FRII galaxies look very different, with the first population appearing as `edge-darkened', while the second one -- generally more luminous and more extended (up to Mpc scales) than FRIs and presenting small-scale high surface brightness regions (the so-called hot spots) -- as `edge-brightened'. 
The remarkable result obtained by \citet*{fanaroff} was that this morphological classification was strongly related to radio luminosity, with sources fainter than $L_{408 \rm MHz}=10^{25}$ W Hz$^{-1}$ being almost all FRIs, while for $L_{408 \rm MHz}>10^{27}$ W Hz$^{-1}$ the overwhelming majority of the sources showed FRII morphologies. Whether this dichotomy, and radio morphology in general, are also the result of factors other than radio luminosity has been the subject of debate for many years (e.g., \citealt{fabbiano}). In this section we mainly aim at investigating the role of the host galaxies. Eventual differences in their environmental properties will then be discussed in Sect.~\ref{sec:3}.

Early works found that, while FRI galaxies were associated to giant ellipticals, the hosts of the radio-brighter FRIIs presented lower optical luminosities and smaller sizes 
(e.g., \citealt*{lilly, lilly2, owen1, owen2, baum, ledlow1, zirbel1}). Their main conclusion was that, since FRIs were also mostly associated with the dominant galaxy (roughly coinciding with the BCG) in relaxed groups or clusters of galaxies, while FRIIs mainly appeared in disturbed galaxies, then these two classes of sources could not be different evolutionary stages of a single object. One possible explanation for this different behaviour (e.g., \citealt{heckman1}) was attributed to different fuelling mechanisms, with FRII galaxies being the product of collision or merger of galaxy pairs, while FRIs would result from accretion of an intracluster medium. On the other hand, other works (e.g., \citealt*{zirbel, baum}) suggested that intrinsic different properties of the central engine should be the main driver of the difference between FRI and FRII galaxies. On top of this, \citet*{ledlow2} (see also \citealt{ledlow3}) reported one of the first evidence for the existence of powerful ($L_{1.4 \rm GHz} > 10^{24}$ W Hz$^{-1}$) FRIs hosted by spiral galaxies, proving that -- although enormously more common -- not all FRIs reside in elliptical galaxies even in the very local universe. It was then clear that  the distinction between FRI and FRII sources was everything but a clear-cut case.

In a seminal paper, \citet{ledlow1} consider observations taken from the literature in the redshift range $z=0.01-0.5$ and for different environments to investigate the dependence of radio morphology on both radio luminosity and optical luminosity of the hosts. It was found that the FRI/II division was a strong function of the optical luminosity of the host galaxy, with radio-AGN above the dividing line $L_{\rm radio}\propto L_{\rm optical}^{1.8}$ being all FRIIs, while those below it being all FRIs. This implied that FRI galaxies could also be found at radio powers equivalent to FRII sources but in galaxies one or two magnitudes brighter (cf.\ left-hand panel of Fig.~\ref{fig:ledlow}), and more importantly suggested that ``the optical luminosity and the properties of the host galaxy are the most important parameters which affect radio source formation and evolution'', even though no strong dependence of the AGN radio luminosity on the host optical luminosity was reported. On the other hand, in the same work it was also noticed that when considering a more homogeneous and local sample of sources within a cluster environment, some (although very few) FRII galaxies lay below the FRI/FRII dividing line (cf.\ right-hand panel of Fig.~\ref{fig:ledlow}). Not much could be however concluded from this finding, given the general paucity of FRII sources. The plot of \citet{ledlow1} quickly gained high visibility and \citet{ghisellini} converted it into an AGN-power vs black hole mass relation, with the dividing line between FRI and FRII sources representing a constant ratio between AGN power and Eddington luminosity of the black hole. This suggested the FRII-FRI dichotomy to be determined by a change in the accretion mode, from a radiatively efficient to an inefficient one, possibly in connection with the ageing of the FRII population. We will get back to the ageing issue later in this Section. 

\begin{figure*}
\centering
 \includegraphics[scale=0.23]{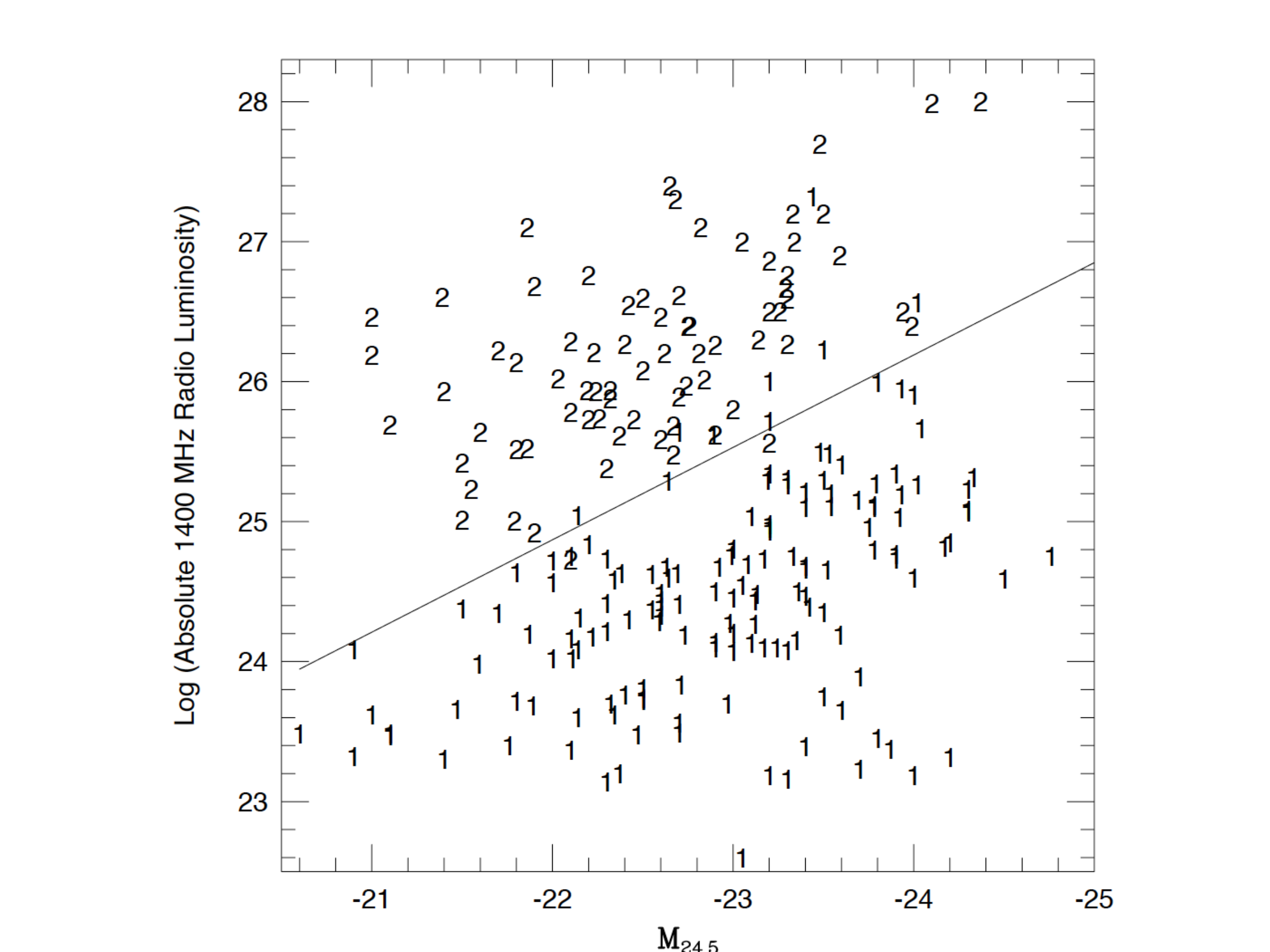}
 \includegraphics[scale=0.23]{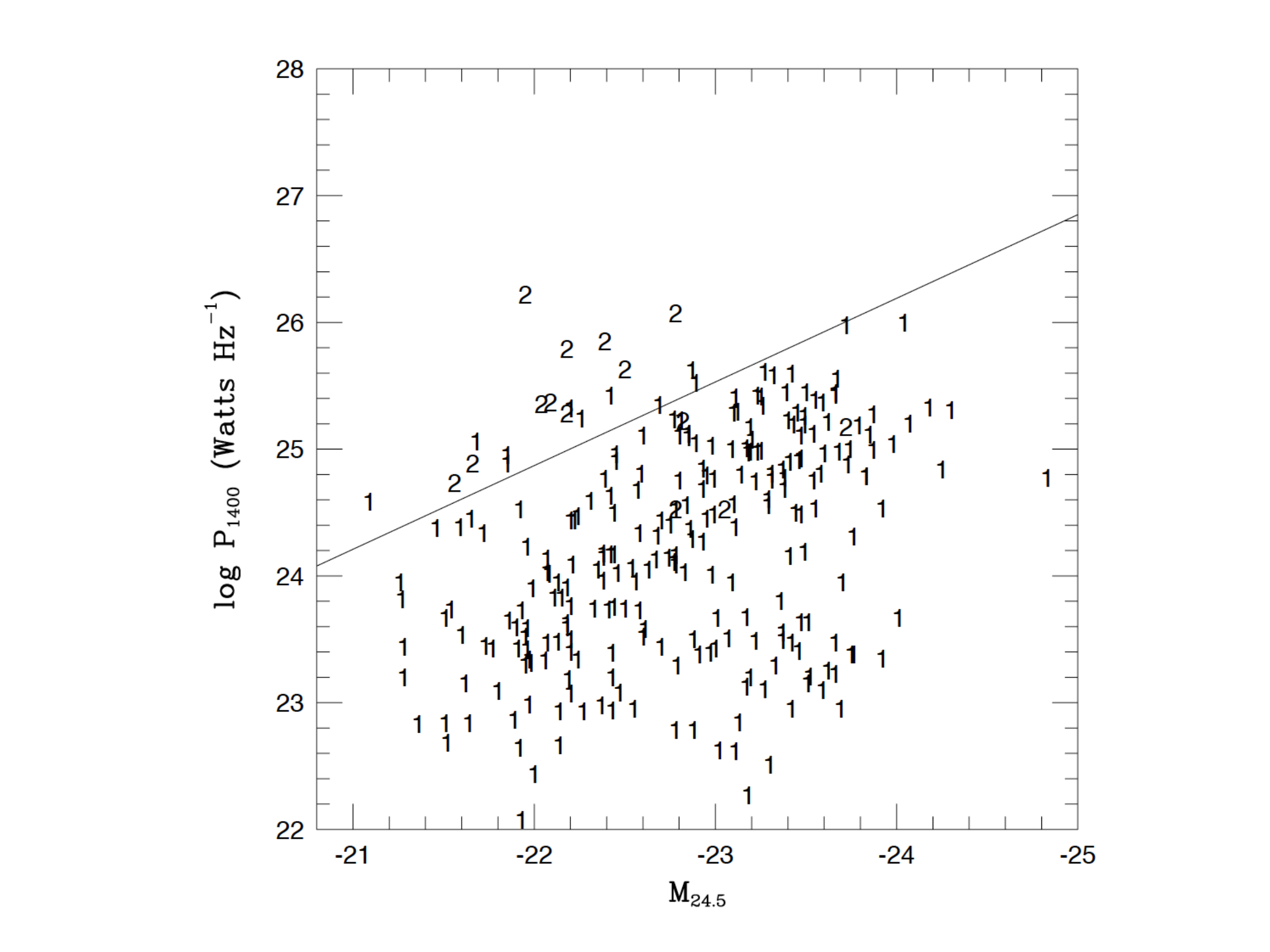}
\caption{Left-hand panel: FRI/II diagram for a sample of objects taken from the literature. Right-hand panel: FRI/II diagram for radio-AGN within the \citet{ledlow1}, $z<0.25$ cluster sample. Figures from \citet{ledlow1}.}
\label{fig:ledlow}       
\end{figure*}

As already noted by the authors themselves, the work of \citet{ledlow1} though suffered from a number of biases due to the fact that they used a highly heterogeneous sample of sources taken from the literature. As a result of that, the population of FRI galaxies lay at redshifts $z<0.1$, while FRIIs were generally found beyond $z=0.25$. Due to a strong redshift-luminosity correlation, this generates a Malmquist bias (only brighter sources are visible at higher redshifts) which is very difficult to disentangle from physically-based effects (e.g., \citealt{singal}). \citet{best5} repeated the \citet{ledlow1} analysis for their much more homogeneous sample of $\sim 1000$ local FIRST and NVSS galaxies endowed with a SDSS counterpart which exhibited extended radio morphologies. Although most of the sources above the \citet{ledlow1} limit were indeed FRII galaxies, while below it they were more likely FRIs, a very large overlap between these two populations in the region below the dividing line was observed. \citet{best5} showed that FRIIs in the overlapping region were on average smaller in the radio maps than FRIs. No further difference was instead found in terms of black-hole or host galaxy, apart from a tendency for the hosts of FRIs to be more extended. According to \citet{best5}, this led to the conclusion that more massive and more extended galaxies are more likely to disrupt the radio jets, therefore leading to an FRI morphology. 

\citet{lin2} also consider a homogeneous sample of $\sim 1000$, FIRST \& NVSS $z<0.3$ radio-AGN classified according to both their radio morphology and nuclear emission line activity. Similarly to the \citet{best5} (but see also \citealt{wing}) results, they find that FRI and FRII sources overlap in their host galaxy properties, except for the sub-class of very extended and high emission-line luminosity FRIIs which are hosted by lower mass galaxies, live in relatively sparse environments, and likely have higher accretion rates onto their central supermassive black hole. Based on these results, \citet{lin2} discuss the possibility of radio morphologies being determined by the joint effect of nuclear accretion (playing the primary role) and environment, with bright FRIIs originating from high-accretion rates and FRIs from low-accretion rates within dense galactic structures. 

An entirely different point of view was instead provided by \citet{sadler} who analyse 202, $<z=0.058>$ radio-AGN selected at 20 GHz with counterparts from the 6dFGRS. Although with some overlap between the two classes of sources (46 FRIs and 16 FRIIs), \citet{sadler} indeed confirm the \citet{ledlow1} relation to hold even at high frequencies. Furthermore, these authors observe that the FRIIs and FRIs belonging to their sample exhibit marked differences in terms of MIR/WISE (\citealt{wright}) colours of the host galaxies, as $\sim 93$\% of FRIs were found within the population of early-type galaxies, while $\sim 93$\% of FRIIs were found to be associated with late-type (i.e., star-forming galaxies and quasars) galaxies. This clearly indicated a near-complete dichotomy between the hosts of FRI and FRII sources. 

The reason for such a disagreement between the findings of \citet{sadler} and those of e.g., \citet{best5} and \citet{lin2} is unclear, as all of them refer to homogeneous samples of local galaxies so that systematic biases should be minimal. Part of the difference could be due to the different flux limits (40 mJy for the AT20G survey used by \citet{sadler}  vs 2.5 mJy for NVSS) and to slightly different classification schemes adopted in the different works. On the other hand, another possibility could be the selection at different frequencies, since it is known that -- unlike those carried out at $\nu \le 1.4$ GHz -- high-frequency surveys preferentially include flat-spectrum sources (e.g., \citealt{dezotti, massardi}). To test for this effect, we can refer to the results of \citet{vardoulaki} who investigate the FRI-FRII dichotomy by considering 59 FRII, 39 FRI and 32 hybrid FRI/FRII galaxies spread between redshift 0 and $\sim 2.5$ (peak in the distribution at $z\sim 1$), selected at  3 GHz in the COSMOS field (\citealt{smolcic3}). However, despite a selection frequency higher than 1.4 GHz (although not by a large amount), at variance with \citet{sadler}, no difference either in the hosts or in the environments (cf.\ Sect.~\ref{sec:3.2}) of these two populations was reported. It should however be kept in mind that the \citet{sadler} analysis was carried out for very local sources, while that of  \citet{vardoulaki} embraces objects up to $z\sim 2.5$, therefore including any possible cosmological evolution of the FRI and FRII populations.
 Interestingly enough though, \citet{vardoulaki} find that as many as $\sim 50$\% of the sources in their sample classified as FRIIs at 3 GHz are recognised as FRIs at 1.4 GHz. The authors attribute this effect to the combination of higher sensitivity and resolution of the 3 GHz sample with respect to those of other 1.4 GHz surveys, or alternatively to synchrotron ageing/losses, although the small frequency difference disfavours this second option. Whatever the reason, given the unexplained disagreement between the FRI vs FRII results obtained at 20, 3 and 1.4 GHz, this is an issue that should be carefully better investigated with further data to come.

\begin{figure*}
\centering
\includegraphics[scale=0.4]{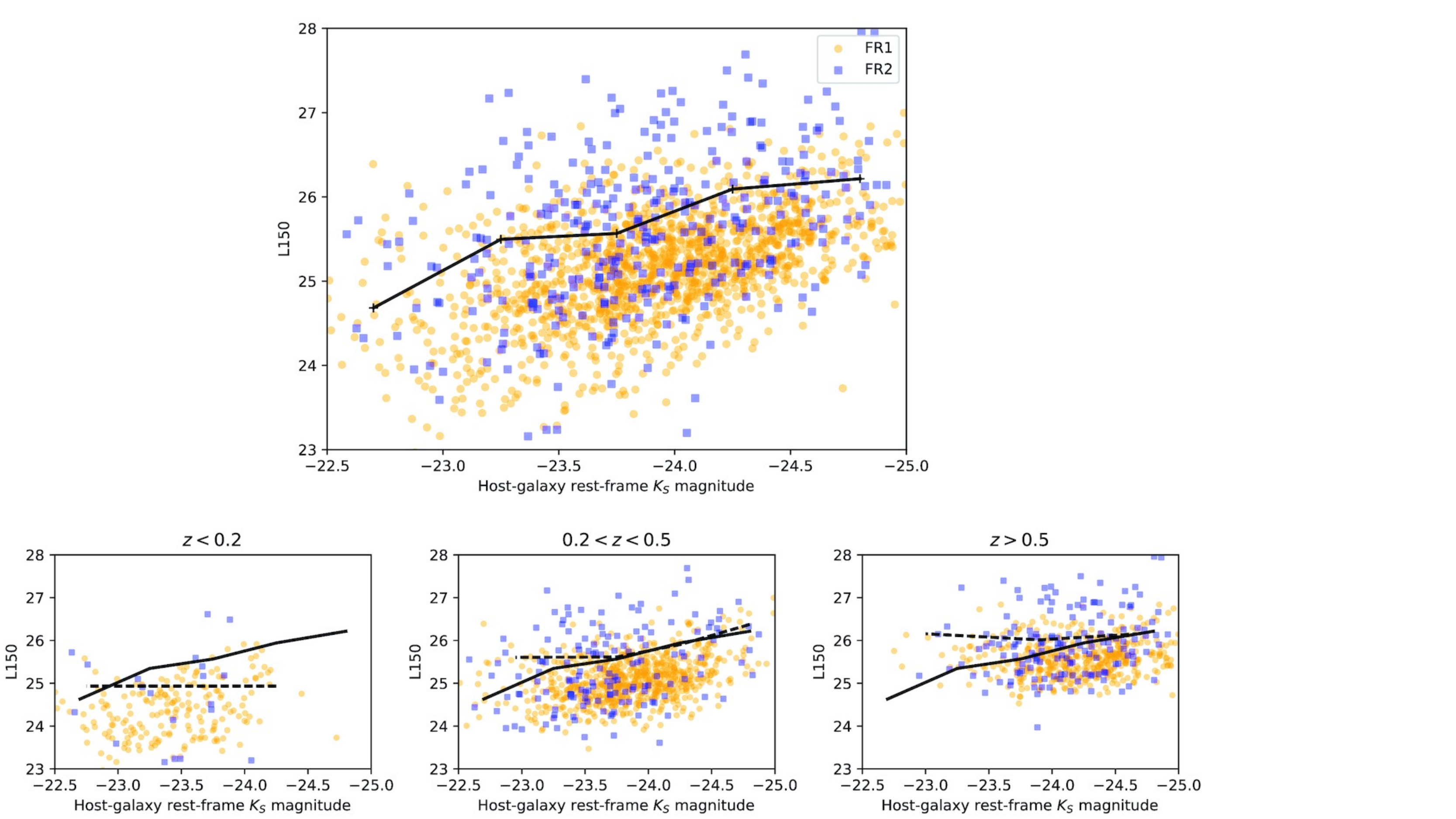}
\caption{Top: the relationship between radio morphology, radio luminosity, and host-galaxy magnitude (a `Ledlow \& Owen' plot). The black line indicates the luminosity above which the normalized probability of finding an FRII exceeds that of finding an FRI. Bottom row: the same sample split into three redshift bins, with dashed lines indicating the break luminosity determined for each redshift slice, and the solid lines showing the full-sample relation as in the upper panel. Figures from \citet{mingo1}.}
\label{fig:mingo}       
\end{figure*}

\citet{mingo1} provide the most comprehensive analysis to date of the FRI-FRII dichotomy based on both radio luminosity and host properties. Their sample include 5805 extended radio-AGN from the LoTSS DR1. 1213 FRIs and 345 FRIIs were subsequently identified via automatic tools complemented by visual inspection at $z<0.8$, where this redshift limit was chosen to minimise selection biases, including that of host galaxy coverage. Their analysis clearly shows that there is a huge overlap between the properties of the two FRI and FRII populations. Indeed, while about only 10\% of FRIs lie above the radio luminosity break identified by \citet*{fanaroff}, as many as 63\% of FRIIs are found below that limit. FRII galaxies appear to fill the entire \citet{ledlow1} plane, just as a non-negligible fraction of FRIs is located on its top half (top panel of Fig.~\ref{fig:mingo}). The relevant sub-population of low-luminosity FRIIs (i.e., FRII galaxies below the \citealt{ledlow1} dividing line) was further compared to its high-luminosity counterpart matched in redshift and distribution of physical sizes, finding that both samples share the same properties, except for host $K$-band luminosities (proxy for stellar mass): low-luminosity FRIIs are observed to inhabit fainter (i.e., smaller) galaxies than those hosting higher-luminosity FRIIs. Based on this result, the authors conclude that the majority of radio-faint FRIIs are indeed sources inhabiting relatively sparse environments that prevent their radio jets from being disrupted, at variance with what is expected to happen to FRI sources which are mainly found to reside within overdense structures. \citet{mingo1} explain the general absence of low-luminosity FRIIs from early works as due to the combination of high flux limits of the e.g., 3CR survey (\citealt{bennett}, see \citealt*{laing} for an updated version) on which these previous studies were mostly based, together with the rarity of such sources in the local universe (bottom panels of Fig.~\ref{fig:mingo}).

\subsubsection{HERGs vs LERGs and the FRI-FRII dichotomy} 
So far we have only concentrated on radio morphology, with the distinction into FRI and FRII classes of sources. However, as already seen in Sect.~\ref{sec:2.1}, radio-AGN can also be sub-divided into HERGs and LERGs. To a zero-th order approximation, FRI galaxies are closely connected to LERGs while FRIIs are strictly related to HERGs. Indeed, in most cases, FRIs and LERGs share the same host properties just as FRIIs and HERGs do, with FRIs and LERGs generally appearing within massive red and passive ellipticals, while FRIIs and HERGs within relatively lower-mass, bluer star-forming galaxies (e.g., \citealt{Baldi, smolcic, buttiglione, jannsen}; \citealt*{best6}; \citealt{hardcastle2}). However, differences exist, in the sense that it is not unlikely to find mixed populations of sources, especially FRIIs which are also LERGs (e.g., \citealt{laing1, tadhunter, hardcastle5, sadler, capetti1, mingo2}). It is then worth investigating the properties of FRI and FRII galaxies also in connection with their HERG/LERG distinction. 

For instance, it has been known for some years (e.g., \citealt*{chiaberge, chiaberge1}; \citealt{baldi2}) that the FRI-FRII distinction is not univocally connected with the optical properties of the nuclei of these sources, since while FRIs of all luminosities are a substantially homogeneous population -- with optical/NIR and radio core luminosities being strongly correlated -- FRIIs show a much more complex behaviour, with $\sim 30-50$\% of them exhibiting nuclear properties typical of FRI galaxies. This indicates a common structure of the central engine despite the differences in radio morphology and radio power. More recently, \citet{buttiglione} presented spectroscopy for a $z<0.3$ sample of 104 3CR sources. And while they do reproduce the \citet{ledlow1} result (but with the caveat of sampling radio-bright sources), they also find a large fraction ($\sim 30$\%) of FRII galaxies associated with LERG nuclei, with LERGs spanning all the probed radio luminosity range.  On the other hand, HERGs were only observed in association with radio-bright FRII sources. About $\sim 30$\% of HERGs showed prominent broad lines in their optical spectra, while this does not happen in any of the LERGs. A substantial superposition of the host galaxy luminosities is also reported. Based on these results, \citet{buttiglione} conclude that the LERG vs HERG dichotomy is dictated by different accretion modes set by the initial temperature of the infalling gas (hot vs cold, see also \citealt*{best6}). 

\citet*{capetti,capetti1} present two catalogues of 219 FRI and 122 FRII galaxies selected at $z< 0.15$ from the \citet*{best6} sample. In agreement with some of the previous works, they found that, while the population of FRI hosts is remarkably homogeneous, being composed by all luminous red early-type galaxies, spectroscopically classified as LERGs and containing large mass black holes, FRIIs do behave very differently. First of all, in the \citet{capetti1} sample, only 10\% of FRIIs are HERGs, all the remaining sources being classified as LERGs. Furthermore, 75\% of them lie below the dividing line identified by \citet{ledlow1}. FRII LERG hosts also exhibit properties which are very similar to those of FRIs, being once again all luminous red elliptical galaxies, hosting massive ($M\simgt 10^8\,M_\odot$) black holes. Even the median radio luminosity of FRII LERGs was observed to be only a factor $\sim 3$ higher than that measured in FRIs. On the other hand, in agreement with \citet{lin2}, the few FRII HERGs are found to be very different from the other two sub-populations, since they are associated with lower mass black holes and with hosts which are fainter, optically bluer and MIR redder. At variance with the \citet{ledlow1} but also with the \citet{mingo1} results, the main conclusion of \citet{capetti1} is that radio morphology is not related to different properties of the host galaxy, except in the rase cases of FRII HERGs. Similar results on the independence of radio morphology on the host galaxy properties has been recently obtained by \citet{jimenez}  by comparing their sample of local, small extension ($< 60$ Kpc), FRII galaxies to that of FRI galaxies from \citet{capetti}, as both populations are spectroscopically classified as LERGs and 
found to be hosted in red, early-type galaxies.

\citet*{miraghaei} also consider FRI and FRII galaxies and their sub-division into LERGs and HERGs. The FRI/FRII dichotomy was investigated by studying a sample of 108, $z<0.1$ FRI LERGs and FRII LERGs with the same stellar mass and total radio luminosity distribution, drawn from the FIRST and NVSS catalogues with a flux cut of $F_{1.4 \rm GHz} >40$ mJy. At variance with the results of \citet{capetti1} and \citet{massaro, massaro1}, it is found that FRIs reside in richer environments (cf.\ Sect.~\ref{sec:3}), are hosted by smaller galaxies with higher concentration, higher mass-surface density and higher black hole-to-stellar mass ratio than FRIIs. These findings lead the authors to conclude that their data prefers models employing extrinsic parameters (i.e., jet disruption by the interstellar and intergalactic media) rather than intrinsic ones (e.g., nuclear or jet content) to explain the FRI/FRII dichotomy. At the same time, HERGs are found in more star-forming and disky galaxies, supporting the theory that AGN fuelling source is the main origin of the HERG/LERG dichotomy. 

\begin{figure*}
\centering
\includegraphics[scale=0.4]{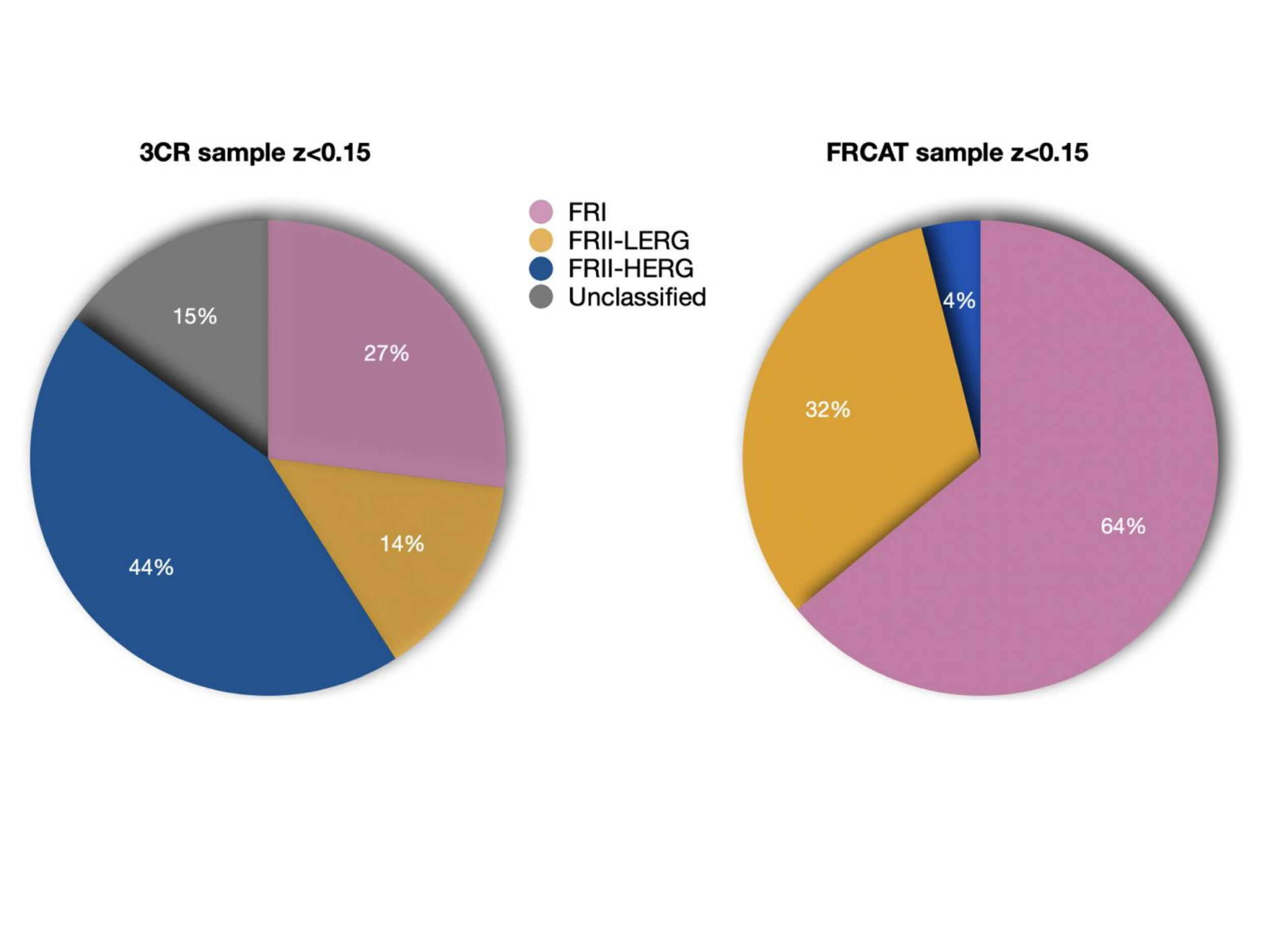}
\caption{Fraction of FRI and FRII radio galaxies with $z < 0.15$ in the 3CR (left panel) and in the \citet{capetti, capetti1} (right panel) catalogues. The relevance of LERGs dramatically increases moving down to fainter sources. Figure from  \citet{grandi}.}
\label{fig:grandi}       
\end{figure*}

A still different point of view is that provided by \citet{macconi} and \citet{grandi} who investigate the FRI-FRII dichotomy from the jet-accretion system point of view, the former authors concentrating on brighter 3CR sources, \citet{grandi} instead re-considering the \citet{capetti, capetti1} samples. After restating that in the local, $z\simlt 0.15$ universe lowering radio fluxes leads to the appearance of a large number of FRII LERGs which were basically unobserved in brighter surveys (cf.\ Fig.~\ref{fig:grandi}), \citet{grandi} find that the majority of the analyzed radio sources (excluding the very few FRII HERGs) are at a late stage of their life, both in terms of host properties (which exhibit evolved stellar populations) and accretion capabilities (more massive black holes producing an inefficient engine). This evidence makes the authors envisage an evolutionary scenario whereby FRII-LERGs are aged FRII-HERGs (see also \citealt{tadhunter1}), even if other explanations such as intrinsic differences related to the black hole properties (e.g., spin or magnetic field at its horizon) cannot be excluded. According to this theory, once the nuclear cold fuel has been consumed, the accretion configuration becomes hot and inefficient, even though the extended radio structures still maintain traces of the past activity. Eventually, the ageing process will move along the lobes and these sources will ultimately turn into FRIs.

At the time of writing, the last word on the FRI-FRII vs LERG-HERG dichotomy is that presented by \citet{mingo2}, who consider a sample of 287 (161 FRIs and 126 FRIIs), $z\simlt 2.5$ sources, selected within the LOFAR Two Metre Sky Survey Deep Fields dataset (\citealt{tasse1, kondapally, duncan, sabater2}) to have large ($>27$ arcsec) separations between their components. Reliable SED information on which to base the LERG/HERG classification is also provided (Best et al. 2022 submitted). 21 sources (out of which 19 FRIIs) are found with sizes grater than 1 Mpc (the so-called giant radio galaxies or GRGs, e.g., \citealt{dabhade}), showing a non-negligible contribution of this sub-population to the extended source radio counts in deep enough radio surveys. In agreement with previous works, 95\% of FRIs are found to be LERGs, while 29\% of FRIIs are HERGs, the remaining being associated to the sub-class of LERGs. According to the authors, the fact that LERGs are the dominant sub-population across all morphological types and radio luminosities argues against either accretion mode or jet power as the cause for large-scale radio-morphology.  Further, by analysing a sample of FRI and FRII galaxies within the same radio luminosity range, it has been observed that FRIIs inhabit less massive galaxies than FRIs. At the same time, HERGs are strongly favoured within systems with high specific star formation rates (sSFRs). Based on these pieces of evidence, \citet{mingo2} conclude that it is the properties of the host galaxy that determine the characteristics of the radio-AGN inhabiting its center, with mass mainly controlling the FRI-FRII dichotomy and gas availability (as inferred from the sSFR) in charge of the LERG vs HERG distinction. 

With the data currently at hand, which also includes evidence for a class of hybrid double sources, with a FRI jet on one side and a FRII lobe on the other (e.g., \citealt*{harwood} and references therein) and also examples of FRI galaxies with clear quasar nuclei (e.g., \citealt*{heywood}), it is clear that we are far from reaching a consensus on the physical processes that determine the different morphologies in radio-AGN. 
In particular, two distinct scenarios for the FRI-FRII dichotomy have emerged. The first one attributes differences in radio morphology to the large-scale properties (i.e., host galaxy and environment) of radio-AGN which are expected to influence the interaction of the radio jet with the external medium (e.g. \citealt{ledlow1, zirbel1}; \citealt*{kaiseraltro, wing}; \citealt{miraghaei, mingo1, mingo2}). According to this scenario, differences between HERGs and LERGs would instead be dictated by different fuelling mechanisms, with HERGs powered by accretion of cold gas, provided by e.g., a recent merger with a gas-rich galaxy (e.g., \citealt{jannsen}), while LERGs accrete hot intergalactic gas from dense environments at a low rate, with fuelling from major mergers being strongly disfavoured by recent observations (e.g., \citealt*{ellison}).\\
On the other hand, other works do not observe any difference in the host and/or environmental properties of FRI and FRII galaxies except in the rare cases of FRII HERGs (e.g., \citealt{lin2, capetti1, jimenez, massaro, massaro1, vardoulaki}), so that an alternative mechanism for their large-scale radio behaviour has to be invoked.  In this second scenario, ageing processes can be thought as the main driver for the observed morphological differences, with an evolutionary pattern that proceeds from FRII HERGs that switch from efficient to inefficient accretion due to gas starvation and transform themselves into FRII LERGs, sources that still maintain their large radio structures thanks to the past nuclear activity at high efficiency (e.g., \citealt{ghisellini, tadhunter1, macconi, grandi}). The switch off/change in accretion mode will eventually show in the radio morphology with the delay needed to reach Kpc-to-Mpc distances, and the source will ultimately turn into an FRI galaxy.

\subsection{Are radio-AGN all hosted in red and dead galaxies?}
\label{sec:2.3}
As we already saw in Sect.~\ref{sec:2.1}, at low -- $z\simlt 1$ -- redshifts, galaxies hosting radio-AGN are dominated by well evolved stellar populations (age $\sim 8-14$ Gyr -- e.g., \citealt{nolan, best4}), exhibit a general absence of ongoing star-formation (e.g., \citealt{siebenmorgen, dicken1}) or -- when present --  normally lower than the hosts of their radio-quiet counterparts (e.g., \citealt{hardcastle1, chen, virdee, gurkan1, pace}), have rather low dust/gas masses (e.g., \citealt*{knapp}), and lie in the same region of the fundamental plane as normal inactive ellipticals (e.g., \citealt{mclure}). 
Although differences according to radio-AGN type have been found  -- with HERGs being on average associated to a factor $\sim 2-3$ higher star-forming activity than LERGs (e.g., \citealt{hardcastle2, gurkan1}) -- the above pieces of evidence have prompted the introduction of the so-called \emph{radio-mode} feedback in theoretical simulations in order to halt star-formation in massive galaxies and produce the locally observed `read and dead' population of ellipticals (e.g., \citealt{croton, bower, best9, fanidakis, weinberger, smolcic6}), both in terms of physical properties and number density. According to this paradigm, the accretion of matter onto a radio-active AGN does not lead to a powerful radiative output, but rather to the production of highly energetic jets that can heat (or even expel) the cold gas within galaxies and suppress star-formation. 

Cosmological evolution for these sources is however expected from theoretical grounding, and since long-standing evidence suggests that the most of the hosts of radio-AGN have evolved only passively since at least $z\sim 1$ (e.g., \citealt{lilly1, best, mclure1, jarvis}), this means that their epoch of formation must be pushed back to higher redshifts, likely earlier than the so-called `cosmic noon' at $z\sim 2$ when both galaxies and AGN experienced the peak of their activity (cf.\ \citealt{madau} for a review on this topic). As a consequence of this scenario, both the gas mass and star forming activity within the hosts of radio-active AGN should be strongly increasing functions of redshift as one approaches $z\simgt 2$.

Indeed, observational evidence for increased star-forming activity with look-back time is presented already between $z<0.3$ and $z\sim 0.6$ by e.g., \citet{chen} for their sample of 1600, $M_*> 10^{11.4}\,M_\odot$ radio-active AGN selected from FIRST and NVSS (cf.\ Fig.~\ref{chen}). Strong evidence for powerful star-formation was also provided by rest-frame UV observations of the hosts of 13 high-redshift ($2\simlt z \simlt 4$) radio galaxies (\citealt{pentericci, pentericci1}, see \citealt*{miley1} for a review) which showed morphologies characterised by clumpy structures -- likely the sites of ongoing star-formation -- or by observations of the $z=3.8$, 4C 41.17 radio galaxy which returned estimates for the rate of ongoing massive star formation activity, $\sim 10^3\,M_\odot$ yr$^{-1}$ (\citealt{dey}). And, as already discussed in Sect.~\ref{sec:2.1}, also the works of e.g., \citet{smolcic1}, \citet{williams} and \citet{lin3} hint to a switch in the host properties of radio-AGN towards more star-forming systems at $z\simgt 1$. 

\begin{figure}
\centering
\includegraphics[scale=0.4]{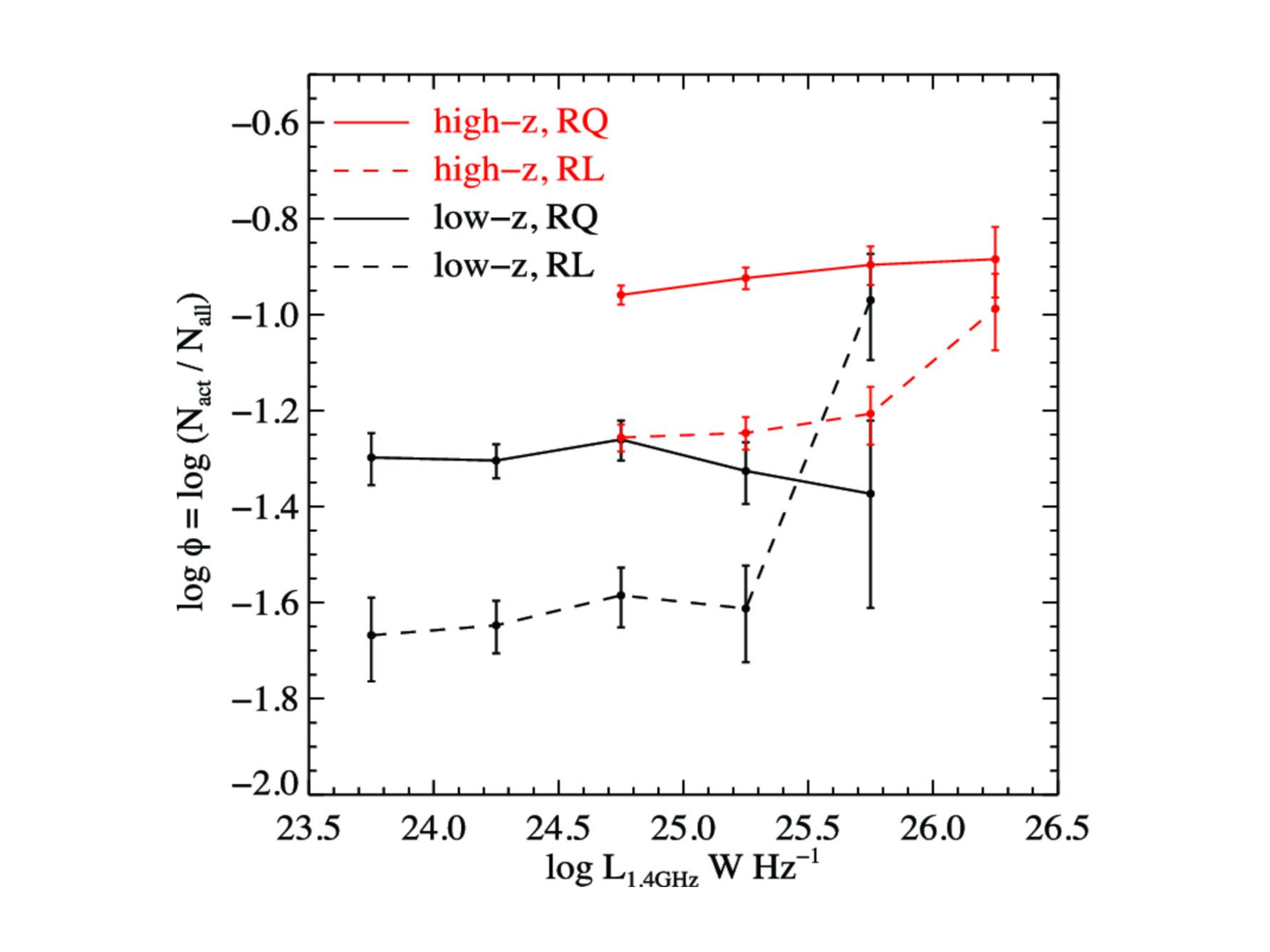}
\caption{Fraction of actively star-forming galaxies as a function radio luminosity. The dashed and solid lines respectively indicate radio-AGN (RL) and the radio-quiet (RQ) control samples, matched in redshift, stellar mass, spatial location and velocity dispersion. High-redshift data (represented in red) refer to $z\sim 0.6$, while low-redshift data to $z< 0.3$.
Figure from \citet{chen}.}
\label{chen}   
\end{figure}
 
However, it has been only with the advent of facilities such as \textit{Spitzer} (\citealt{werner}), \textit{Herschel} (\citealt{pilbratt}) and the Sub-millimetre Common User Bolometer Array (\textit{SCUBA},  \citealt{holland}) which allowed to systematically sample the infrared (both near and far) and sub-millimetre skies to high sensitivity levels, that it has become more and more clear that -- especially at high, $z >1.5$, redshifts -- radio-active AGN often cohabit with episodes of intense star-formation.
\citet{Archibald} observed in the sub-millimetre 47 bright ($L_{151\, \rm MHz}\simgt 10^{25}$ W Hz$^{-1}$ sr$^{-1}$) radio galaxies in the redshift range $0<z<5$, finding that their typical sub-millimetre luminosities (and hence dust masses)  were a strongly increasing function of redshift, with $L_{850 \mu \rm m}\propto (1+z)^3$, in line with the hypothesis of enhanced star-formation within radio-AGN hosts at earlier cosmological epochs. \citet{reuland} expanded on the work of \citet{Archibald} 
by adding 24 radio-AGN, many of which at $z>3$. They confirmed the \citet{Archibald} conclusion for a strong redshift evolution of the sub-millimetre luminosity of these sources, found to be associated with intense star-forming activity, with rates of up to a few thousand $M_\odot$ yr$^{-1}$. \citet{seymour} instead used \textit{Spitzer} observations at all available wavelengths for 69 radio-AGN with $L_{3\, \rm GHz}> 10^{26}$ W Hz$^{-1}$ and $1<z<5.2$. In agreement with the previous works, it was concluded that these sources not only are associated to very massive, $M_*\simeq 10^{11}$--$10^{12}\,M_\odot$, galaxies, but also that most of them can be classified as Luminous Infrared Galaxies (LIRGs) or even as Ultra-Luminous Infrared Galaxies (ULIRGs), given that they exhibit MIR luminosities $> 10^{11}\, L_\odot$. \citet*{maglio8} come to the same conclusion for concurrent AGN and star-forming activity in their sample of $z\sim 1-3$ radio sources observed with \textit{Spitzer} at $24\mu$m  in the First Look Survey field (\citealt{fadda}). \citet{rawlings}  also derive from MIR, \textit{Spitzer} spectroscopy of seven, $1.5<z<2.6$, very powerful ($L_{500 \rm MHz}=10^{27.8}-10^{29.1}$ W Hz$^{-1}$) radio-AGN extremely high star-formation rates, of up to $\sim 10^3\,M_\odot$ yr$^{-1}$, with no significant correlation between radio emission and star formation activity.

\begin{figure}
\centering
\includegraphics[scale=0.4]{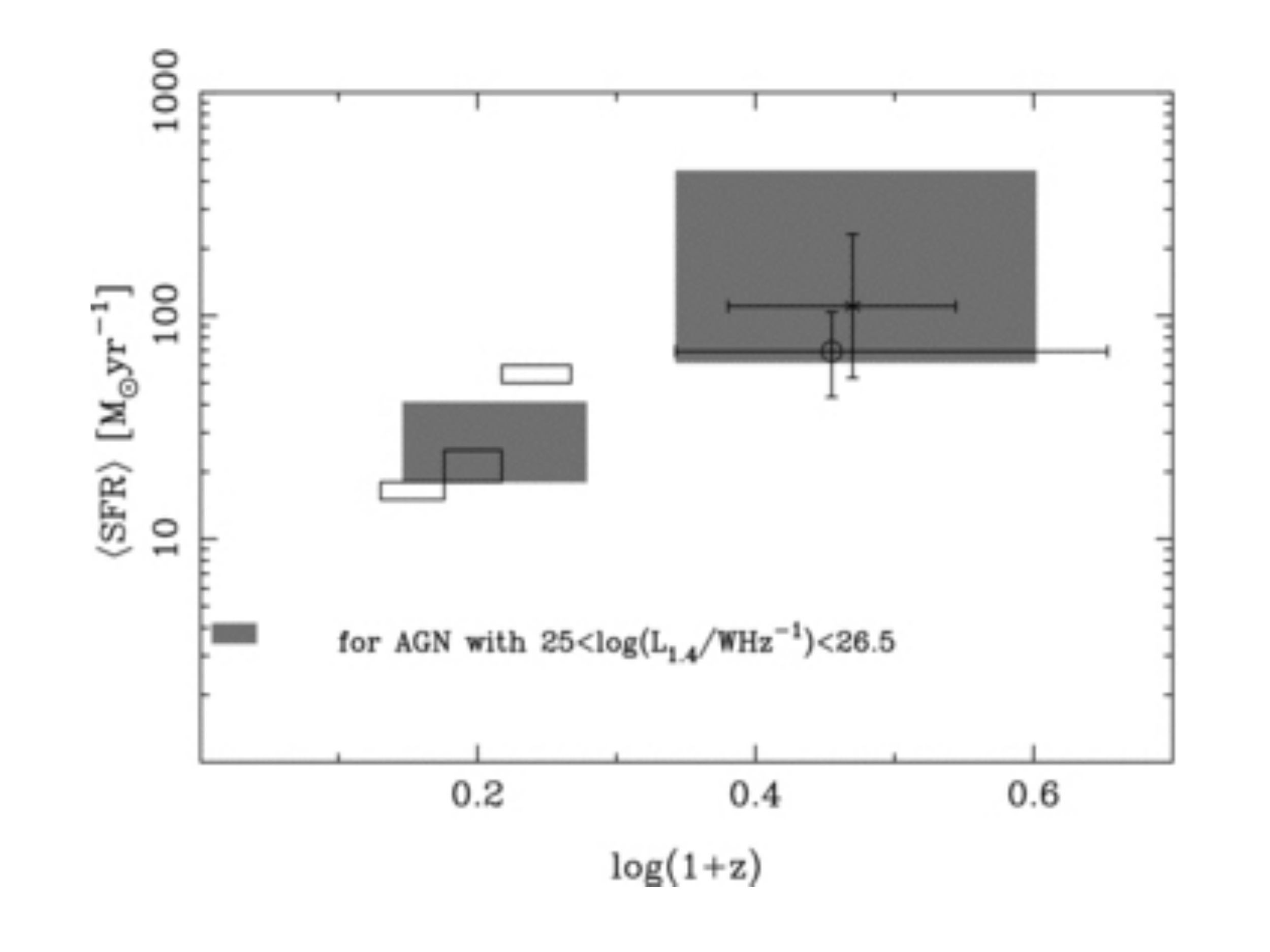}
\caption{Range of mean SFRs plotted as a function of redshift for radio-AGN with $10^{25}\le (L_{1.4\, \rm GHz}/{\rm W Hz^{-1}})\le 10^{26.5}$ (shaded regions). The open rectangles indicate the results from \citet{hardcastle1} using \textit{Herschel} observations of sources with a similar range of radio luminosities. The points with error bars present the approximate mean SFRs of X-ray-selected AGN over the range of redshifts indicated from \citet{lutz} (open circle) and \citet{shao} (asterisk). Figure from \citet{seymour1}.}
\label{seymour_2011}   
\end{figure}

A leap forward in the field though is provided by the work of \citet{seymour1} who use both \textit{Herschel}-SPIRE (\citealt{griffin}) and \textit{Spitzer} data for a sample of 32 powerful radio sources with  $L_{1.4 \rm GHz}=10^{25}-10^{26.5}$ W Hz$^{-1}$ to derive their total (i.e., [8-1000] $\mu$m) infrared luminosities. Data were subdivided into moderate-redshift ($0.4<z<0.9$) and high-redshift ($1.2<z<3$) sub-samples and then compared to local ($z<0.1$) 3CRR sources with available \textit{Spitzer} information (\citealt{dicken}).  By also taking into account 250$\mu$m non-detections, \citet{seymour1} estimate the total mean star-formation rates for the three sub-samples to be:  $\sim 3.7$, $29.5\pm11.6$ and $581\pm143\, M_\odot$ yr$^{-1}$ respectively at low-, moderate- and high-redshift (cf.\ Fig.~\ref{seymour_2011}). This implies an extremely steep increase of the star-formation activity associated with radio-AGN, ${\rm SFR} \propto (1+z)^{4.2\pm 0.8}$, with a degree of evolution larger than what observed for the general population of star-forming galaxies. According to the authors, this might indicate that at least some of the inferred star formation may be directly associated with the AGN radio outputs, although no dependence of the star-formation activity on radio luminosity is reported for these sources. \citet{drouart} also considered [3.6--870] $\mu$m multi-wavelength information for a sample of 70, $L_{3 \rm GHz}>10^{26}$ W Hz$^{-1}$  radio-AGN spanning the range $1< z< 5.2$, finding that almost all the hosts are ULIRGs with total infrared luminosities $L_{\rm IR}>10^{12}\, L_\odot$, corresponding to SFRs of up to $5000\,M_\odot$ yr$^{-1}$. Again, no connection was found between IR and AGN luminosity. These authors also report higher levels of star-forming activity in radio-AGN at $z>2$ with respect to their lower redshift counterparts, as well as a faster growth in terms of stellar assembly when compared to the population of inactive, $z\simgt 2$, star-forming galaxies (cfr. \citealt{weinmann} and references therein). A very similar result is obtained by \citet{podigachoski} who concentrate on a complete sample of 64, $z>1$, 3CR sources. The median infrared luminosity for these radio-AGN is estimated from \textit{Herschel} and \textit{Spitzer} observations to be $\langle L_{\rm IR}\rangle \sim 2\cdot 10^{12}\, L_\odot$, corresponding to SFRs $\sim 100-1000\,M_\odot$ yr$^{-1}$. Also, in agreement with \citet{seymour1} and \citet{drouart}, no strong correlation between AGN activity and luminosity due to star formation is found.

These results provide clear, breakthrough evidence for intense star forming activity ongoing within galaxies hosting a radio-AGN at high, $z\simgt 1$, redshifts.
However,  they are all limited to the bright end of the luminosity distribution at both radio and FIR wavelengths, and therefore by construction miss the bulk of the population of radio-AGN, both in terms of faint radio emission and low, $\simlt 100\,M_\odot$ yr$^{-1}$, star-formation rates. 
\citet{maglio13, maglio14, maglio15}  fill this gap by considering radio-AGN selected in the deepest cosmological fields known to date (COSMOS -- \citealt{maglio13, maglio15} and GOODS North, GOODS South and Lockman Hole -- \citealt{maglio14}). These fields not only are covered with extremely deep 1.4 GHz and \textit{Herschel} observations (fluxes down to $\sim 6$ $\mu$Jy and 0.9, 0.6 and 1.3 mJy, respectively at 1.4 GHz and 70, 100 and 160 $\mu$m), but can  also rely on a huge amount of multi-wavelength information. Furthermore, since observations on COSMOS are somehow shallower but obtained on a larger area than those on the other three fields, the combination of them allows for the first time to probe, with a good statistical significance, both the mid- and faint-luminosity regimes. The total number of radio-AGN selected on the basis of their radio luminosity (cf.\ Sect.~\ref{sec:1}) on the four fields is $\sim 800$, 325 of them with a FIR counterpart in the deep PACS Evolutionary Probe (PEP -- \citealt{lutz1}) maps. The percentage of FIR identifications greatly varies as a function of FIR depth across the fields, from 39\% in the shallower COSMOS, to 72\% in GOODS North. No dependence is instead found on either radio flux or redshift, at least for $z\simlt 3.5$. In agreement with all previous studies, the authors find that the hosts of these sources all exhibit very large stellar masses, with 90\% of them with $M_*>10^{10.5}\,M_\odot$ and 50\% with $M_*>10^{11}\,M_\odot$.  Also, irrespective of the level of radio-AGN activity, FIR emission was proved to stem from star-forming processes ongoing within the host galaxies and not observed to affect the AGN output, at least at radio wavelengths. More importantly, the \citet{maglio13, maglio14, maglio15} analyses clearly indicate that the level of star-forming activity within the hosts of radio-AGN presents a dramatic increase with look-back time, and that relatively powerful radio-AGN much more likely cohabit with a star-forming event at the earliest epochs: indeed, if the probability for FIR emission from the $z <1$ host of a $L_{1.4 \rm GHz}\sim 10^{25.5}$ W Hz$^{-1}$ sr$^{-1}$ ($L_{1.4 \rm GHz}\sim 10^{23.5}$ W Hz$^{-1}$ sr$^{-1}$) AGN is $\sim 0$\% ($\sim 20$\%), this becomes $\sim 40$\% ($\sim 50$\%) at $z> 2$ (cf.\ the top panel of Fig.~\ref{manu}). Furthermore, the deep data available for the Lockman Hole and the two GOODS fields shows that virtually \emph{all} radio-AGN beyond $z\sim 1$ co-exist with a star-formation event of intensity as low as ${\rm SFR} \sim 10\,M_\odot$ yr$^{-1}$, which proceeds independently of AGN radio luminosity (cf.\ bottom panels of Fig.~\ref{manu}). These findings put together then imply that negative feedback is present in radio-AGN hosts \emph{only} in the local, $z<1$ universe, and with an effectiveness that strongly decreases for decreasing AGN radio power. 

\begin{figure*}
\centering
\includegraphics[scale=0.25]{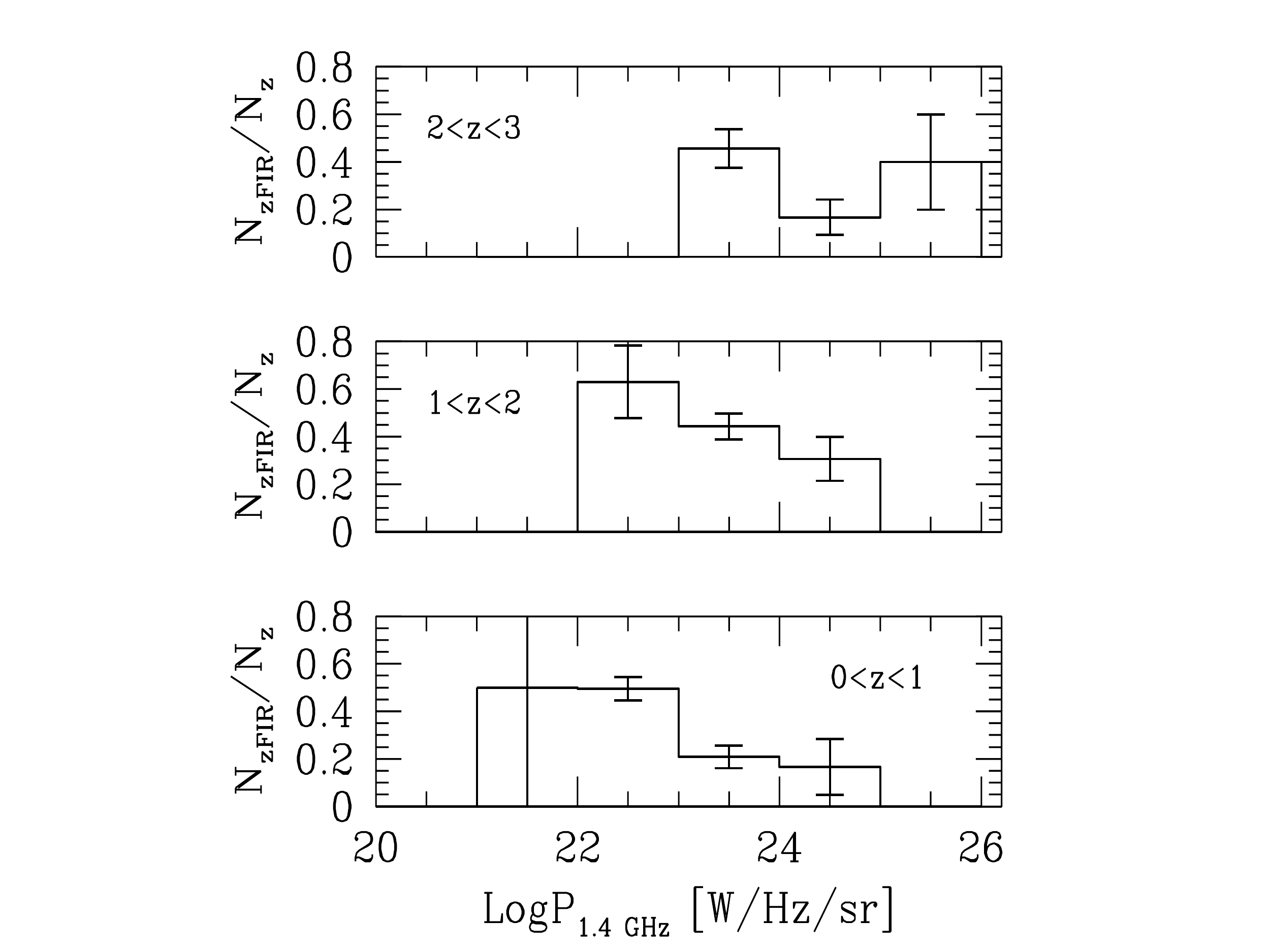}
  \includegraphics[scale=0.22]{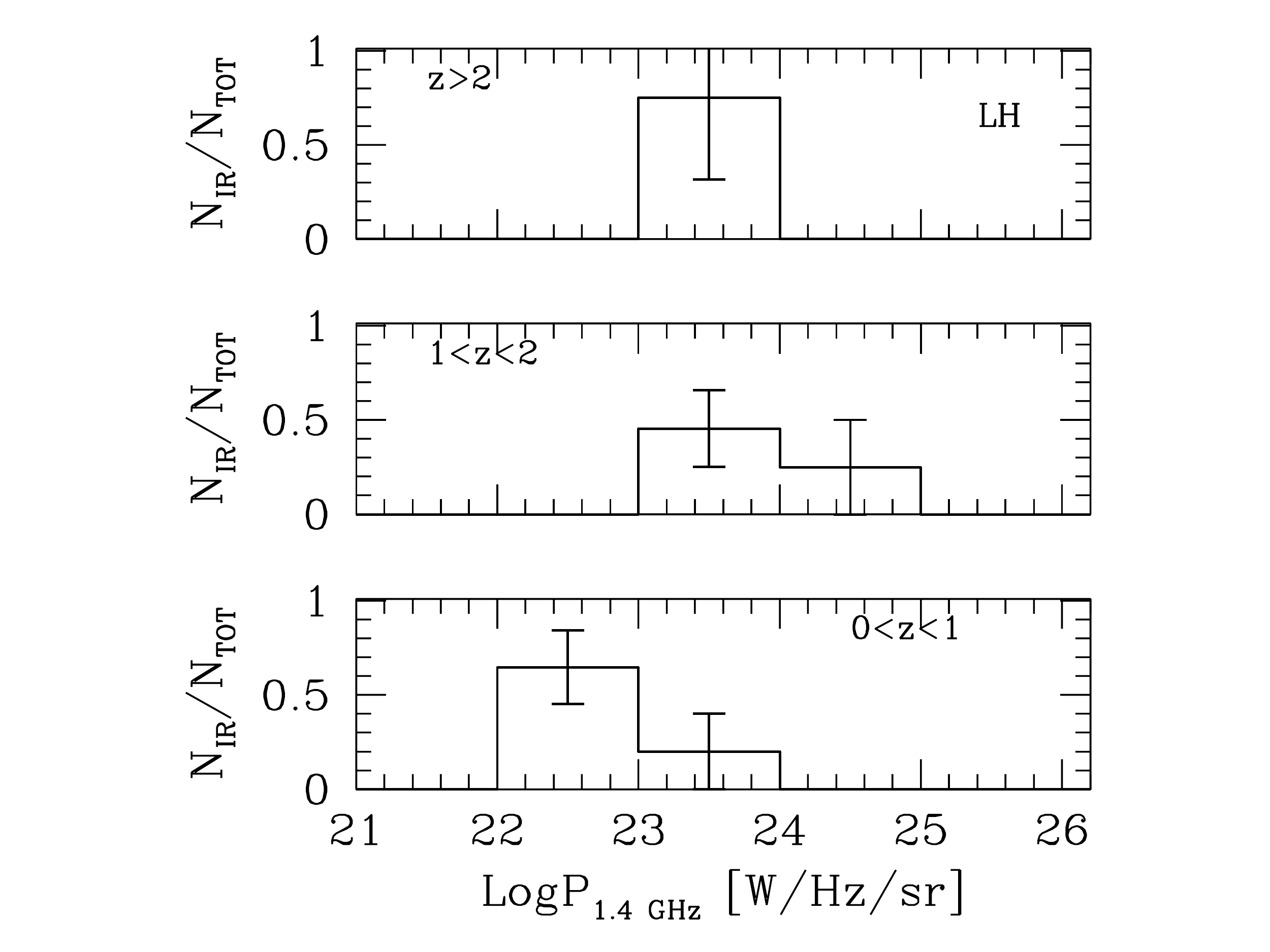}
 \includegraphics[scale=0.23]{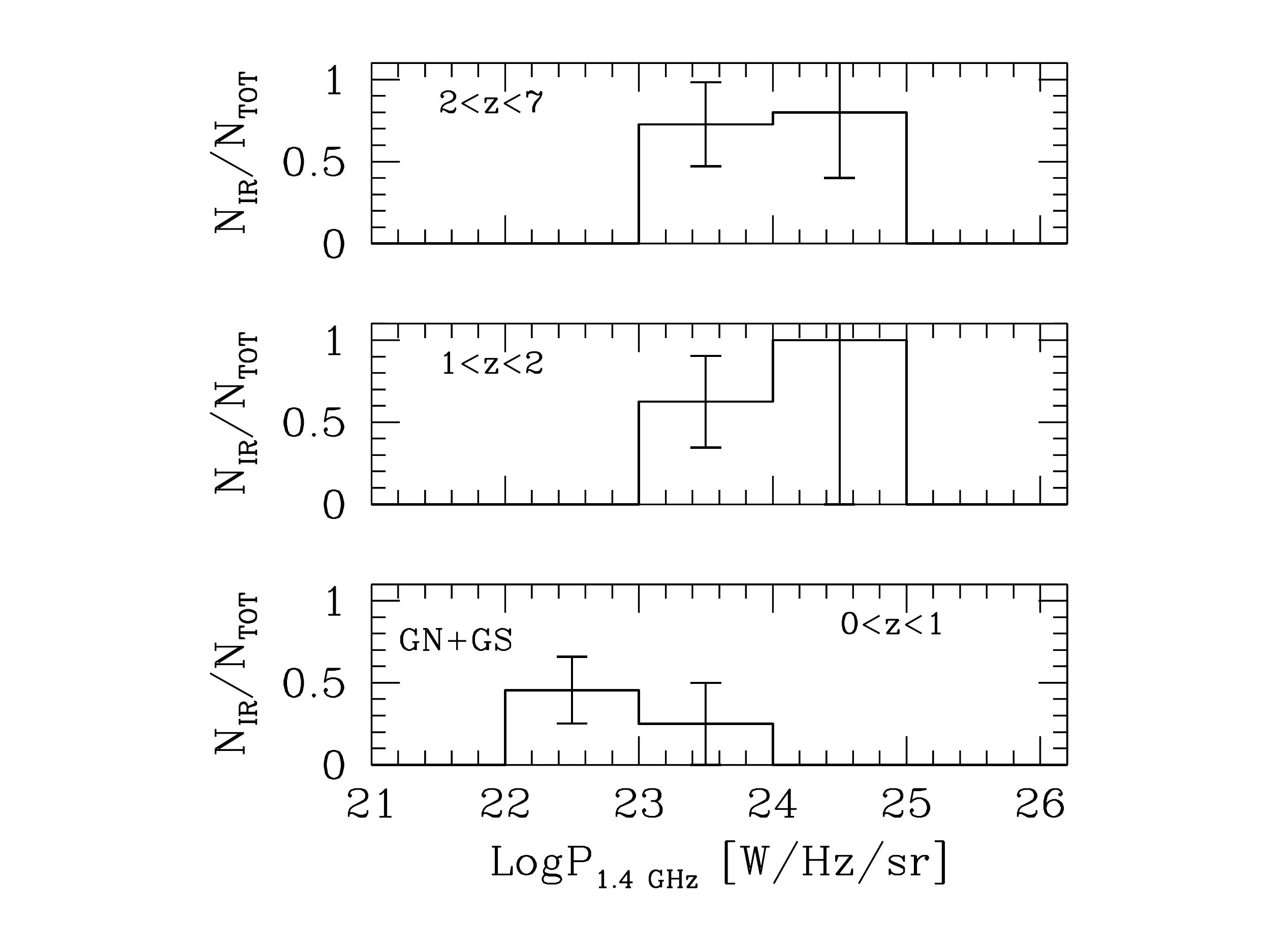}
\caption{Fraction of FIR emitters amongst radio-selected AGN as a function of radio luminosity at different cosmological epochs. The top panel refers to the COSMOS field, while the labels `LH' and `GN+GS' respectively to the Lockman Hole and the combined GOODS North and GOODS South. Figures from \citet{maglio14} and \citet{maglio15}.}
\label{manu}   
\end{figure*}

As a last step, it also worth highlighting a number of works that have recently made use of IR information to investigate the star formation histories of samples of radio-AGN not selected uniquely in radio luminosity as those presented so far, and that further provide comparisons between the star-forming activities of radio-active and radio-quiet populations. \citet{delmoro} find that $\sim 86\%$ of their 51, $0<z<3$, AGN presenting a radio excess with respect to the IR-radio relation (cf.\ Sect.~\ref{sec:2.4}) have a far-infrared spectrum dominated by star-formation, although their specific star formation rates (sSFRs) are on average lower than those observed for X-ray selected AGN hosts. On the other hand, \citet{kal} find that in the redshift range $0<z<4$, radio-active quasars and their radio-quiet counterparts exhibit on average the same star formation activity, except at low optical luminosities, whereby the SFRs of radio-active quasars are about four times as large as those of radio-quiet ones. Also, a strong positive correlation between radio and FIR luminosity is observed for the former population. The authors argue that this is possibly due to powerful radio jets boosting the star formation activity by compressing the intergalactic medium (\emph{positive feedback} -- e.g., \citealt{silk}), especially in radio-active quasars with low nuclear accretion rates. This result was further expanded by \citet{kal1} who analysed a sample of 173, $z\sim 1$, AGN sub-divided into radio-active quasars (74 objects), radio-quiet quasars (72 objects) and radio-active galaxies (27 objects). Clear evidence for enhanced star-forming activity (a factor of $\sim 2.5$) in radio-active quasars with respect to radio-quiet ones matched in black-hole mass and bolometric luminosity and also with respect to radio galaxies is reported, independent of other AGN and host galaxy properties. Given these differences between radio-active quasars and galaxies, the authors argue for a mechanism which works differently according to jet power, with a threshold -- also dependent on galaxy mass -- that determines when radio-AGN feedback switches from enhancing star formation (by compressing gas) to suppressing it (by heating or ejecting gas). 

Based on the above discussion, the picture that emerges from IR observations of radio-active AGN can then be summarized as follows. In the local, $z\simlt 1$, universe star-formation within the hosts of radio-AGN is generally suppressed, and these galaxies appear as passively evolving, massive red ellipticals. However, at higher redshifts and up to the earliest epochs probed by the observations, not only star-formation is enhanced with a growth rate that is a strongly increasing function of $z$, but there are indeed cases at $z\simgt 1$ whereby star-formation activity appears to be boosted with respect to galaxies that do not host a radio-AGN (e.g., \citealt{drouart, kal, kal1}).  These findings, which hold irrespective of the depth of both radio and IR observations, point to a scenario whereby at early -- $z \simgt 1$ -- epochs radio activity of nuclear origin does not quench star formation in the host galaxy, but either favours (\emph{positive feedback}) or proceeds parallel to it, with the two processes evolving independently thanks to the large reservoirs of gas present within massive high-redshift galaxies.  According to this second theory, the main responsible for the quenching of the massive galaxy population in the nearby universe would be intense star formation which rapidly consumes the available gas, possibly then followed by the action of the radio jets, that at $z \simlt 1-1.5$ succeed at removing the residual gas (e.g., \citealt{falkendal}). More work and data are needed to assess the validity of either scenarios and also to understand the drastic change in feedback behaviour when moving from $z\sim 1.5$ down to the local universe.

\subsection{The IR-radio relation in star-forming galaxies and its cosmological evolution}
\label{sec:2.4}
As already mentioned in the Introduction, radio-emitting star-forming galaxies (SFGs) at low-to-intermediate redshifts do not represent any special category different from star-forming galaxies selected at other wavelengths. Furthermore, from an observational point of view, it has been known since the 1970s that in this population of sources the emission at radio and infrared (IR) wavelengths are strongly correlated.  First statistically significant evidence in this direction was provided by \citet{Dejong} and \citet*{helou} who both relied on the data acquired by the IRAS satellite (\citealt{neugebauer}) to report a remarkably tight, linear correlation between far-infrared (FIR) and radio (1.4 GHz) flux densities, extending over three orders of magnitude. This held for a joint sample of blue spirals in the Virgo cluster and in the field, as well as for a subset of galaxies with starburst nuclei (\citealt{helou}, cf.\ Fig.~\ref{helou}) and also for a heterogeneous sample of 91 sources which included normal spiral galaxies, irregulars, dwarfs and galaxies containing an AGN (\citealt{Dejong}). 

\begin{figure}
\centering
\includegraphics[scale=0.4]{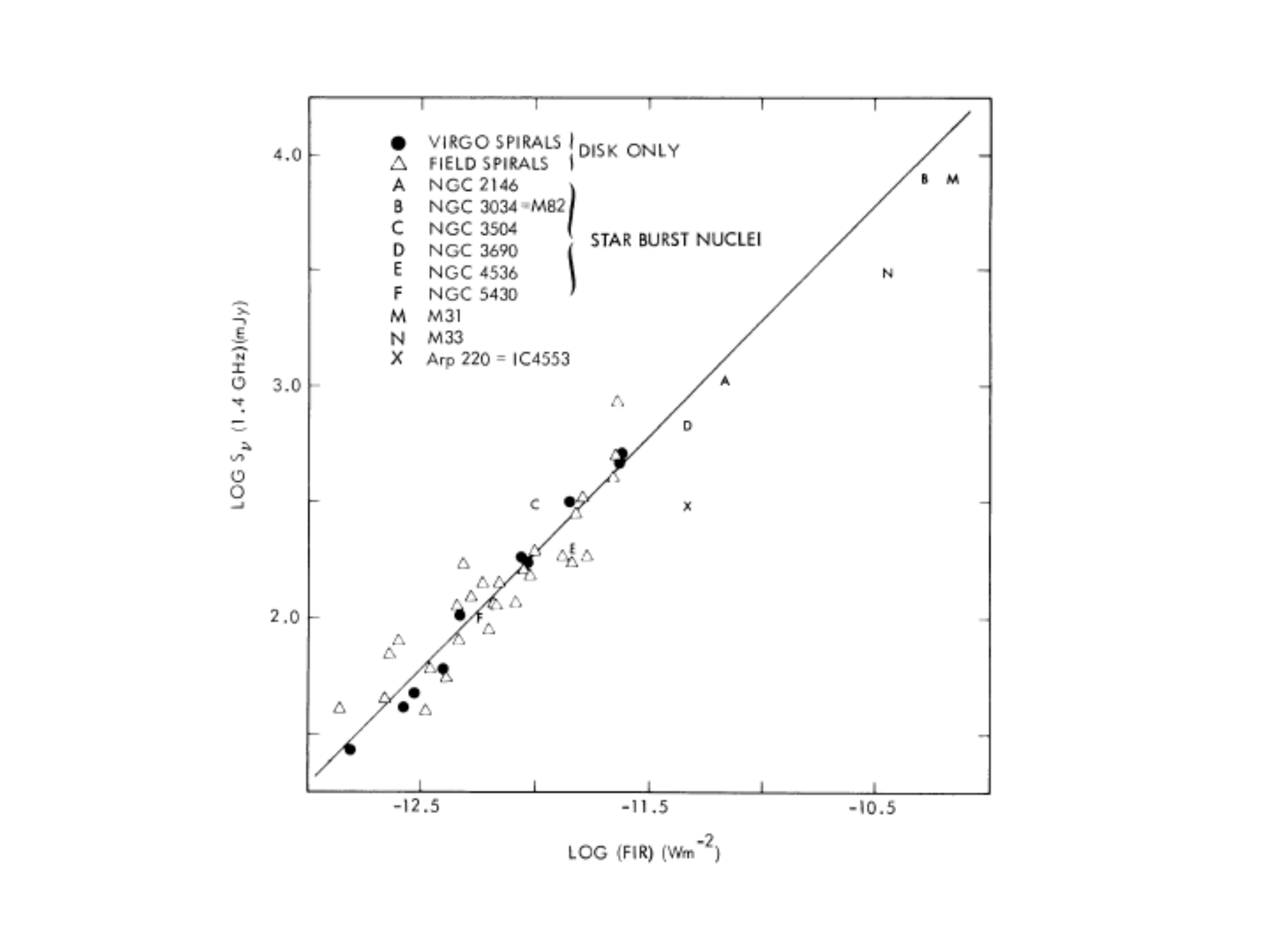}
\caption{Comparison of observed FIR and 1.4 GHz fluxes for galaxies with extended emission and no compact nuclear component detected at
1.4 GHz, both in the Virgo Cluster ({\it circles}), and in the field ({\it triangles}). Nine other galaxies with starburst nuclei are
also shown for comparison. Figure from \citet*{helou}.}
\label{helou}   
\end{figure}

This incredibly tight relation which -- we stress again -- was obtained regardless of the galaxy type as long as it hosted ongoing star formation, was parametrized by the FIR-radio luminosity ratio $q_{\rm FIR}$, where
\begin{equation}
q_{\rm FIR}={\rm Log}\left(\frac{L_{\rm FIR}[\rm W]}{3.75 \cdot 10^{12} [\rm Hz]}\right) -{\rm Log}(L_{1.4 \rm GHz}[\rm{W Hz^{-1}}])
\label{qir}
\end{equation}
(\citealt{helou}), with $L_{\rm FIR}$ luminosity extending over the far-infrared (rest-frame [42.5-122.5] $\mu$m) range and where  $3.75 \cdot 10^{12}$ Hz is the frequency at 80 $\mu$m, approximately corresponding to the center of the FIR domain. We note that in the literature also the parameters $q_{\lambda}$ -- with $\lambda$ within the IR regime -- and $q_{\rm IR}$ are often used. This latter corresponds to the $q$ value inferred from the luminosity estimated over the whole, [8-1000] $\mu$m, infrared range. As \citet{magnelli} note, these different definitions bear no significant impact, since one has $L_{\rm IR}=1.91^{+0.10}_{-0.05}\cdot L_{\rm FIR}$, and therefore $q_{IR}\simeq q_{\rm FIR} + 2$.

As summarized by \citet{condon}, nearly all the radio emission from non AGN-powered galaxies is synchrotron radiation from relativistic electrons and thermal (free-free) emission from HII regions. Only stars more massive than $\sim 8\,M_\odot$ can produce the Type II and Type Ib supernovae whose remnants are thought to accelerate most of the relativistic electrons in star-forming galaxies, and can also ionize the HII regions. Since both the life-time of massive stars and that of relativistic electrons are relatively short ($\simlt 3\cdot 10^7$ yr for massive stars and $\simlt10^8$ yr for relativistic electrons), this implies that radio observations are powerful probes of very recent star-formation activity taking place within galaxies. On the other hand, FIR luminosity seems to be a good estimator of the bolometric luminosity produced by relatively massive ($M\simgt 5\,M_\odot$) young stars heating the surrounding dust. To a first order, the very tight FIR-radio correlation is then thought to result from the fact that both emissions are produced by the same sources and within the same regions, as the infrared radiation is the thermal re-radiation from HII regions, while that at 1.4 GHz comes from relativistic electrons accelerated in supernova remnants produced by the same population of massive stars that heat and ionize the HII regions (\citealt{harwit}). 

The IR-radio correlation, which is found to extend over five orders of magnitude ($10^{7.5}\simlt L_{\rm FIR}/L_\odot\simlt 10^{13}$) with an extremely small dispersion ($\sigma\simlt 0.2$, e.g., \citealt*{yun}), has been used throughout the years in order for instance to: a) distinguish between SFGs and galaxies powered by a radio-active AGN (e.g., \citealt{donley, park, delmoro, bonzini, delvecchio}); b) use the radio continuum to estimate the star-formation rate of galaxies in a way which is unaffected by the presence of dust (e.g., \citealt{condon, murphy1, murphy2, davies}); c) estimate distances and temperatures of high-redshift sub-millimetre galaxies (e.g., \citealt{carilli, chapman}). We also note that, although the correlation was originally found at 1.4 GHz, recent works investigated its validity also at lower (150 MHz; e.g., \citealt{gurkan2}) and higher (3 GHz; e.g., \citealt{delhaize}) radio frequencies, finding again -- at least in the local universe -- a tight correlation between radio and FIR luminosities. In particular, by using relatively local ($z\simlt 0.35$) LOFAR-150 MHz and \textit{Herschel}-250$\mu$m observations, \citet{gurkan2} observe a slightly steeper slope than what obtained at 1.4 GHz (cf.\ Fig.~\ref{gurkan}). The authors argue that this is due to thermal radiation becoming less important at lower radio frequencies. 

\begin{figure}
\centering
\includegraphics[scale=0.4]{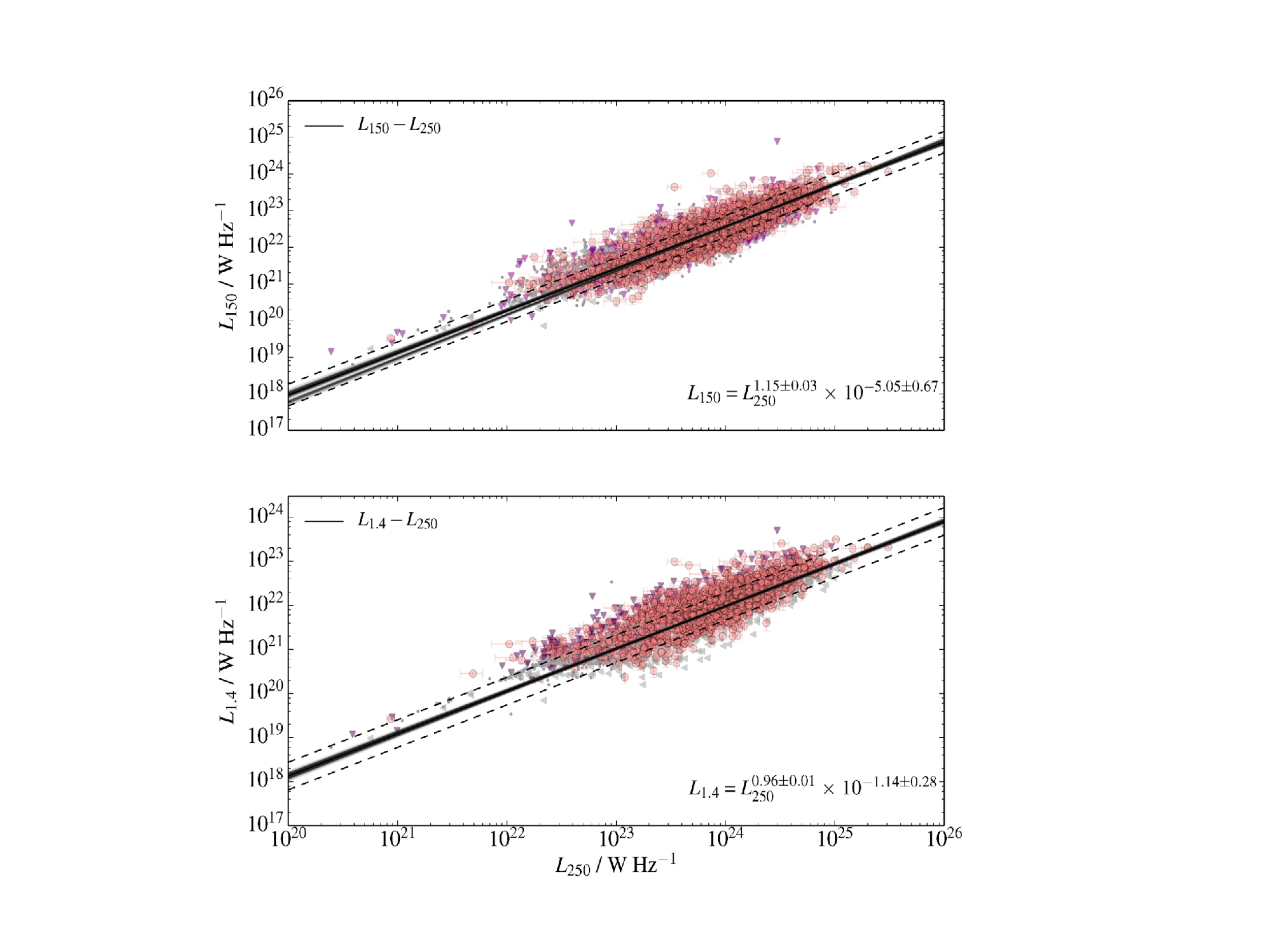}
\caption{Distribution of \textit{Herschel} 250 $\mu$m luminosities of SFGs as a function of their LOFAR 150-MHz luminosities. Dashed lines indicate the 1$\sigma$ intrinsic dispersion of the fit. Salmon circles show 3$\sigma$ detections; LOFAR 150 MHz 3$\sigma$ limits are indicated by purple down-pointing triangles and 3$\sigma$ limits on Herschel 250$\mu$m luminosities are indicated by grey left-pointing triangles; grey dots indicate 3$\sigma$ upper limits on both quantities. A tight relationship between  $L_{\rm 250\mu m}$ and $L_{\rm 150 GHz}$  is clearly seen. Bottom: Distribution of \textit{Herschel} 250 $\mu$m luminosities of SFGs as a function of their FIRST 1.4 GHz luminosities. Symbols and lines as for the top panel. Figure from \citet{gurkan2}.}
\label{gurkan}   
\end{figure}

However, despite the above explanations and results, many unknowns remain, both from the theoretical point of view and also from observations. Indeed, the "calorimeter theory" (\citealt{voelk}) which suggests that the IR-radio relation holds because galaxies are both electron calorimeters and UV calorimeters, hence the total radio and IR outputs remain proportional independent of variations within the galaxy, requires galaxies to be optically thick to UV light from the young high-mass stars and thus does not hold for
optically thin galaxies. Also, a number of works seem to converge at indicating a change in the slope of the IR-radio relationship at low luminosities (e.g., \citealt{yun, bell, best3, gurkan2}). This effect could be expected if small galaxies were unable to prevent synchrotron electrons to escape, therefore causing a global deficit of radio emission at a fixed star-formation rate (e.g., \citealt{chi}). However, for instance \citet{gurkan2} find the opposite effect, with low-SFR objects ($\simlt 1\,M_\odot$ yr$^{-1}$) exhibiting more radio emission than what would be expected from extrapolations of the trend at higher SFRs.  

Possibly the most important open question associated to the IR-radio correlation is its eventual redshift evolution, as erroneous estimates at high redshifts directly translate into erroneous estimates of the star-formation activity of galaxies throughout the cosmic epochs, therefore hampering the huge potential of present and future radio surveys such as SKA to investigate the history of formation and evolution of galaxies since the cosmic dawn. 
This redshift evolution would indeed be expected, as e.g., electrons might lose energy by inverse Compton interactions with the cosmic microwave background whose
energy density scales as $(1+z)^4$, resulting in a lower level of radio emission in galaxies at higher redshifts. Also, synchrotron emission is proportional to the square of the magnetic field strength, so any cosmological evolution of this latter quantity should affect the constancy of the $q_{\rm IR}$ parameter at earlier epochs (e.g., \citealt{murphy}). 
Furthermore, changes in the galaxy spectral energy distributions (SEDs) may also be expected due to evolution in dust properties and metallicity (e.g., \citealt{chapman, Amblard}). 

The possibility for a redshift dependence of $q_{\rm IR}$ has started being investigated only in the last 15 years or so, thanks to the advent of IR satellites such as \textit{Spitzer} (\citealt{werner}) and \textit{Herschel} (\citealt{pilbratt}) which overcame the IRAS limitations of observing just the local universe.
\citet{Appleton} use \textit{Spitzer} data at 24$\mu$m and $70\mu$m from the First Look Survey to analyse the behaviour of the monochromatic quantities $q_{24 \rm \mu m}$ and $q_{70 \rm \mu m}$. By relying on spectroscopic redshifts for $\sim 500$ objects, these authors find that $q_{70 \rm \mu m}$ stays invariant up to $z\sim 1$. The same conclusion was reached for $q_{24 \mu \rm m}$ up to $z\sim 2$, although with a larger spread that could be attributed to variations in the galaxy SEDs throughout the probed population. Similar results were obtained by \citet*{vlahakis}, \citet{Ibar}, \citet{garn} for sources selected at 610 MHz and \citet{jarvis1}, although on a more limited, $0<z<0.5$, redshift range. \citet{jarvis1} also report no dependence of the IR-radio relation on the radio luminosity of the sources -- as long as they probed the non-AGN regime, $L_{\rm 1.4 GHz}<10^{23}$ W Hz$^{-1}$ -- and a factor $\sim 2$ tighter relationship between radio and 250$\mu$m luminosities than what inferred for the whole $[8-1000]$ $\mu$m range. In agreement with e.g., \citet{vlahakis}, it is argued that this could be
obtained if the longer wavelength emission, i.e., in the sub-millimetre regime, were not just produced by massive star formation and thus did not trace the same physical mechanism(s) as the radio or the MIR-to-FIR emission.  No evolution was either observed by \citet{chapman1} for their sample of $\sim 70$ sub-millimetre galaxies. On the other hand, \citet{kovacs}  -- by considering a smaller sample of 15 sub-millimetre galaxies endowed with 350$\mu$m fluxes -- infer in the redshift range $1\simlt z\simlt 3$, $q_{\rm IR}\sim 2.07$, value which is sensibly lower than that reported for the local universe ($\sim 2.3$ -- e.g., \citealt{condon, yun, mauch}). 

\citet{ivison} find no evolution in the values of $q_{\rm 250\mu{\rm m}}$ obtained between $z\sim 0$ and $z\sim 2$  for a sample of $\sim 20$ sources observed with the Baloon-borne Large Aperture Space Telescope (BLAST; \citealt{truch}). However, when they instead consider a sample selected at $24\mu$m, with stacked FIR information subdivided into redshift bins, they find that $q_{\rm IR}$ is a steadily decreasing function of redshift in the whole range probed by their analysis, with $q_{\rm IR}\propto (1+z)^{-0.15\pm 0.03}$. The \citet{ivison} sample suffered the limitations borne by BLAST in terms of low sensitivity and large beam Full Width Half Maximum (FWHM). In order to overcome this problem, \citet{ivison1} consider 65 star-forming galaxies selected in the GOODS North field by \textit{Herschel} at 250$\mu$m and with available 1.4 GHz information. By also making use of the enormous wealth of multi-wavelength information available in that field, they find a very mild evolution of $q_{\rm IR}\propto (1+z)^{-0.08\pm 0.07}$ in the redshift range $0<z<2$, result which is confirmed by a larger sample of $24\mu$m-selected, $L_{\rm FIR}$-matched galaxies. However, when the most local $q_{\rm IR}$ value is discarded, the redshift dependence becomes much more appreciable, with a trend similar to that reported by \citet{ivison}, $q_{\rm IR}\propto (1+z)^{-0.26\pm 0.07}$. 

The investigation of the cosmological evolution of  $q_{\rm IR}$ was subsequently repeated for most of the cosmological deep fields observed by \textit{Spitzer} and \textit{Herschel}, returning somehow discording results. \citet{sargent} considered a homogeneous sample of $\sim 4000$ sources with comparable IR luminosities over the last 10 Gyr, selected at $24\mu$m in the COSMOS field and, by also taking into account flux limits in their analysis, conclude that there is no evolution in the $z=0-2$ range. A similar result was obtained by \citet{mao} in the Extended Chandra Deep Field South (ECDFS) for a sample of 70$\mu$m \textit{Spitzer}-selected sources, while instead negative evolution was found for $q_{70\mu \rm m}$ by \citet{bourne} in the very same ECDFS field by applying stacking analyses to a mass-limited/NIR-selected sample of star-forming galaxies. The main point that emerged from all these studies was that the reported differences were mainly to be attributed to selection effects since, while the radio selection favours higher $L_{\rm radio}$/SFR ratios, the opposite is true for the FIR and sub-millimetre selections which favour high SFRs. The bias associated with optical and NIR selection is less straightforward; however such a selection is found to under-represent dust-enshrouded, high-SFR galaxies. 

\citet{magnelli} bring the IR-radio investigation to much more solid statistical grounding by analysing the variation of the $q_{\rm IR}$ parameter for a very large sample of $\sim$ 300,000 galaxies selected on the basis of their stellar masses. All the best known deep extragalactic fields such as GOODS North, GOODS South, ECDFS and COSMOS were considered, and data included both deep \textit{Herschel} FIR (100, 160, 250, 350 and 500 $\mu$m) and deep radio 1.4 GHz VLA and 610 MHz GMRT observations. FIR and radio images were then stacked at the positions of this mass-selected sample, complete down to $M_*= 10^{10}\,M_\odot$ and across $0<z<2.3$. 
By doing this, they found $q_{\rm FIR}=(2.35\pm 0.08) \cdot (1+z)^{-0.12\pm 0.04}$, where the $z\sim 0$ value perfectly agrees with local observations (e.g., \citealt{yun}) and where the same degree of evolution holds for both normal star-forming galaxies and starbursts, i.e., galaxies with a largely enhanced specific star-formation rate. The authors exclude such an evolution to be due to large variations with redshift of the radio spectral index of the probed population(s) since these are not observed in their data. 

The issue of the evolution of the IR-radio correlation has moved in the recent years to radio surveys performed at frequencies different from the "standard" 1.4 GHz, in order to assess its universal validity. \citet{calistro} use LOFAR-150 MHz data for 810 star-forming galaxies from the Bo\"otes field.  At this frequency, they estimate 
an evolution of the $q_{\rm IR}$ parameter which goes as $\propto (1+z)^{-0.22 \pm 0.05}$, with an intrinsic scatter of $\sigma \sim 0.53$, in the redshift range $0<z<2.5$. The authors stress that the observed cosmic evolution of the IR-radio relation must be taken into account for the future use of radio luminosity to estimate unbiased SFRs at high $z$. 

Roughly at the same time, \citet{delhaize} and \citet{molnar} analyze the behaviour of the IR-radio correlation for sources in the VLA-COSMOS 3GHz Large Project. More in detail, \citet{delhaize} -- accounting for non-detections in the radio or infrared bands by means of the doubly-censored analysis presented in \citet{sargent} -- find that the parameter $q_{\rm IR}$ decreases with increasing redshift as $\propto (1+z)^{-0.19\pm 0.01}$. This was in agreement with most of the recent works, even though the authors could not rule out that unidentified AGN contributions only to radio wavelenghts may be steepening the observed trend.

We note that, at variance with the local result obtained for low-frequency-LOFAR sources (e.g., $q_{\rm IR}=1.72\pm 0.04$, \citealt{calistro}), the $z\sim 0$ value of $q_{\rm IR}=2.88\pm 0.03$ at 3 GHz is higher than that obtained at 1.4 GHz (e.g., \citealt{yun}). This trend can be recovered if one assumes for SFGs a single power-law radio emission with a spectral index $\alpha\sim 0.73$ in the whole 150 MHz -- 3 GHz range (e.g., \citealt{calistro}). \citet{delhaize} indeed conclude that the choice for the value of the average radio spectral index directly affects the normalisation of the $q_{\rm IR}$ parameter, as well as its estimated trend with redshift. At the same time, an increasing fractional contribution to the observed 3 GHz flux by free-free emission from star-forming galaxies may also affect the derived evolution. Based on their results, the authors conclude that imperfect $K$-corrections in the radio regime might be responsible for the decreasing trend of the IR-radio relation with redshift.

\begin{figure}
 \centering
\includegraphics[scale=0.4]{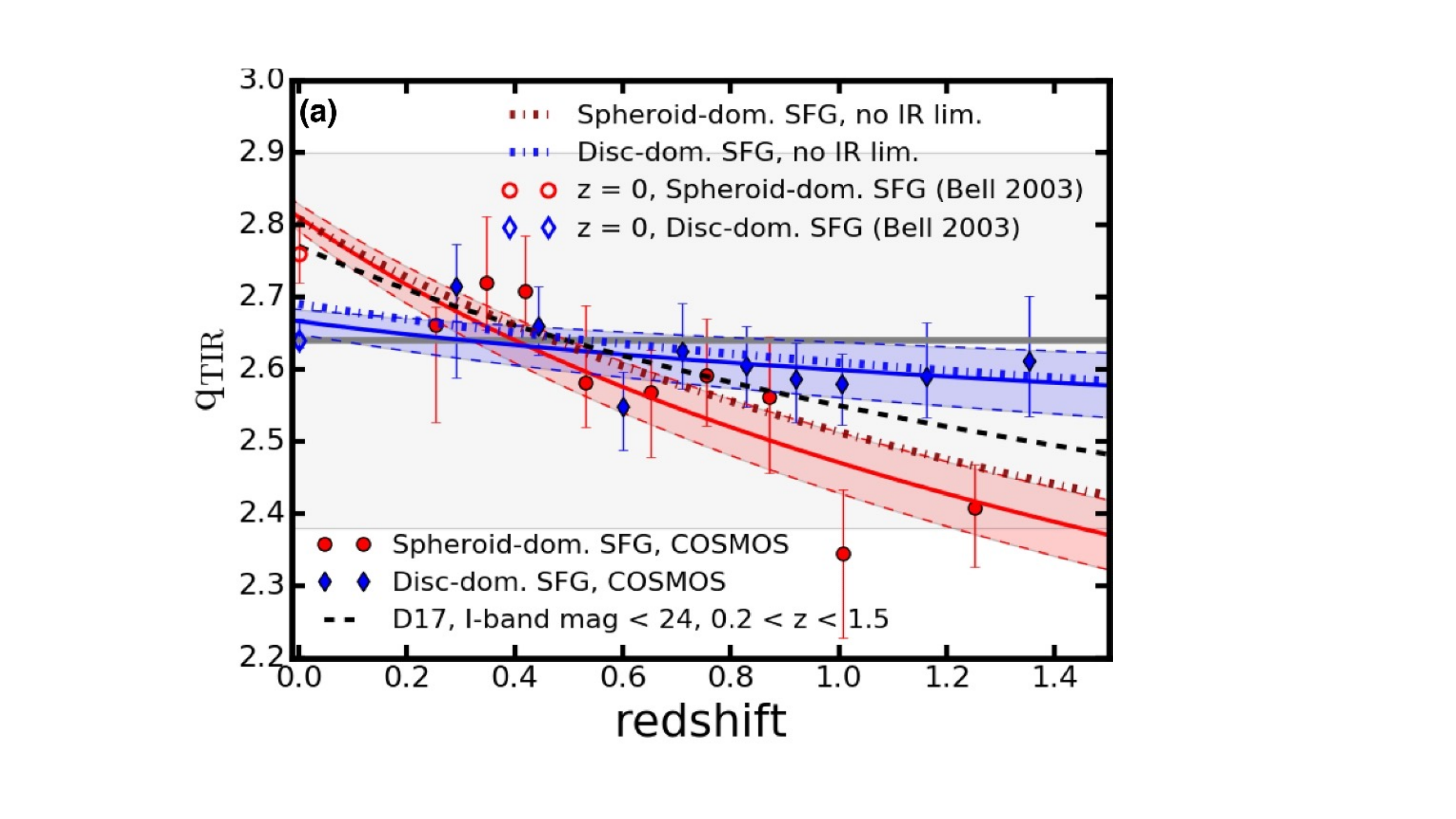}
\caption{Redshift evolution of the IR-radio correlation $q_{\rm TIR}\equiv q_{\rm IR}$, for disc- and spheroid-dominated star-forming galaxies (blue and red symbols, respectively). The local $z = 0$ measurements (open symbols) are based on a morphologically-selected subset of the \citet{bell} sample. Shaded regions bordered by dashed lines show the upper and lower limits of the 1$\sigma$ confidence interval of the fit. The black dashed line is the evolutionary trend found for $0.2 < z < 1.5$ SFGs with I-band magnitude $< 24$ in the sample of \citet{delhaize}. Figure from \citet{molnar}.}
\label{molnar}   
\end{figure}

\citet{molnar} refine the analysis of \citet{delhaize} and investigate the $q_{\rm IR}$ evolution for their sample of 3 GHz-selected SFGs subdivided by morphology into spheroid- and disc-dominated galaxies. These authors find that the spheroid-dominated population follows a declining trend for $q_{\rm IR}$ versus $z$, similar to that measured in the most recent studies. However, for disc-dominated galaxies, where radio and IR emission should be linked to star formation in a more straightforward way, they only measure a very small -- if any -- change (cf.\ Fig.~\ref{molnar}). This suggests that low-redshift calibrations of radio emission as a star formation rate tracer may remain valid out to at least $z\simeq 1-1.5$ for pure star-forming systems, and that the different redshift evolution of $q_{\rm IR}$ for the spheroid- and disc-dominated samples is mainly due to an increasing radio excess in spheroid-dominated galaxies at $z\simgt 0.8$, fact which hints at some AGN activity in these systems, as already discussed in Sects.~\ref{sec:1} and \ref{sec:2.3}. 

\begin{figure}
 \centering
  \includegraphics[scale=1.0]{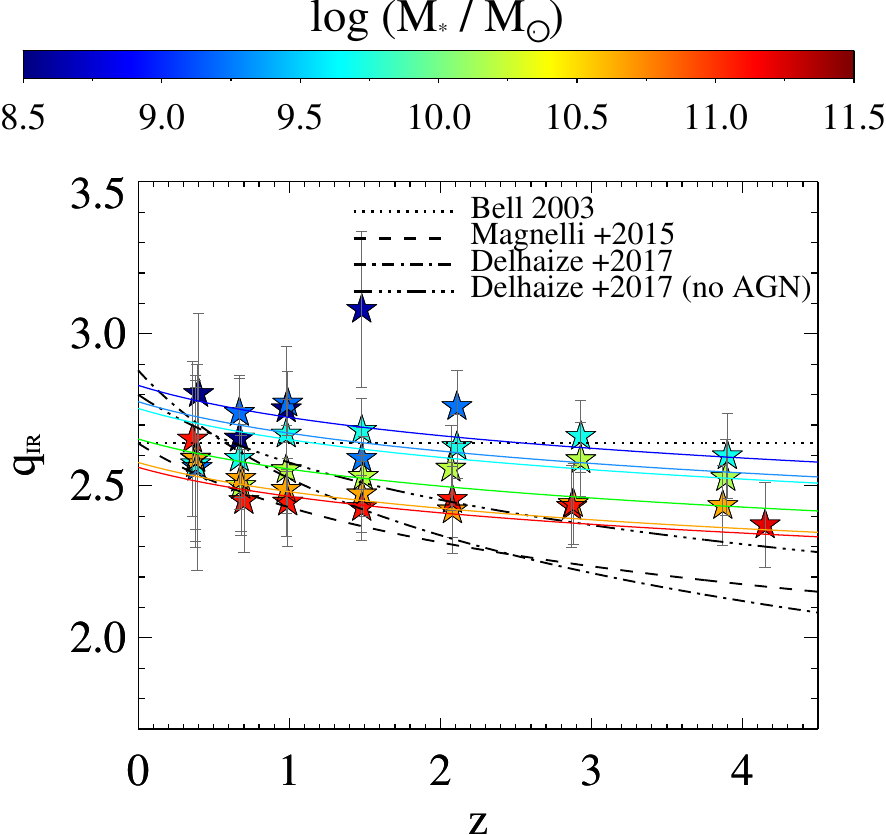}
\caption{Intrinsic $q_{\rm IR}$  evolution as a function of redshift (x-axis) and stellar mass $M_*$ (colour bar). For comparison, other trends with redshift are taken from the literature (black lines): \citeauthor{bell} (\citeyear{bell}, dotted); \citeauthor{magnelli} (\citeyear{magnelli}, dashed); \citeauthor{delhaize} (\citeyear{delhaize}, dot-dashed) and their AGN-corrected version after removing 2$\sigma$ outliers (triple dot-dashed lines). Figure from \citet{delvecchio1}.}
\label{delvecchio}   
\end{figure}

Another step forward in the comprehension of the relation between radio and IR luminosities and of its cosmic evolution is provided by the work of \citet{delvecchio1}. These authors again use the 3GHz-selected sample of \citet{smolcic3} to calibrate for the first time the $q_{\rm IR}$ parameter both as a function of redshift and stellar mass. Starting from a sample of $\sim 400,000$ -- $M_*> 10.5\,M_\odot$, $0.1<z<4.5$ --  star-forming galaxies selected in the deep COSMOS UltraVISTA catalogue from \citet{muzzin}, and relying on radio and IR stacking techniques, these authors conclude that the main driver of the observed $q_{\rm IR}-z$ evolution is the stellar mass, with more massive galaxies systematically displaying lower $q_{\rm IR}$ (cf.\ Fig.~\ref{delvecchio}). 
A secondary, weaker dependence on redshift is also observed, leading to a total functional form of the kind:  $q_{\rm IR}(M_*,z)= (2.626\pm 0.024) (1+z)^{-0.023 \pm 0.008}$--$(0.148\pm 0.013) (\log M_*/M_\odot- 10)$. No dependence is instead observed on SFRs.

On the other hand, \citet{bonato}, when using LOFAR-selected star-forming galaxies from the Lockman Hole field, do find a dependence of the SFR/$L_{150\, \rm MHz}$ relation on stellar mass, but only for $z\simlt 1$ and weaker than what obtained by \citet{delvecchio1} (decrement of $\simlt 0.1$ as opposed to $\sim 0.23$ for increasing masses), while  \citet{smith} report for LOFAR DR1 observations of the ELIAS-N1 field a linear SFR/$L_{150\rm MHz}$ relation with very little (possibly differential due to mass dependencies) cosmological evolution at least up to  $z\sim 1$. 
A linear relation between SFR and 1.4 GHz luminosity is also observed by \citet{tisanic} by using a local, $z\simlt 0.1$,  250$\mu$m-selected sample from the \textit{Herschel} Astrophysical Terahertz Large Area Survey (H-ATLAS;  \citealt{eales1}) with both radio detections and stacking information from FIRST and NVSS. This is however at variance with the results of \citet{molnar1}, who further claim no redshift dependence for the $q_{\rm IR}$ parameter, since this would merely arise as the consequence of selection effects, whereby different parts of the non-linear IR-radio correlation are probed for varying redshift ranges and sample depths. From the above discussion it is clear that, despite all the progress made in the recent years, a convergence on $q_{\rm IR}$ -- especially for what concerns its cosmological evolution -- is still far from being reached. 

\section{The environments of radio-active AGN and star-forming galaxies}
\label{sec:3}
This Section presents results related to the environmental properties of radio-AGN and -- to a lesser extent -- of radio-emitting star-forming galaxies. We have chosen to divide them into three categories: those obtained with clustering studies (Sect.~\ref{sec:3.1}), those derived by pinpointing radio-AGN within known structures (Sect.~\ref{sec:3.2}) and those stemming from searching for over-densities around known radio-AGN (Sect.~\ref{sec:3.3}). As already discussed in the Introduction, the choice is dictated by the fact that these different approaches present different points of view that then need to be integrated with each other in order to return a complete and coherent picture of the large-scale behaviour of radio galaxies. 

\subsection{Methods for the detection of environment 1: clustering analyses}
\label{sec:3.1}

\subsubsection{Useful definitions}
\label{sec:3.1.1}

This paragraph is devoted to define a number of quantities which will be used throughout the section.

\begin{description}
\item[\textbf{a) The spatial two-point correlation functions}]
The \emph{spatial two-point auto-correlation} (or more simply, just \emph{correlation}) \emph{function} $\xi(r,z)$ is one of the most commonly used tools to investigate the large-scale structure traced by a population of sources. 
It is defined as the excess probability with respect to a Poissonian (random) distribution of finding two galaxies at a distance $r$ from each other within the volume element $\delta V$. By following \citet{Peebles}, this can be expressed as:
\begin{equation}
\delta P= \bar{N} \left[1+\xi(r,z)\right] \delta V,
\label{xi}
\end{equation}
where $\delta P$ is the probability and $\bar{N}$ is the mean number density of galaxies, and where we have made it explicit that the quantity $\xi$ depends on both scale $r$ and redshift $z$. In most cases of astrophysical (i.e., non-cosmological) interest, $\xi(r,z)$ can be parametrized as a power-law $\xi(r,z)=\left[r/r_0(z)\right]^{-\gamma}$, with slope $\gamma$  and correlation (or clustering) radius (or length) $r_0$. Higher values of $r_0$ imply that the analysed sources are more strongly clustered. 

Closely related to the two-point correlation function is the \emph{two-point cross-correlation function $\xi_{\rm AB}(r,z)$} which compares objects coming from two different datasets A and  B and is defined as the joint probability of finding an object from A in the volume element $\delta V_{\rm A}$ and object from B in the volume element $\delta V_{\rm B}$ at a distance $r$ (\citealt{Peebles}):
\begin{equation}
\delta P_{\rm AB}=\bar{N}_{\rm A}\bar{N}_{\rm B} [1+\xi_{\rm AB}(r,z)]\delta V_{\rm A}\delta V_{\rm B},
\end{equation}
where $\bar{N}_{\rm A}$ and $\bar{N}_{\rm B}$ are the mean number densities of the two sets of objects. The cross-correlation function can also be fitted as a power-law in the same manner as the auto-correlation function discussed above.\\
From a practical point of view, we note that cross-correlation analyses are necessary whenever the sample of interest is too small to allow for meaningful statistical studies of the auto-correlation function 
so that a second, much larger, dataset has to be considered in order to beat the uncertainties. This is indeed the reason why this method was so widely used in the early works that tackled the issue of Large-Scale Structure as traced by radio sources.

\item[\textbf{b) The angular two-point  correlation function}]
In the absence of a complete set of redshift information as is the case for many large-area surveys, one can rely on the study of the projected distribution onto the sky of a population of sources via the two-point angular correlation function $\omega(\theta)$. This is defined as  the excess probability with respect to a Poissonian distribution of finding two galaxies at an angular distance $\theta$ from each other within the solid angle element $\delta \Omega$. In a very similar manner to Eq.~(\ref{xi}), the above definition can be expressed as:
\begin{equation}
\delta P_{\Omega}=\bar{N}_\Omega \left[1+\omega(\theta)\right] \delta \Omega,
\label{w}
\end{equation}
with $\bar{N}_\Omega$ mean surface density over the region that subtends the solid angle $\Omega$, and $\delta\Omega$ surface area element. As it was for the spatial correlation function, also $\omega(\theta)$ can be expressed as a power-low in most of the cases of astrophysical interest and parametrized as $\omega(\theta)=A\theta^{1-\gamma}$, where $\gamma$ is the same as in the definition of $\xi$ and $A$ is the amplitude.

\item[\textbf{c) The Limber equation}]
The Limber equation (\citealt{limber, Peebles}) provides the standard way to relate the angular two-point correlation function to the spatial two-point correlation function, once the redshift distribution of the sources under consideration is known. It can be expressed as:
\begin{equation}
\omega(\theta)=2\frac{\int_0^\infty \int_0^\infty   F^{-2}(x) x^4 \Phi^2(x) \xi(r,z) dx \;du}{\left[  \int_0^\infty F^{-1}(x) x^2 \Phi(x) dx \right] ^2},
\label{limber}
\end{equation}
where $x$ is the comoving coordinate, $F(x)$ gives the correction for cosmic curvature, and the selection function $\Phi(x)$ satisfies the relation
\begin{equation}
\bar{N}_\Omega=\int_0^\infty \Phi(x)F^{-1}(x) x^2 dx=\frac{1}{\Omega}\int_0^\infty N(z) dz,
\end{equation}
in which $\bar{N}_\Omega$ is as in Eq.~(\ref{w}) and $N(z)$ is the redshift distribution of the sources defined as the number of objects in a given survey within the shell ($z$, $z+dz$). If $\xi(r,z)$ can be parametrized as a power-law of the form $\xi(r,z)=\left[r/r_0(z)\right] ^{-\gamma}$, where all the dependency on $z$ is absorbed by the clustering length $r_0(z)$, then Eq.~(\ref{limber}) reduces to an integral over $z$ (cf.\ \citealt{maglio2})
\begin{equation}
\omega(\theta)=\frac{H_\gamma \int_0^\infty N(z)^2 [x(z) \theta]^{1-\gamma} r_0(z)^\gamma F(z) (dx/dz) \; dz}{\frac{c}{H_0}\left[ \int_0^\infty N(z) dz\right] ^2},
\end{equation} 
with $H_\gamma=\Gamma[1/2]\Gamma[(\gamma-1)/2]/\Gamma[\gamma/2]$, $c$ speed of light and $H_0$ Hubble constant, expressed as $h \;\cdot$ 100 km s$^{-1}$ Mpc$^{-1}$.

\item[\textbf{d) The bias parameter}]
The differences in the clustering properties of different classes of extragalactic
sources and between these and the background (dark) matter distribution motivate the introduction of
the bias parameter $b$, defined as (cf.\ \citealt{kaiser}):
\begin{equation}
b^2=\frac{\xi_{\rm gal} }{\xi_{DM}},
\label{b}
\end{equation}
where $\xi_{\rm gal}$ is the spatial correlation function of a population of galaxies/AGN and $\xi_{\rm DM}$ that of the dark matter. 
In the case of cross-correlations between a population A and a population B, the bias $b_{\rm AB}$ can be written as:
\begin{equation}
b_{\rm AB}^2=\frac{\xi_{\rm AB} }{\xi_{DM}}=b_{\rm A}\cdot b_{\rm B}, 
\label{bc}
\end{equation}
where $b_{\rm A}$ and $b_{\rm B}$ are expressed as in Eq.~(\ref{b}).

\begin{figure}
\centering
\includegraphics[scale=0.5]{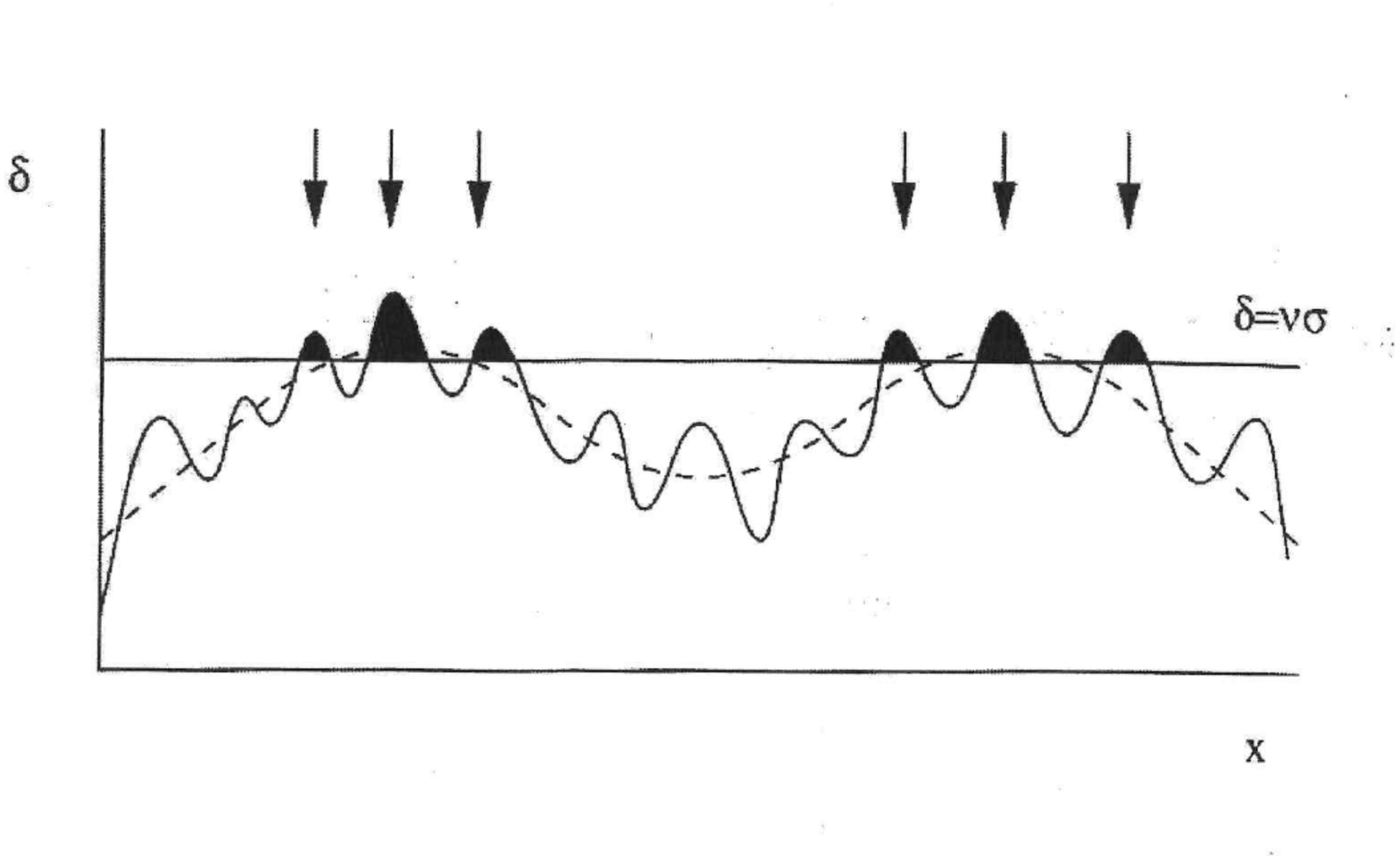}
\caption{The high peak bias model. If we decompose a density field into a fluctuating component on galaxy scales, together with a long-wavelength "swell" (broken curve), then those regions of density that lie above the critical density threshold for collapse of $\delta=\nu\sigma$, with $\sigma$ variance of the density field on a scale $R$ corresponding to a mass $M$, will be strongly clustered. If proto-galaxies are assumed to form at the sites of these high peaks (shaded and indicated by arrows), then this population has a non-uniform (i.e., biased) spatial distribution even before dynamical evolution of the density field. Figure from \citet{peacock1}.}
\label{high_peak}       
\end{figure}

From the theoretical point of view, the paradigm of \emph{biased galaxy formation} states that the efficiency of galaxy formation is higher in dense environments since an object of a given mass will collapse sooner if it lies in a region of large-scale over-density, leading to an enhanced abundance of such objects with respect to the mean (``natural bias'' -- \citealt{white}). These rare high peaks of a Gaussian random field will then be significantly more clustered than the underlying Gaussian field itself (Fig.~\ref{high_peak}).  \citet{mo} provided an expression for the bias $b$  starting from the extension of the \citet{press} formalism which reads:
\begin{equation}
b(M,z|z_F)=\left(1+\frac{\nu_F^2-1}{\delta_F^2}\right),
\label{bth} 
\end{equation}
where $M$ is the mass of the dark matter haloes, $z_F$ is the redshift of formation of such haloes, and $\nu_F=\delta_F/\sigma(z)$, with $\sigma(z)$ variance of the density field on a mass-scale $M$ at a redshift $z<z_F$ and $\delta_F=\delta_c(1+z_F)/(1+z)$ ($\delta_c$ is the critical linear over-density for spherical collapse). 
From the above equation it is possible to appreciate that $b$ depends not only on halo mass (i.e., more massive haloes cluster more strongly), but also on the epoch when the haloes collapse, being higher for haloes formed at earlier times. Note that there have been suggestions that $b$ could also be stochastic, non-linear and scale-dependent (e.g., \citealt*{dekel, narayanan}), but an in-depth analysis of this topic is outside the scopes of the present review.

Following the \citet{mo} approach, one then has that the more massive haloes (corresponding to rare peaks in the density field) that collapse earlier in time will be more clustered than haloes of lower mass that collapse in more recent epochs. Assuming that all galaxies are hosted within dark matter haloes, this for instance implies that early-type galaxies -- associated with older and more massive haloes -- in general cluster more strongly than late-type ones.  

The above scenario holds for a one-to-one association between galaxies/AGN and dark matter haloes. As in general dark matter haloes are occupied by more than one galaxy/AGN, in more recent years the Mo \& White approach has been implemented with the so-called Halo Occupation framework (e.g., \citealt{scoccimarro}, see \citealt{maglio5} for an application to 2dFGRS galaxies), which relates the clustering properties of a chosen population of extragalactic objects with the way such objects populate their dark matter haloes. Within this scenario, the two-point correlation function $\xi$ can be written as the
sum of two components, $\xi_{1h}$ and $\xi_{2h}$, where the first quantity accounts for pairs of galaxies residing within the same halo, while
the second one takes into account the contribution to the correlation function of galaxies belonging to different haloes. 
However, since most of the clustering studies presented in this review do not have enough signal-to-noise to investigate the properties of the one-halo term $\xi_{1h}$ and, from a more physical point of view, since for the overwhelming majority of radio-AGN the approximation one source per halo is a good one (e.g., \citealt{maglio6}), in the following discussion we will stick to the \citet{mo} formalism.

The main point to be taken away from this discussion is that the biased galaxy formation framework has the important advantage of relating the clustering properties of a selected population of sources with the dark matter content of the structures that host them, allowing to provide statistical estimates for their masses. From an operational point of view, the rule of thumb is that, at fixed $z$, sources more strongly clustered (i.e., with higher values for $r_0$ in the parametrization of $\xi(r)$ as a power-law) will present larger values of the bias parameter $b$ (cf.\ equation \ref{b}) and therefore (via equation \ref{bth}) will be hosted by more massive haloes. 
\end{description}

\subsubsection{Early works on radio-galaxy cross-correlations}
For many years after the discovery of radio-emitting sources, their distribution was thought to be isotropic (e.g., \citealt{Webster, Webster1}). 
First attempts to investigate the Large-Scale Structure traced by radio-AGN were presented in the work of \citet{seldner}. These authors considered radio sources from the 4C catalogue (\citealt{pilkington}) and correlated them with galaxies from the Lick catalogue (\citealt{seldner2}), finding that the shape of the cross-correlation function was the same as that of the galaxy-galaxy correlation function, but presented an amplitude which was about 10 times higher. By also interpreting some tentative results on the auto-correlation function of radio sources, \citet*{seldner} concluded that radio-AGN are not randomly distributed, but rather tend to be found amongst strongly clustered galaxies. This result was confirmed by the work of \citet{longair}, who further reported a distinct behaviour in faint ($L_{178\, \rm MHz}\sim 10^{22}$--$10^{24}\, \mathrm{W\, Hz^{-1}\ sr^{-1}}$) and powerful ($L_{178\, \rm MHz}\sim 10^{24}$--$10^{25}\, \mathrm{W\, Hz^{-1}\ sr^{-1}}$) radio galaxies, whereby the latter population (with the noticeable exception of classical double-lobed AGN) prefers to reside in regions of space where the amplitude of the spatial cross-correlation function is about four to five times that of inactive galaxies, while weak radio emitters show no tendency to belong to overdense structures. \citet{Masson} also observed the same effect in the 6C catalogue (\citealt{masson1}), while \citet*{seldner1} speculated on the possibility that this result could be due to faint radio sources residing at higher redshifts than brighter ones. 

Some years later, \citet*{prestage} used cross-correlation techniques to derive the environmental properties of 118 radio-AGN taken from a number of surveys, amongst which the $S_{2.7 \rm GHz}\ge 2$ Jy Parkes (\citealt*{wall}) catalogue. For the first time, sources were divided according to their morphology. It was found that, while compact radio sources did not appear in regions of enhanced galaxy density, the same was not true for extended radio-AGN, especially FRI galaxies which were typically found within overdense structures. 

The first evolutionary studies were those presented by \citet*{yates}, who analysed the environments of 25 radio galaxies drawn from the 3C (\citealt{edge}) and Parkes (\citealt{bolton}) samples, with flux densities $>1$ Jy and redshifts between $z=0.15$ and $z=0.8$, although only 11 sources had $z>0.35$. These authors found that the most luminous, $\langle L_{178\, \rm MHz}\rangle=10^{27.1}$ W Hz$^{-1}$, sources at $z\sim 0.5$ lay in rich environments, while the same population at $z\sim 0.3$ occupied structures which were roughly three times sparser. However, given the limited size of their sample, it was not possible to conclude whether the observed trend was due to cosmological or luminosity evolution.
\citet*{ellingson} instead performed a comparative study between the environments inhabited by radio-active and radio-quiet quasars in the redshift range $0.3<z<0.6$, finding that the latter population resided in structures which were less rich than those occupied by radio-active quasars. This discovery made the authors conclude that radio-active and radio-quiet quasars did not belong to the same parent population simply observed with a different viewing angle, but were different physical objects. Furthermore, based on the morphology-density relation by \citet{dressler}, \citet{ellingson} speculate on the possibility that radio-active quasars could be hosted by elliptical galaxies, while radio-quiet quasars by spirals. Finally, in agreement with the results by \citet{yates}, they also report an evolution of the environmental properties of radio-active quasars, since a larger number was observed to inhabit regions of enhanced density at the highest redshifts probed by their analysis. No relationship was instead found between environmental properties and radio luminosities.

\subsubsection{The correlation functions of radio sources}
Although some more studies (mostly of small high-redshift samples) kept relying on the cross-correlation function between radio-AGN and galaxies (e.g., \citealt{best1, mclure2, hickox, ramos,Lindsay1}), the advent in the 1990s of deep ($\sim$ mJy level), large-area radio surveys such as the Faint Images of the Radio Sky at Twenty centimeters (FIRST -- \citealt{becker}) and the NRAO VLA Sky Survey (NVSS -- \citealt{condon1}) allowed to probe samples of objects about $10^3-10^4$ times more numerous than what was possible in the past. This in turn allowed for the first time studies of the environmental properties of extragalactic radio sources by means of their auto-correlation function. And, even more importantly, for the first time these studies permitted to derive cosmological information from the distribution of radio galaxies all over the sky. 

\begin{figure}
\centering
\includegraphics[scale=0.4]{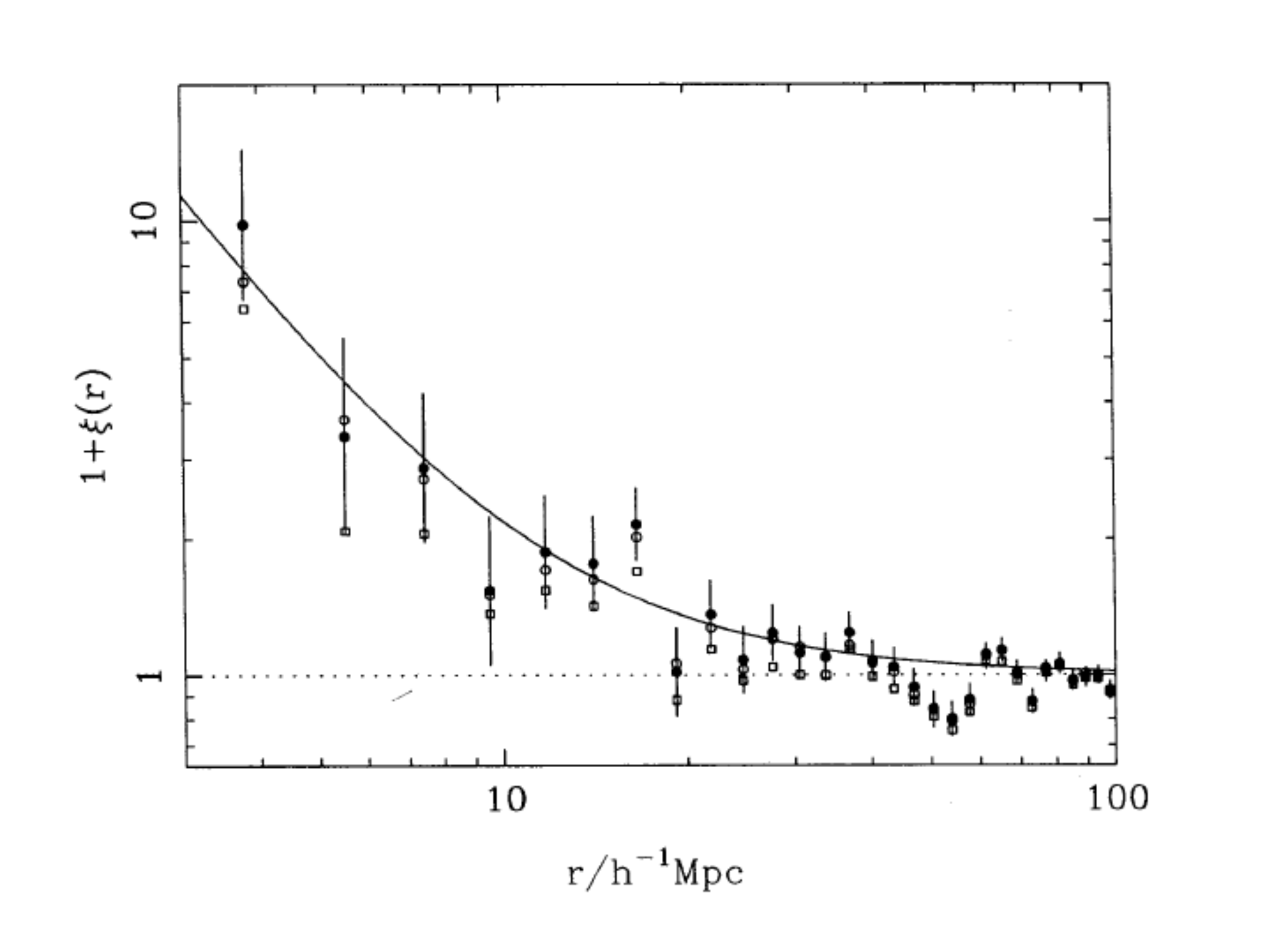}
\caption{The spatial two-point correlation function $\xi$ for the sample of \citet{peacock}. The curve shows the best-fit model $\xi=(r/r_0)^{-1.8}$, with $r_0=$11 Mpc/h. Figure from \citet{peacock}.}
\label{peacock}       
\end{figure}

First works in this direction were presented by \citet{peacock} who analysed the spatial two-point correlation function for a sample of 310 local ($z<0.1$) and bright ($F_{1.4 \rm GHz}> 0.5$ Jy) radio galaxies for which spectroscopic redshifts were available, finding that it showed a power-law shape that could be parametrized  as $\xi(r)=\left(r/r_0\right)^{-\gamma}$, with $\gamma\sim 1.8$ and $r_0\sim 11$ Mpc/h (cf.\ Fig.~\ref{peacock}). Noticeably, while the value of $\gamma$ resembled that found for correlation function analyses of the more general galaxy population within the same redshift range, that of the correlation length $r_0$ was about two times higher ($\sim 5.4$ Mpc/h -- e.g., \citealt{baugh}) and closer to the value found for clusters of galaxies ($\sim 15$ Mpc/h -- e.g., \citealt{dalton}), in agreement with the early findings of \citet*{seldner}.

But it was really with the work of \citet{cress} that correlation studies of radio sources entered a new era. Indeed, these authors  investigated the clustering properties of $\sim 138,000$ FIRST radio galaxies down to 1.4 GHz fluxes of 1 mJy, finding that the angular (since no 3D information for most of the sources considered in the Cress et al. analysis was available) correlation function could be parametrised as $\omega(\theta)=A \theta^{1-\gamma}$, with $A\sim 3\cdot 10^{-3}$ and $\gamma \sim 2.1$. By also considering only sources fainter than 2 mJy, \citet{cress} found that $\omega(\theta)$ was shallower than what observed for the whole population, and interpreted this flattening as due to the contribution of star-forming galaxies, expected to dominate the radio counts below the mJy level (e.g., \citealt{prandoni1}).

One problem that affected the \citet{cress} analysis was the presence of extended and multi-component radio-AGN within their sample which, when overlooked or erroneously combined together, produced spurious signal (excess if considering each radio component as a single source or defect in the case all sources within a chosen distance from each other are combined together regardless of their physical association) in the correlation function on scales smaller than $\sim 0.2^\circ$. This issue was tackled in better detail by \citet{maglio}, who provided a simple, but effective recipe for combining multi-component sources together based on the brightness and relative distance of the various members of a single object. The formula read:
\begin{equation}
\theta_{\rm link}=100\left(\frac{F_{\rm TOT}}{100}\right)^{0.5} \rm arcsec,
\label{eq:perc}
\end{equation}
where $F_{\rm TOT}$ is the the sum of the fluxes of the components of each apparent multiple source. Equation \ref{eq:perc} was then combined with a criterion for the ratio of the fluxes of the components  to be around 1, expected for real aggregations since fluxes in this case are correlated. After a careful investigation, \citet{maglio} then chose to combine pairs of sources only if their distance $\theta_{\rm link}$ was below that identified by equation \ref{eq:perc} and if their fluxes differed by a factor less than 4 (i.e., $1/4\le f_1/f_2\le 4$, with $f_1$ and $f_2$ fluxes of two components of a single radio-AGN). Another difference between the work of \citet{maglio} and that of \citet{cress} was that the former authors considered a brighter flux limit of 3 mJy for the sources to be considered in their analysis: this was due to the known incompleteness of the FIRST catalogue below those flux levels. Such a sample, corrected for flux incompleteness and the presence of multicomponent sources, comprised 86,074 objects and was then used to investigate the distribution of the second (variance) and third (skewness) moments of the radio galaxy distribution. 
The results from the variance $\mu_2$ could be then connected to the angular two point correlation function via the relation (\citealt{Peebles}):
\begin{equation}
\mu_2=\bar{N}_\Omega+(\bar{N}_\Omega)^2\Psi_2,
\end{equation}
where $\bar{N}_\Omega$ is the mean count of radio sources in the solid angle $\Omega$, and the reduced variance $\Psi_2$ is expressed as:
\begin{equation}
\Psi_2=\frac{1}{\Omega^2}\int w(\theta) d\Omega_1 d\Omega_2.
\end{equation}
\begin{figure}
\centering
\includegraphics[scale=0.4]{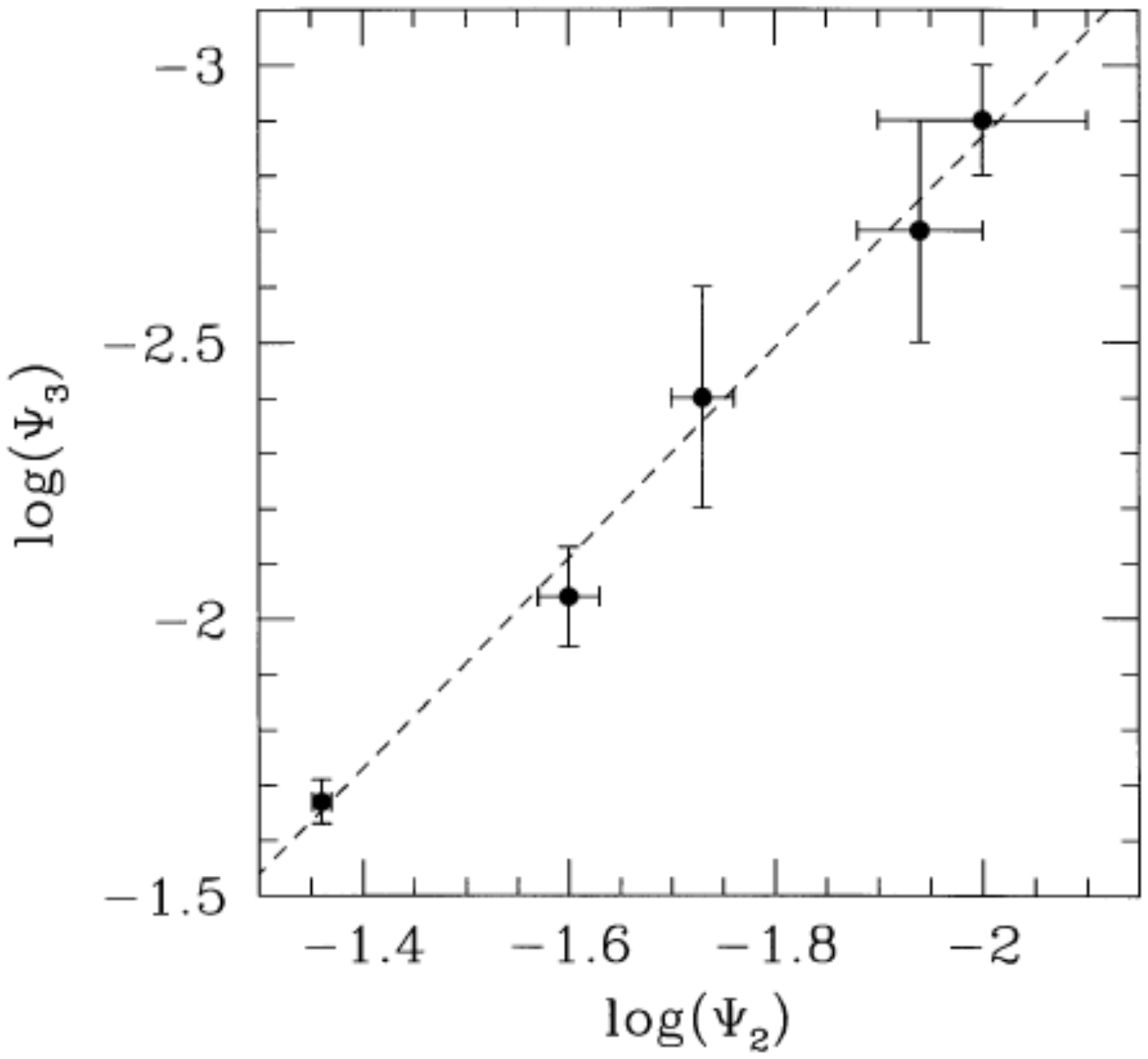}
\caption{Normalized skewness $\Psi_3$ vs normalized variance $\Psi_2$. Errors are estimated from the variance in four random subsets. The dashed line represents the best-fit to the data  obtained for a functional form $\Psi_3= S_3(\Psi_2)^\alpha$, with $\alpha=1.9\pm 0.1$ and $S_3=9\pm 3$. Figure from \citet{maglio}.}
\label{maglio_1998}       
\end{figure}
From the above equations, \citet{maglio} obtained for the amplitude of the two-point correlation function $\omega(\theta)=A \theta^{1-\gamma}$, the value $A\sim 10^{-3}$, in fair agreement with the results of \citet{cress}, but derived a steeper slope, $\gamma\sim 2.5$. Furthermore, by investigating the third moment of the radio galaxy distribution, \citet{maglio} could measure for the first time in an extragalactic radio sample \emph{positive} skewness $\Psi_3$ in the distribution of the sources and assess that variance and skewness were related through the functional form $\Psi_3= S_3(\Psi_2)^\alpha$, with $\alpha\simeq 2$ and $S_3=$constant (cf.\ Fig.~\ref{maglio_1998}). This result has important cosmological implications, since it meant that the large-scale clustering of radio galaxies was in agreement with the gravitational instability picture for the non-linear growth of perturbations from a primordial Gaussian field. 

These findings were subsequently used by \citet{maglio1} to put constraints on the composition and redshift distribution of radio galaxies at the mJy level and also to investigate how their spatial distribution traces that of the underlying dark matter within the biased galaxy formation scenario (e.g., \citealt{mo}, see Sect.~\ref{sec:3.1.1}). It was found that observations could only be matched if one assumed two populations, a less numerous and more local one made of star-forming galaxies whose spatial distribution follows that of the dark matter, and a more numerous population of radio-AGN, much more strongly clustered and presenting a largely biased distribution with the bias parameter $b$ (defined in equation \ref{b}) increasing with look-back time.

A few years later, \citet{Blake} and \citet{Blake1} analyzed the clustering properties of 361,644 radio objects brighter than $F_{1.4 \rm GHz}=15$ mJy in the NVSS survey. These authors did not correct for the issue of multi-component sources, but rather parametrized the two-point correlation function of radio galaxies as the sum of two power-laws: one at small angular scales that mainly described the spurious clustering signal induced by multiple-component and extended sources, and one on scales larger than $\sim 0.2^\circ$ that would supposedly only take into account the cosmological signal produced by radio-AGN. By doing this, they found an amplitude on large angular scales $A\sim 10^{-3}$ and a slope $\gamma\sim 1.8$, values that remained unchanged when raising the flux threshold. Conversely, \citet{overzier} -- by re-analysing the clustering properties of radio sources in the FIRST and NVSS surveys --  found that the amplitude of the projected correlation function increased from $A\sim 10^{-3}$ for objects fainter than 40 mJy to $A\sim 7\cdot 10^{-3}$ at 200 mJy. These authors interpreted their results at high fluxes, mostly dominated by FRII galaxies, as evidence for the fact that these sources preferentially inhabit more massive structures than fainter radio galaxies. By using the redshift distribution models proposed by \citet{dunlop}, \citet{overzier} estimate a correlation length $r_0\sim 14$ Mpc/h (value which depends on the adopted assumption for the cosmological evolution of clustering), in rough agreement with the results of \citet{peacock}.
 
All the aforementioned works contributed in a substantial way at proving that extragalactic radio sources could provide an invaluable contribution to the investigation of the Large-Scale Structure of the Universe. However, better information on their physical and environmental properties as obtained from clustering studies has only been made possible with the combination of radio data with observations at different wavelengths, especially for what concerns the determination of redshifts. To this aim, \citet{maglio6} presented the clustering properties of a local, $z<0.3$, 
$F_{1.4\rm GHz}\ge 1$ mJy sample of 761 FIRST radio galaxies with optical spectra and redshift information from the 2dF Galaxy Redshift Survey (2dFGRS -- \citealt{colless, maglio4}). 
On the basis of such spectra, radio sources could be divided into star-forming galaxies (225 objects) and AGN (536 objects). An analysis of the spatial two-point correlation function indicated that these sources presented clustering properties which were similar to those obtained for a general population of inactive galaxies drawn from the same 2dFGRS parent catalogue (\citealt{hawkins, madgwick}): $r_0\sim 7$ Mpc (h is fixed here to the value of 0.7) and $\gamma\sim 1.6$. However, when the authors only included \textit{bona-fide} AGN, both the correlation length and the slope increased to $r_0\sim 11$ Mpc and $\gamma\sim 2$. No difference was found between faint ($L_{1.4 \rm GHz} \le 10^{22}$ W Hz$^{-1}$ sr$^{-1}$) and bright AGN-fuelled sources. Comparisons with physical models for the clustering properties of galaxies (i.e., \citealt{mo}) further showed that radio-AGN reside within dark matter haloes more massive than $\sim 10^{13.4} \rm\,M_\odot$, value which is higher than what found for radio-quiet quasars in the same 2dFGRS survey (\citealt*{porciani}), and compatible with a scenario that associates radio-AGN with protocluster/cluster-like structures. Taken these results at face value, \citet{maglio6} suggest that radio-quiet quasars and radio-AGN are two different populations, with more massive dark matter haloes required for an AGN to trigger radio activity. Similar values for the halo masses of radio-selected AGN have been obtained by  \citet{mandelbaum} and \citet{hickox}. \citet{mandelbaum} also observe radio-AGN to be associated with more massive dark matter haloes than those hosting both inactive galaxies and optically-selected AGN with the same stellar mass content, finding which implies that radio-activity of AGN origin is enhanced within overdense structures.  \citet{hickox} instead report a higher clustering signal (and therefore larger halo masses) for their radio-selected AGN when compared to those derived for the two other populations of AGN respectively selected in the X-ray and \textit{Spitzer}-IRAC bands ($M_{\rm halo}^{\rm radio} \sim 10^{13.5}\,M_\odot$ vs $M_{\rm halo}^{\rm X} \sim 10^{13}\,M_\odot$ and $M_{\rm halo}^{\rm IR} \simlt 10^{12}\,M_\odot$).

\citet{brand} investigated the clustering properties of 206, $z<0.5$, $F_{1.4\rm GHz}\ge 3$ mJy sources drawn from the Texas-Oxford NVSS Structure (TONS) radio galaxy redshift survey, finding a correlation length $r_0\sim 9$ Mpc for $\gamma$ fixed to the standard 1.8 value. Large variation in the amplitude of the two-point correlation function -- possibly due to cosmic variance -- was observed amongst the various fields of the TONS survey. \citet{wake} instead analysed a sample of 250, $z\sim0.55$ radio galaxies drawn from the FIRST and NVSS surveys with spectroscopic redshifts determined from the 2SLAQ Luminous Red Galaxy (LRG) redshift survey (\citealt{cannon}). They found that the correlation length was $r_0\sim 12$ Mpc/h ($\gamma\sim 1.75$), higher than the value obtained for the population of radio-quiet LRGs ($r_0\sim 9$ Mpc/h) matched in redshift and optical luminosity. These results implied radio-active LRGs to reside within haloes of typical masses $\sim 10^{14}\,M_\odot$/h, higher than the value of $\sim 6\cdot 10^{13}\,M_\odot$/h obtained for radio-quiet LRGs, and called for an enhanced probability for LRGs to become radio-emitters if hosted within (more) massive structures. A very similar result was obtained by \citet{Fine} who matched NVSS and FIRST radio galaxies with a larger catalogue of LRGs endowed with photometric redshifts. As in \citet{maglio6}, no difference in the clustering properties was found for sources of different radio luminosity. \citet{donoso1} also concentrated on samples of radio-AGN and phometrically-identified LRGs. However, at variance with \citet{Fine}, they did find a dependence on radio luminosity on scales below $\sim 1$ Mpc, and also reported different clustering behaviours for the two sub-classes of radio-galaxies and radio-active quasars, with radio-galaxies clustering more strongly than radio-active quasars matched both in black hole mass and radio luminosity. This result led the authors to conclude that -- in disagreement with AGN unification models (e.g., \citealt{antonucci}) -- radio galaxies and radio-active quasars were powered by different physical mechanisms. 

\citet{Lindsay} considered $F_{1.4\rm GHz}\ge 1$ mJy FIRST radio sources with spectroscopic and photometric information from the Galaxy and Mass Assembly (GAMA -- \citealt{driver}) survey, finding for the two-point correlation function at $z\sim 0.48$ a correlation length $r_0\sim 8$ Mpc/h and a slope $\gamma\sim 2.15$, with a value for the bias $b(z=0.48)\sim 2$. \citet*{Lindsay1}  instead concentrated on 766 radio sources brighter than $F_{1.4 \rm GHz}=90 \mu$Jy drawn from the VIDEO survey (\citealt{jarvis2}) and, by cross-correlating this population with $K_s\le 23.5$ galaxies observed within the same field, found that their clustering signal sensibly increased with look-back time. This in turn implied an increment of the bias function $b(z)$ from the value of $\sim 0.6$ at $z\sim 0.3$ to $\sim 8$ at a redshift $z\sim 2.2$. \citet{Lindsay1}  also report a noticeable dependence of the clustering strength on the radio luminosity of the sources even though, with the luminosity range probed by their work, their finding simply restated the well-known effect of (radio brighter) radio-AGN preferentially residing in denser environments than (radio fainter) star-forming galaxies.

\begin{figure*}
  \centering
\includegraphics[width=0.49\textwidth]{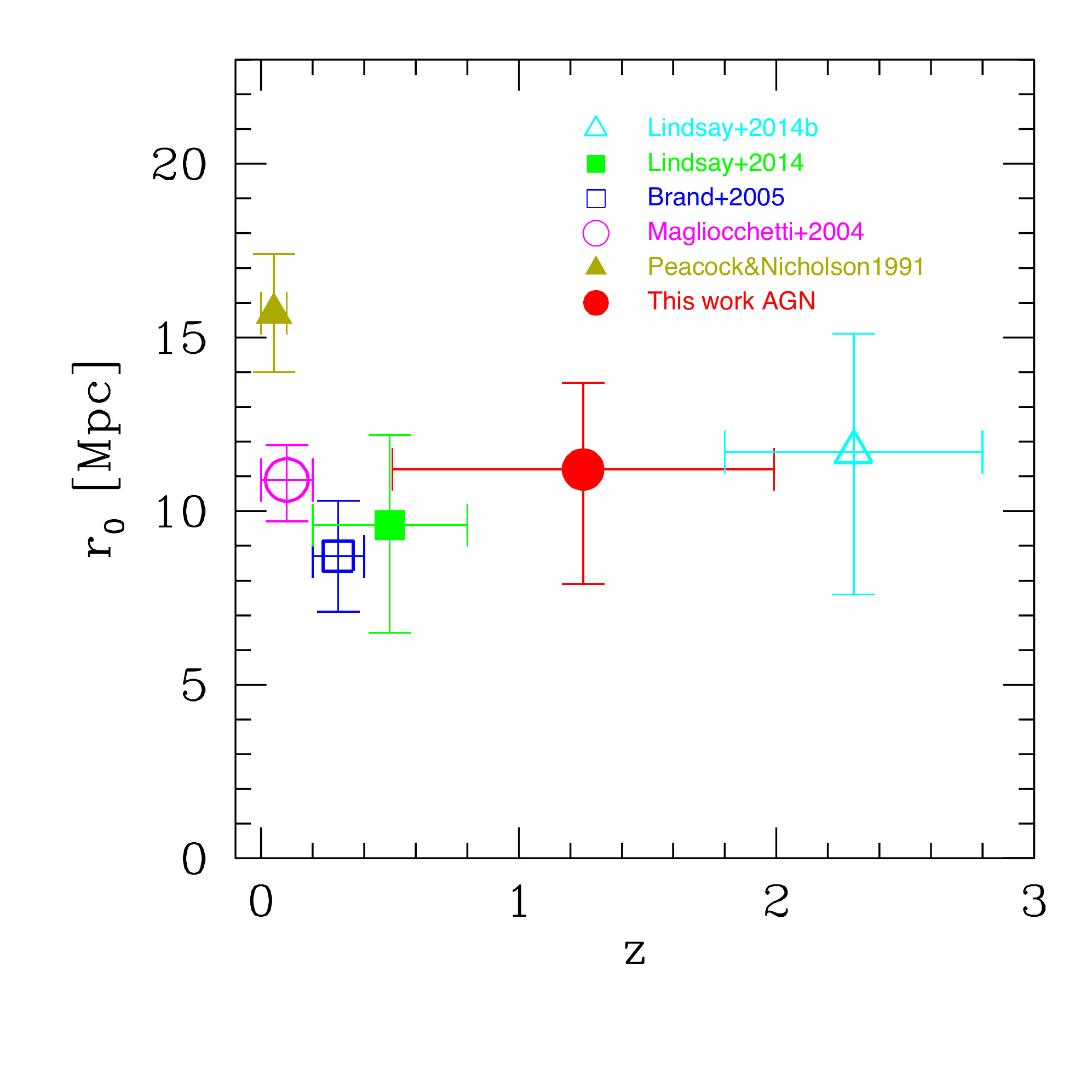}
\includegraphics[width=0.49\textwidth]{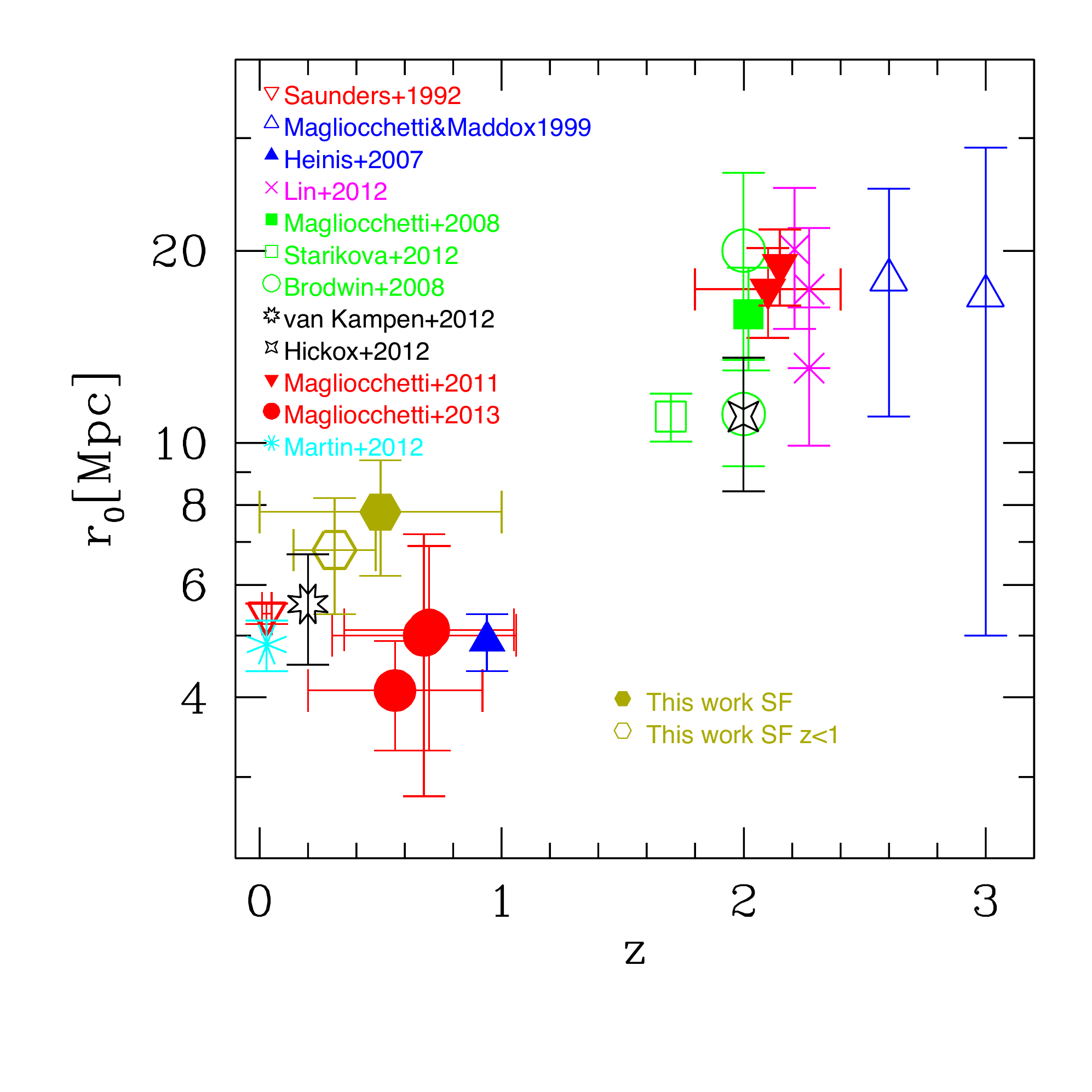}
\caption{Redshift evolution of the clustering length $r_0$ in the case of radio-selected AGN (left-hand panel) and star-forming galaxies (right-hand panel). The left-hand panel compares the results from the work of \citet{maglio16} with those of \citet{peacock}, \citet{maglio6}, \citet{brand}, \citet{Lindsay} and \citet{Lindsay1} . The right-hand panel instead includes star-forming galaxies selected in different ways: Far-Infrared (\citealt{saunders, maglio10, maglio11, hickox1, vankampen}), UV (\citealt{maglio2, heinis}), Mid-Infrared (\citealt{brodwin, maglio9, starikova}), BzK method (\citealt{lin}), HI emission (\citealt{martin}) and radio at 1.4 GHz (\citealt{maglio16}, where the entire population of star formers is represented by the full hexagon, while the empty one is for more local, $z < 1$, sources). Figure from \citet{maglio16}.}
\label{maglio_2017}       
\end{figure*}

\citet{maglio16} considered $\sim 1000$ radio sources from the VLA-COSMOS survey (\citealt{bondi}) endowed with photometric redshifts. By relying on their radio-luminosity as it was done in \citet{maglio13, maglio14, maglio15} (cf.\ Sect.~\ref{sec:1}), these objects were classified as AGN (644 sources) and star-forming galaxies (247). Their clustering properties returned values for the correlation length of $\sim 7.8$ Mpc for star-forming galaxies and $\sim$11 Mpc for radio-AGN ($h=0.7$). This implied radio-AGN to be hosted within dark matter haloes more massive than $\sim 10^{13.6}\, M_{\odot}$, in full agreement with the results obtained both locally by \citet{maglio6} and \citet{mandelbaum} and in the more distant universe by \citet{hickox}, \citet{hatch1} and \citet{Allison} (cf.\ Sect.~\ref{sec:3.3}), and in line with the idea that radio-active AGN preferentially reside in group/cluster-like structures. $M_{\rm halo} > 10^{12.7}\,M_\odot$ was instead found for star-forming galaxies (group which -- we remind -- given the adopted selection method also includes radio-quiet AGN) at $z <0.9$. 
These authors took a step further and compared the above results with information on the stellar mass content of the galaxies harbouring a radio-active source as derived for the COSMOS population by \citet{laigle}. This allowed to estimate the relationship between dark and luminous matter in both populations: for radio-AGN it was found $\langle M_*\rangle/M_{\rm halo}< 10^{-2.7}$ (here $\langle M_*\rangle$ is the average stellar mass), while for star-forming galaxies they obtained $\langle M_*\rangle/M_{\rm halo}< 10^{-2.4}$ at all redshifts and $\langle M_*\rangle/M_{\rm halo}< 10^{-2.1}$ at $z <0.9$, result that clearly showed the cosmic process of stellar build-up as one moves towards the more local universe. Lastly, comparisons of the observed abundance of radio-AGN with that predicted by cosmological models for dark matter haloes more massive than the value of $10^{13.6}\,M_\odot$  obtained from the above clustering measurements, indicated that about one in two $z<2.3$ haloes is associated with a radio-emitting black hole. This led to estimates for the life-time of the AGN radio-active phase of $\sim 1$ Gyr, possibly suggesting evidence for a recurrent phenomenon, whereby each halo undergoes multiple episodes of radio-AGN activity between $z=2.3$ and $z=0$.

\citet{maglio16} further provide a compilation of the results obtained to date for what concerns the large-scale properties of AGN active in the radio band (mainly at 1.4 GHz) and star-forming galaxies. Figure~\ref{maglio_2017} reports the various values taken from the literature for the correlation length $r_0$ of radio-AGN of different luminosity (left-hand panel) and star-forming galaxies selected at all wavelengths (right-hand panel). Once again, radio-AGN were classified as such on the basis of their luminosity as in \citet{maglio13, maglio14, maglio15}.
Note that the results from \citet{wake}, \citet{donoso1} and \citet{Fine} were not included in the AGN compilation as they refer to sub-populations of radio-active LRGs which are known to present enhanced clustering strengths due to the properties of their parent population made of very massive galaxies.

Comparisons of the various works clearly show that:

\begin{enumerate}
\item In the case of radio-AGN there is no dependence of their clustering signal on either radio luminosity or redshift. This  implies similar environmental properties (and therefore similar hosts) at all different radio luminosities and -- possibly more importantly -- no evolution of such properties throughout cosmic epochs, from $z \simeq 2.5-3$ down to the very local universe.\\
\item In the case of galaxies which form stars at relatively high rates, ${\rm SFR} \simgt 30\,M_\odot$ yr$^{-1}$, there is no difference between the environmental properties of objects selected at different wavelengths, including the radio band. This also holds irrespective of the star-forming rate, as long as it is above the aforementioned value. The striking feature highlighted by clustering studies of star-forming galaxies is that in the relatively local universe these sources are all hosted by relatively small dark-matter haloes, of the order of $\sim 10^{12}\,M_\odot$ or less, while beyond $z\sim 1.5$ vigorous star-formation is only found within very massive, $M_{\rm halo} > 10^{13.5}\,M_\odot$ structures. Taking these results at face value implies that star-forming galaxies at high and low redshifts belong to two different populations, with intense star-forming activity at $z \simgt 1.5$ preferentially happening in protocluster- or cluster-like structures (\citealt{maglio12}).
\end{enumerate}
  
\begin{figure}
\centering
\includegraphics[scale=0.4]{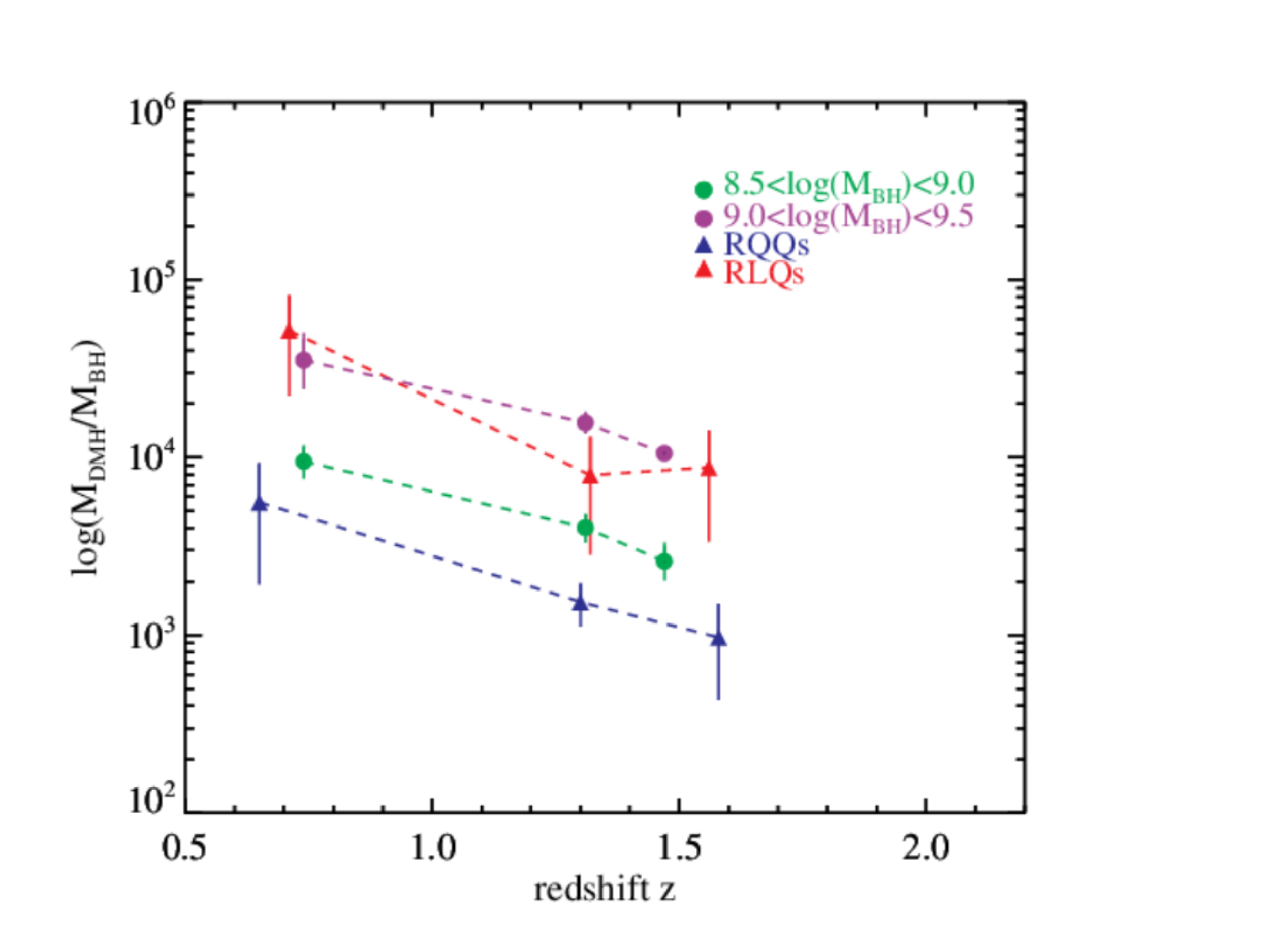}
\caption{Ratio between dark matter halo (DMH) masses and average virial black hole masses for the quasar samples in \citet{retana} as a function of redshift. The labels RLQs and RQQs respectively correspond to radio-active and radio-quiet quasars. Figure from \citet{retana}.}
\label{retana_fig}       
\end{figure}

\citet{retana} also used the large-scale structure results obtained from large samples of radio-active and radio-quiet quasars selected in the $0.3<z<2.3$ range to infer information on the physics of these two populations. They find that radio-active quasars are more strongly correlated at all redshifts, with mean dark matter host masses of $\sim 5 \cdot 10^{13}\,M_\odot$/h, to be compared with those of radio-quiet quasars for which they find $\sim 2 \cdot 10^{12}\,M_\odot$/h. Intriguingly, the large-scale properties of quasars are also found to depend on their black hole mass, with quasars powered by more massive black holes clustering more strongly than quasars harbouring less massive nuclei (cf.\ Fig.~\ref{retana_fig}). For black hole masses above $10^9\,M_\odot$, radio-quiet quasars are found to have the same clustering properties and therefore the same environmental properties of radio-active quasars. This suggests that for the onset of the radio-active phase, black holes need to be at least more massive than the aforementioned value (also see e.g., \citealt{laor, mclure4, metcalf}). 

\citet{hale} were the first to present statistically significant clustering estimates for a population of radio sources selected at frequencies different from 1.4 GHz. These authors consider 2937 objects observed at 3GHz in the COSMOS field (\citealt{smolcic3} ) and divide them into star-forming galaxies and radiatively efficient (roughly corresponding to HERGs) and inefficient (roughly corresponding to LERGs) AGN on the basis of a number of criteria, such as AGN emission in other bands. By restricting their analysis to $z<1$, these authors find for star-forming galaxies values for the halo mass $\sim 3-5 \cdot 10^{12}\,M_\odot$/h, while for AGN $M_{\rm halo}\sim 1-2 \cdot 10^{13}\,M_\odot$/h, in line with previous studies. By further analyzing the two populations of high-efficiency and low-efficiency radio-AGN separately, it was found that low-efficiency AGN cluster slightly more strongly, and therefore inhabit slightly more massive haloes than high-efficiency AGN ($M_{\rm halo}\sim 3-4 \cdot 10^{13}\,M_\odot$/h vs $M_{\rm halo}\sim 1-2 \cdot 10^{13}\,M_\odot$/h). 
A dependence of the clustering strength on the radio luminosity of the AGN was also reported, whereby higher-luminosity sources seem to be hosted by more massive haloes than lower-luminosity ones.

Other recent works which use observations at frequencies different from 1.4 GHz are those of \citet*{rana} and \citet{chakraborty}. The first paper uses GMRT observations at 150 MHz of about 90\% of the whole sky (TGSS survey) and therefore can only provide information on projected quantities, which nevertheless show that the 2D clustering amplitude increases with radio flux threshold, hinting to the conclusion that brighter objects are hosted by more massive haloes. It should however be noted that the clustering signal as measured from TGSS exhibits an unexplained excess mostly on large ($\theta\simgt 0.3^\circ$) scales with respect to that derived from NVSS (\citealt{dolfi}). The second paper instead concentrates on the much smaller region of ELIAS-N1 and can therefore rely on the availability of multi-wavelength observations.
\citet{chakraborty} then consider $\sim 2500$ radio sources observed with uGMRT in the 300-500 MHz band. About 60\% of them are endowed with photometric redshifts and objects were subdivided into AGN and star-forming galaxies on the basis of their radio luminosity as in \citet{maglio13, maglio14, maglio15}. Their analysis of the clustering properties of these sources returns values for the correlation length of $r_0\sim 4.6$ Mpc/h and $r_0\sim 7.3$ Mpc/h, respectively for star-forming galaxies and AGN. We note that these values are in excellent agreement with those of \citet{maglio16} and \citet{hale} obtained for sources selected at higher radio frequencies.

At the time of writing, observations stemming from LOFAR \citep{vanhaarlem} have started providing breakthrough insights on the radio phenomena as seen at low frequencies. The first results on the large-scale properties of LOFAR galaxies are provided in \citet{siewert}, where the authors investigate the first and second moments of the radio galaxy distribution, finding for sources from the DR1 brighter than $F_{144\, \rm MHz}=2$ mJy an angular correlation function described by a power-law with amplitude $A\sim 5 \cdot 10^{-3}$ and slope $\gamma \sim 1.7$. More works in this direction are those of \citet{Alonso}, who cross-correlate LOFAR data with CMB lensing in order to calibrate the (unknown) redshift distribution of LOFAR sources and of \citet{tiwari} who provide the first results for the angular power spectrum.
As it was in the 1990s for FIRST and NVSS, the main limitation of these studies performed with LOFAR observations is the lack of reliable redshift information for the majority of the sources. This issue is going to be overcome soon with DR2, especially since multi-wavelength data for the deep fields will also be released (Best et al. 2022 submitted). 

\subsection{Methods for the detection of environment 2: direct pinpointing of sources within known structures} 
\label{sec:3.2}
A second method used to investigate the environmental properties of radio galaxies is based on direct searches for these sources within known structures. In the case of virialized groups and clusters with available X-ray information, this approach also has the advantage of investigating their thermal properties in the presence or absence of a radio-AGN and therefore to study the interaction between radio emission and intracluster medium (ICM).\\

After some early works  aimed at assessing the effects of dense/cluster-like environments on radio-AGN activity that involved a relatively limited number of sources (e.g., \citealt*{jaffe}; \citealt{Auriemma, lari, fanti}) to our knowledge \citet{ledlow1} were the first ones to provide statistical robust results on the recurrence of radio-AGN within known structures. This was obtained by performing a 1.4 GHz survey of 188, $z\le 0.09$, radio-AGN within Abell clusters and by computing their radio and bivariate radio/optical luminosity functions. Such functions were then compared to those of other samples of radio sources not residing in rich clusters in order to determine the effects of the environment on radio-AGN evolution. By doing so, \citet{ledlow1} found no significant dependence on the local galaxy density, evidence that made the authors conclude that the radio luminosity function is representative of the intrinsic properties and evolution of radio galaxies as one class, independent of the the large-scale structure within which these sources reside.

\begin{figure*}
  \centering
\includegraphics[width=0.49\textwidth]{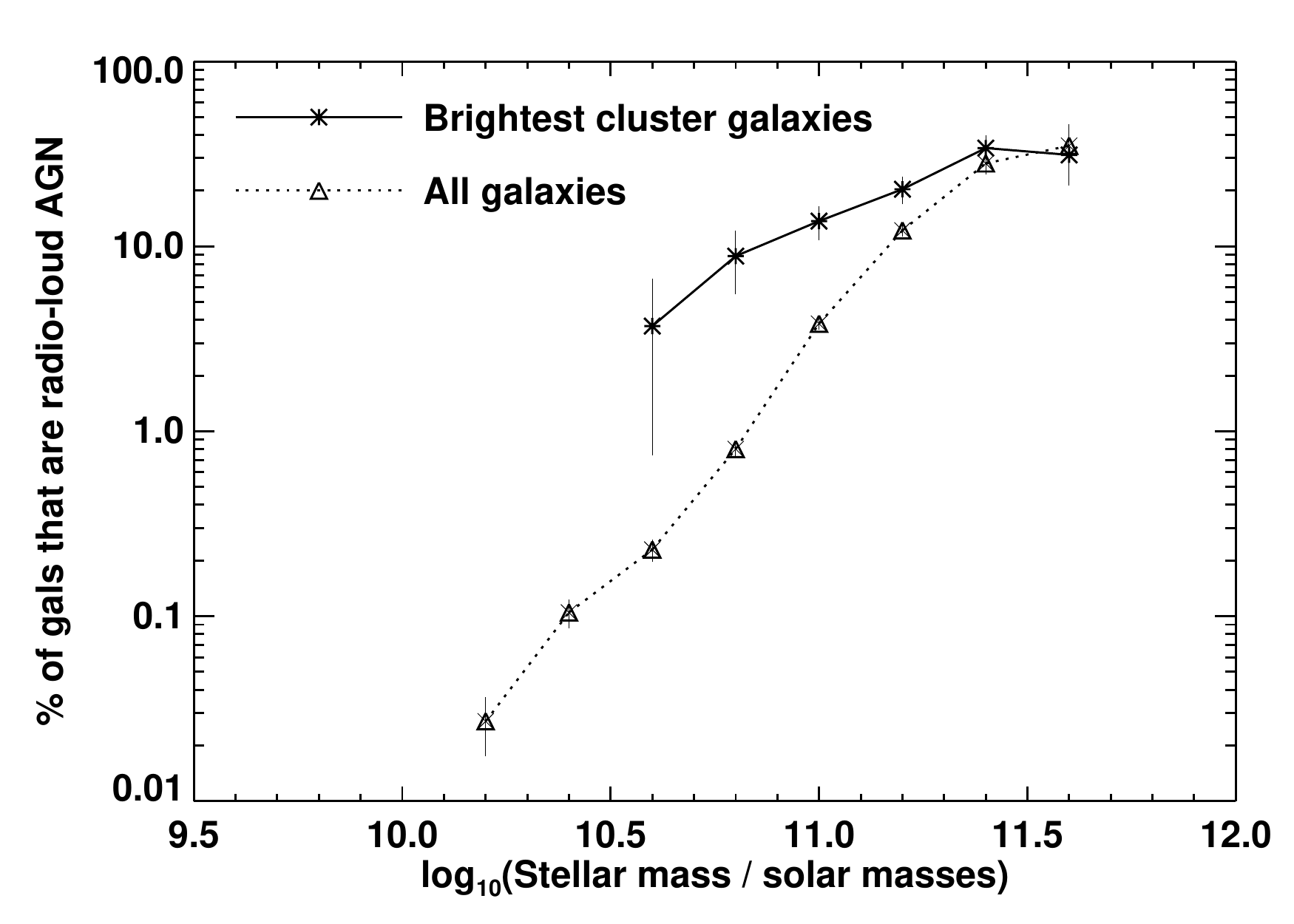}
\includegraphics[width=0.49\textwidth]{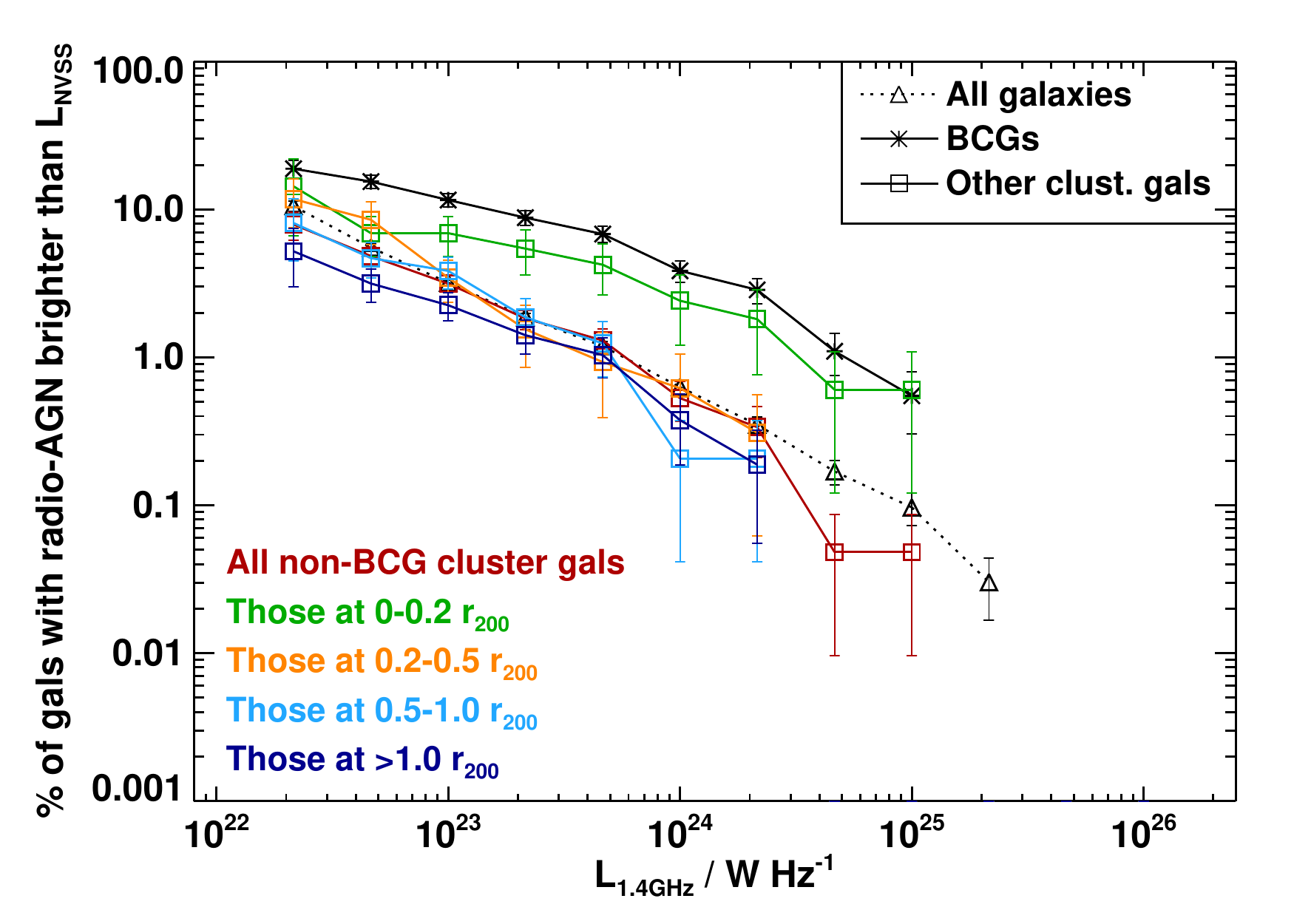}
\caption{Left-hand panel: the percentage of galaxies that host a radio-AGN, as a function of stellar mass, for all galaxies and for BCGs. BCGs are more likely to be radio-active than other galaxies of the same stellar mass, and the two relations have different slopes. Right-hand panel: the percentage of galaxies which host a radio-AGN brighter than a given radio luminosity for non-BCG cluster galaxies (red line) as compared to the all galaxy (dotted black line) and BCG (solid black line) samples. The other coloured lines indicate the equivalent relations for different subsets of the non-BCG cluster galaxies with different distances from the center. The radio-AGN fraction is boosted for galaxies within 0.2 $r_{200}$ of the cluster center, but retains the same overall shape.
Figures from \citet{best8}.}
\label{best_2007}       
\end{figure*} 
 
A decade later, \citet{best8} used a sample of 625, $z<0.1$ groups and clusters of galaxies selected from the Sloan Digital Sky Survey (\citealt{miller, Vonderlinden}) and cross-correlated them with the NVSS and FIRST radio samples in order to study the properties of radio-AGN hosted by the Brightest Cluster Galaxies (BCGs). In their work they show that the fraction of BCGs that are radio-active AGN increases linearly with (stellar) mass up to a plateau level of about $\sim 20-30$ per cent reached for their highest mass range which is about $M_*\sim 10^{11.5}\,M_\odot$ (cf.\ left-hand panel of Fig.~\ref{best_2007}), and that BCGs are more likely to host a radio-AGN than other galaxies of the same stellar mass. This trend is independent of the velocity dispersion (i.e., the mass) of the clusters, as is the radio luminosity distribution, which is observed to be the same for clusters of different masses. Interestingly enough, the same behaviour is also observed for non-BCG cluster galaxies but only as long as they reside in the proximity of the cluster center ($r<0.2 \; r_{200}$, cf.\ right-hand panel of Fig.~\ref{best_2007}, where $r_{200}$ is the radius within which the average density equals 200 times the critical density $\delta_c$ for spherical collapse, see Sect.~\ref{sec:3.1.1}). The above findings led the authors to conclude that radio activity of AGN origin is associated with the cooling of gas from the hot envelopes of elliptical galaxies and, in the case of central cluster galaxies, also from the intracluster medium. This is because accretion of hot gas from a strong cooling flow is able to explain both the boosted likelihood of BCGs hosting radio-active AGN, and the different slopes of the mass dependencies of the radio-AGN fractions for BCGs and other galaxies.
  
Indeed, by then, evidence for a connection between radio-AGN activity and thermal state of groups and clusters of galaxies had already started to accumulate, although it was only based on a handful of observations (e.g., \citealt{smith1, kraft, croston}). 
\citet*{croston1} were the first ones to sistematically investigate a well-defined and homogeneous sample of 30 local groups with available X-ray information (\citealt{osmond}), reporting evidence for the gas properties of groups containing radio-active galaxies (about 60\% of them) to differ from those not hosting them, in the sense that radio-active groups were likely to be hotter at a given X-ray luminosity. They attributed this effect to the heating induced by radio galaxies.
 
 \begin{figure}
\centering
\includegraphics[scale=0.35]{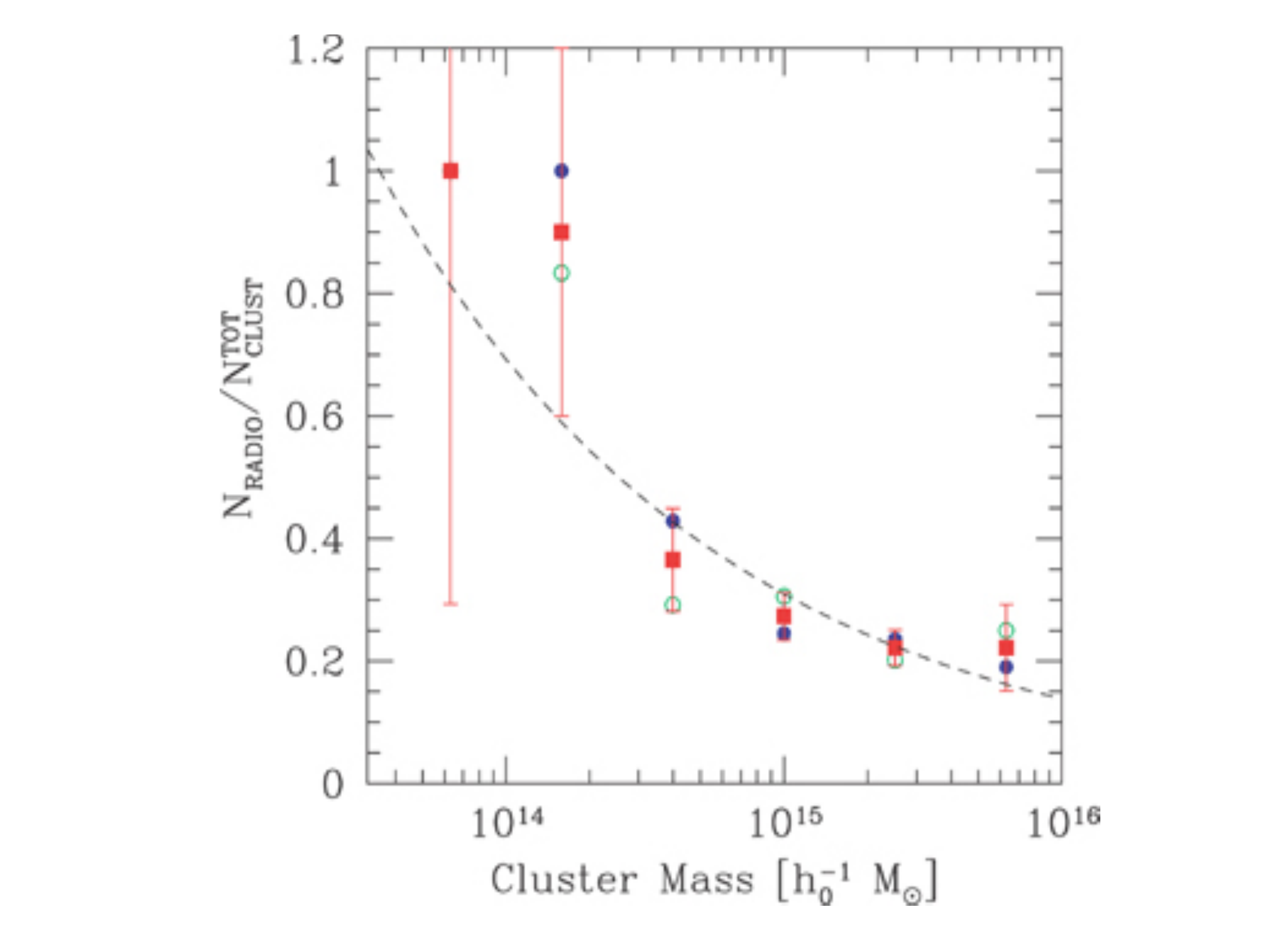}
\caption{Ratio between number of clusters with a central radio counterpart brighter than 3 mJy in the NVSS maps and the whole $F_X\ge 3 \cdot 10^{-12}$ erg s$^{-1}$ cm$^{-2}$ cluster population as a function of cluster mass. Open (green) circles are for NORAS objects, filled (blue) dots for REFLEX, while (red) squares and associated error bars represent the combined REFLEX + NORAS dataset. The dashed line indicates the best fit to the data. Figure from \citet{maglio7}.}
\label{bruggen}       
\end{figure} 

\citet{maglio7} extended the analysis of \citet{croston1} and \citet{best8} to clusters of higher mass and with known X-ray information, again in order to assess the importance of (central) AGN radio emission on the thermal properties of the ICM. This was done by considering combined observations from the REFLEX (\citealt{bohringer1}) and NORAS  (\citealt{bohringer2}) cluster surveys with NVSS for 145, $z<0.3$, X-ray selected clusters brighter than $3 \cdot 10^{-12}$ erg s$^{-1}$ cm$^{-2}$ that showed central radio emission 
 above 3 mJy. For cluster masses $M_{\rm CL}\le 10^{14.5}\,M_\odot$, 11 clusters out of 12 (corresponding to 92 per cent of the systems) were found to be inhabited by a central radio source. This fraction decreases for higher masses as $\propto M_{\rm CL}^{-0.4}$ (cf.\ Fig.~\ref{bruggen}). 
 \begin{figure}
\centering
\includegraphics[width=0.49\textwidth]{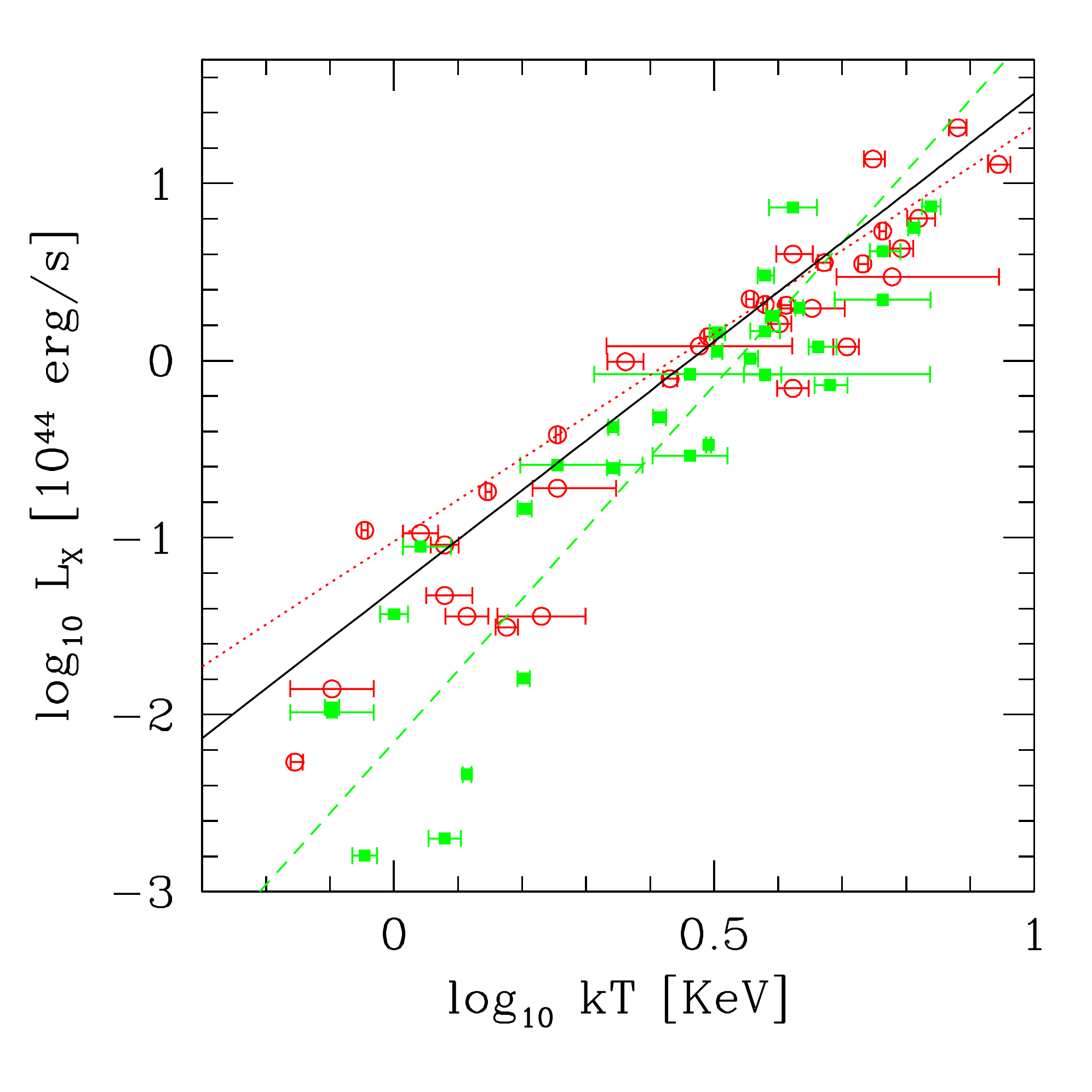}
\includegraphics[width=0.49\textwidth]{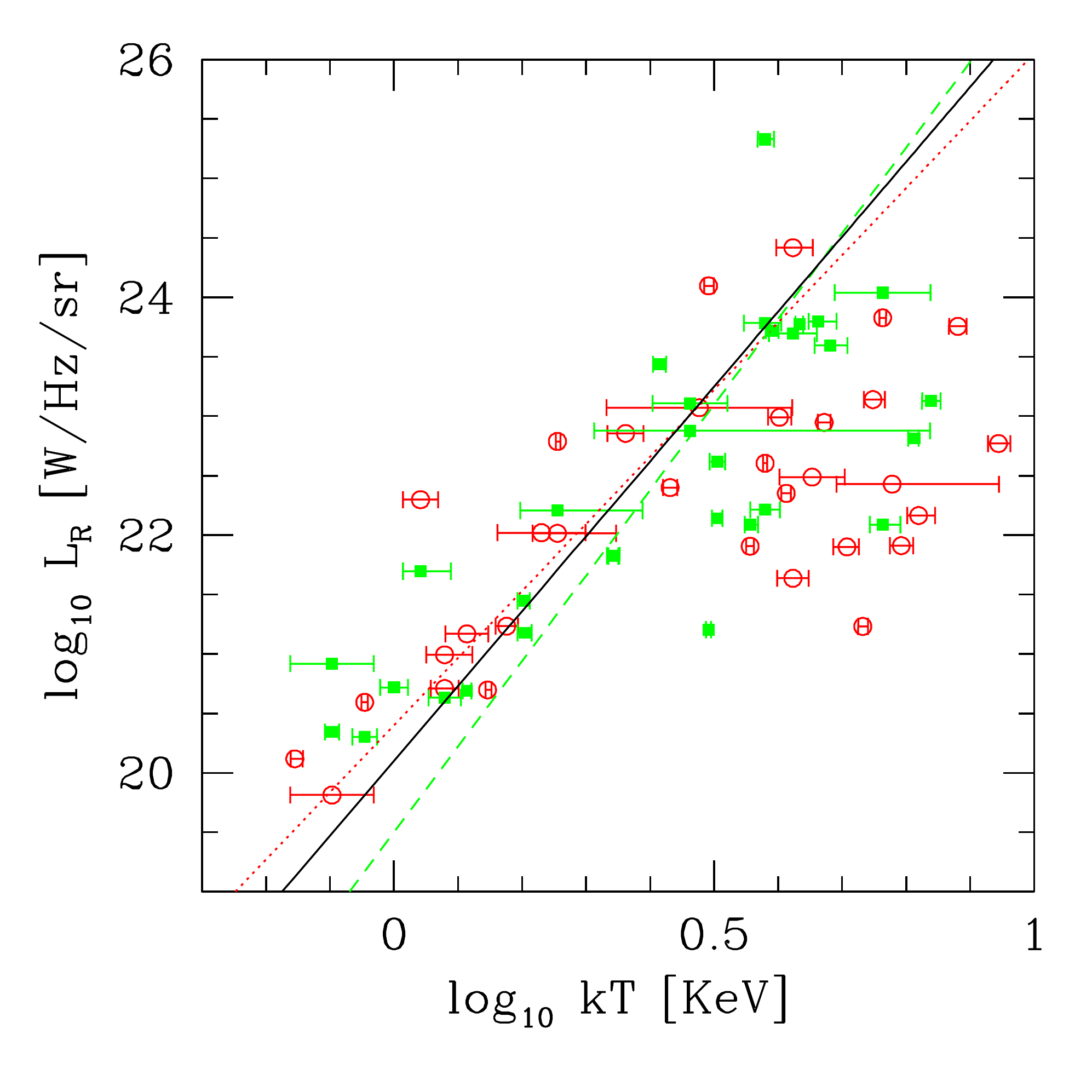}
\caption{Left-hand panel: X-ray luminosity versus cluster temperature for those X-ray-selected clusters brighter than $3 \cdot 10^{-12}$ erg s$^{-1}$ cm$^{-2}$ which exhibit central radio emission above 3 mJy. Open (red) circles represent clusters associated with point-like radio sources, while solid (green) squares indicate clusters inhabited by radio-AGN presenting extended structures. The dashed (green) line indicates the L$_{\rm X}$--kT best fit for the sub-population of extended radio sources, while the dotted (red) one is for point-like radio objects. The solid line is the best fit to the whole radio cluster sample. Right-hand panel: radio luminosity of the sources associated with cluster centres versus cluster temperature. Symbols are as before. Figure from \citet{maglio7}.}
\label{bruggen1}       
\end{figure} 
 By further dividing the sample into clusters harbouring either a point-like or an extended radio-AGN (about 41\% of the analyzed cases, a percentage which is much higher than what is generally found for radio galaxies within the same redshift range and at similar radio flux levels, $\sim 5$\% --  e.g.,  \citealt{maglio4}), these authors observed that the steepening of the X-ray Luminosity-Temperature ($L_X-T$) relation observed in low-temperature clusters (e.g., \citealt*{helsdon}) was strongly associated with the presence of central radio-AGN with extended jets and/or lobed structures. In this latter case, $L_X\propto T^4$, while for point-like sources one recovers the approximately self-similar relation $L_X\propto T^{2}$ (cf.\ left-hand panel of Fig.~\ref{bruggen1}), expected when gravity is the only source of heating. This bimodal behavior suggests a more efficient mode with which extended sources can permeate the surrounding ICM and transfer heat within the cluster. Interestingly, the net effect of extended radio-AGN on the ICM is observed to decrease with increasing cluster mass (i.e., increasing $kT$), and at
larger masses also any relationship between AGN radio luminosity and cluster temperature is lost (cf.\ right-hand panel of Fig.~\ref{bruggen1}). Based on their data combined with Montecarlo simulations, \citet{maglio7} conclude that the presence of radio-AGN with an extended structure is responsible for the overheating of the intracluster gas, with an importance which dramatically increases in low-mass systems. 
At the same time, \citet{lin1} considered a sample very similar to that of \citet{maglio7} and estimated the radio-luminosity function of radio-AGN within $r_{200}$. At variance with \citet{ledlow1}, their results showed that it was about seven times more likely for a galaxy to be radio-active if it resided within a cluster environment rather than in the field, 10 ($\sim 30$) times more likely in the case of associations between BCGs and $L_{1.4\, \rm GHz}>10^{23}$ W Hz$^{-1}$ ($L_{1.4\, \rm GHz}>10^{25}$ W Hz$^{-1}$) AGN. The authors also found that the surface density profiles of the hosts of radio-AGN were much more concentrated than those of inactive galaxies, and that more powerful radio sources were more concentrated than weaker ones. All these pieces of evidence made the authors invoke radio-AGN activity as an important heating source for the intracluster medium, especially within 10\% of the clusters virial radii. Lastly, by converting the estimated radio-luminosity function into a life-time for the radio-active AGN phase, \citet{lin1} derived values of about 0.6 Gyr, in rough agreement, even though on the lower side, with \citet{maglio16}. 
  
\citet{smolcic2} moved the analysis to higher redshift sources and investigated the recurrence of 217 low-power ($10^{23.6} <L_{1.4 \rm GHz}/[{\rm W \;Hz^{-1}}]< 10^{25}$) radio-AGN selected in the COSMOS field out of $z=1.3$ within X-ray selected groups of masses $10^{13.2}< M_{200}/M_\odot<10^{14.4}$ ($M_{200}$ is total mass of the group enclosed within $r_{200}$). In agreement with \citet{best8}, they found a strong enhancement of radio-AGN in group centers ($<0.2 \; r_{200}$). Furthermore, by comparing their radio-AGN to a control sample of inactive galaxies with the same stellar mass and colour distributions, they observed that the fraction of radio-AGN was enhanced by a factor $\sim 2$ in galaxy groups. By using these results combined with Halo Occupation methods (cf.\ Sect.~\ref{sec:3.1.1}), \citet{smolcic2} also estimate the average time a massive red galaxy in a galaxy group is switched on to a radio-active AGN phase to range between 0.6 and 3.5 Gyr, value which depends on the mass of the group (the more massive the group, the longer the active phase), in good agreement with the findings of \citet{lin1} and \citet{maglio16} obtained from clustering methods. It is also worth mentioning that \emph{all} galaxies with stellar masses $M_*> 10^{12}\,M_\odot$ in the \citet{smolcic2} sample are radio-active AGN found to reside within groups.
 
\begin{figure}
\centering
\includegraphics[scale=0.4]{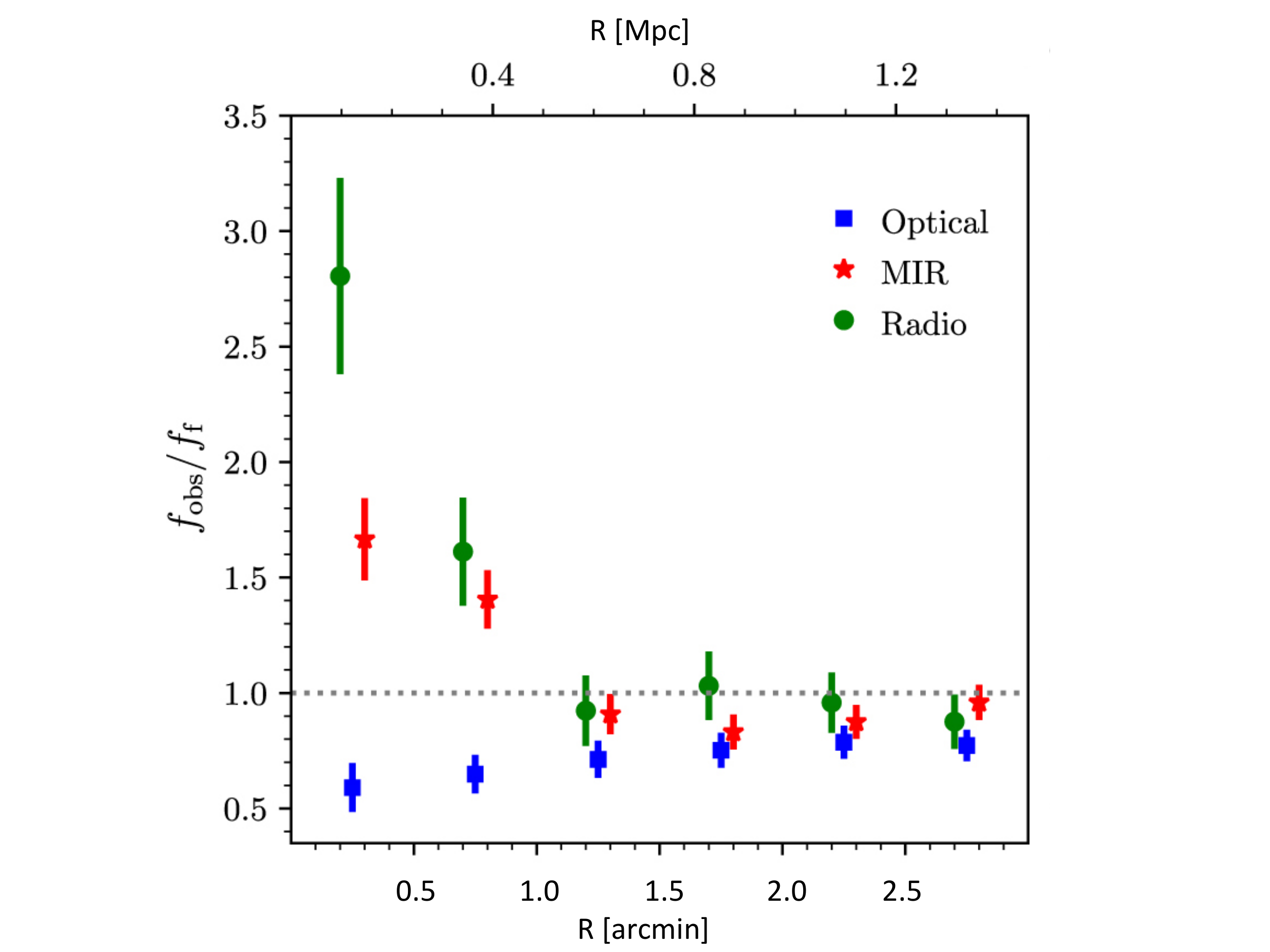}
\caption{Observed AGN fraction in MaDCoWS clusters divided by the field fraction as a function of cluster-centric distance. Deviations from the field level are in the central 1$^\prime$ region and are dependent on the selection method of the AGN catalogue. The bottom x-axis represents the distance from the cluster center expressed in arcmin, while the top x-axis that in Mpc. Figure from \citet{mo1}.}
\label{Mo}       
\end{figure} 

\citet{mo1} instead investigate the interaction between radio-AGN activity and cluster environment for 2300 galaxy clusters (212 with an embedded radio-AGN from the FIRST survey) from the MaDCoWS survey (\citealt{brodwin1}), which selected galaxy over-densities in the WISE bands up to $z\sim 1$. Radio-AGN are observed to be more concentrated towards the cluster centers, in line with the behaviour of MIR-selected AGN but at variance with optically-selected AGN (cf.\ Fig.~\ref{Mo}). Also, the radio-AGN fraction was reported to be more than 2.5 times that of the field, implying -- in agreement with e.g., \citet{best8} and \citet{smolcic2} --  that the centers of clusters favour the triggering of radio emission in AGNs.   No statistically significant change in the AGN fraction as a function of cluster richness was instead found. \citet{mo2} further find that the probability for a cluster to host a radio-active AGN with $L_{1.4\, \rm GHz}> 10^{25}$ W Hz$^{-1}$  is a steep function of redshift, showing a factor $\sim 10$ increase in the range $0\simlt z\simlt 1.2$ (see also e.g., \citealt*{donoso}; \citealt{birzan}), and that such a probability is also strongly dependent on radio luminosity. 

\citet{maglio17} analyse the recurrence of 218, $z\le 1.2$ radio galaxies from the same COSMOS-VLA survey (\citealt{schinnerer, bondi}) considered by \citet{smolcic2}, but within the different components of the cosmic web such as clusters, filaments and the field as reconstructed by the work of \citet{darvish}. It was found that while radio-emitting star-forming galaxies are distributed in a fashion which is similar to the more general population of COSMOS galaxies, radio-AGN tend to be more preferentially hosted by dense environments (20\% vs $\sim 10$\% of the general population). Furthermore, radio-AGN with $L_{1.4 \rm GHz} > 10^{23.5}$ W Hz$^{-1}$ sr$^{-1}$ are twice more likely to be found in clusters with respect to fainter sources ($\sim 38$\% vs $\sim 15$\%, cf.\ left-hand panel of Fig.~\ref{manu_env}), just as radio-selected AGN hosted by galaxies with stellar masses $M_* > 10^{11}\,M_\odot$ are twice more likely  to be found in overdense environments with respect to objects of lower mass ($\sim 24$\%  vs $\sim 11$\%, cf.\ right-hand panel of Fig.~\ref{manu_env}), and to avoid underdense structures more than inactive galaxies with the same stellar mass content. Stellar masses also seem to determine the location of radio-active AGN within clusters as $\sim 100$\% of the AGN coinciding with a BCG have $M_* > 10^{11}\,M_\odot$. No different location within the cluster was instead observed for AGN of different radio luminosities. Furthermore, in agreement with the early work of \citet{tasse}, radio-AGN which also emit in the MIR are observed to present a marked preference to be found as
isolated galaxies ($\sim 70$\% of the sub-sample) at variance with those also active in the X-ray that all seem to reside within over-densities. Based on the above results, \citet{maglio17} advocate for a strong link between processes taking place on sub-pc/pc/Kpc scales and the large-scale -- Mpc -- environment of radio-emitters, irrespective of whether star-forming galaxies or AGN. A strong dependence of the probability for a radio-active AGN to belong to an overdense structure on its radio luminosity ($\sim 10$\% at $L_{150 \rm MHz} =10^{22.5}$ W Hz$^{-1}$ vs $\sim 30$\% at $L_{150 \rm MHz} =10^{26}$ W Hz$^{-1}$) is also found by \citet{croston2}, who use a sample of 8745, $z<0.4$ radio-active AGN taken from the LoTSS DR1 catalogue (\citealt{shimwell, hardcastle3}) matched with two cluster catalogues from the SDSS DR8 (\citealt{wen} and \citealt{rykoff}). These authors further observe that cluster richness and AGN radio luminosity are related, since AGN with $L_{150 \rm MHz} >10^{25}$ W Hz$^{-1}$ are likely to be found in rich cluster environments, and that the location of a radio-AGN within its overdense region is also dictated by its radio luminosity, as the most radio-luminous AGN are typically found close to the centre. In agreement with \citet{maglio17} though, it is concluded that stellar mass is likely to be the dominant cause for the different location preferences of low- and high-luminosity AGN.

\begin{figure*}
  \centering
 \includegraphics[width=0.49\textwidth]{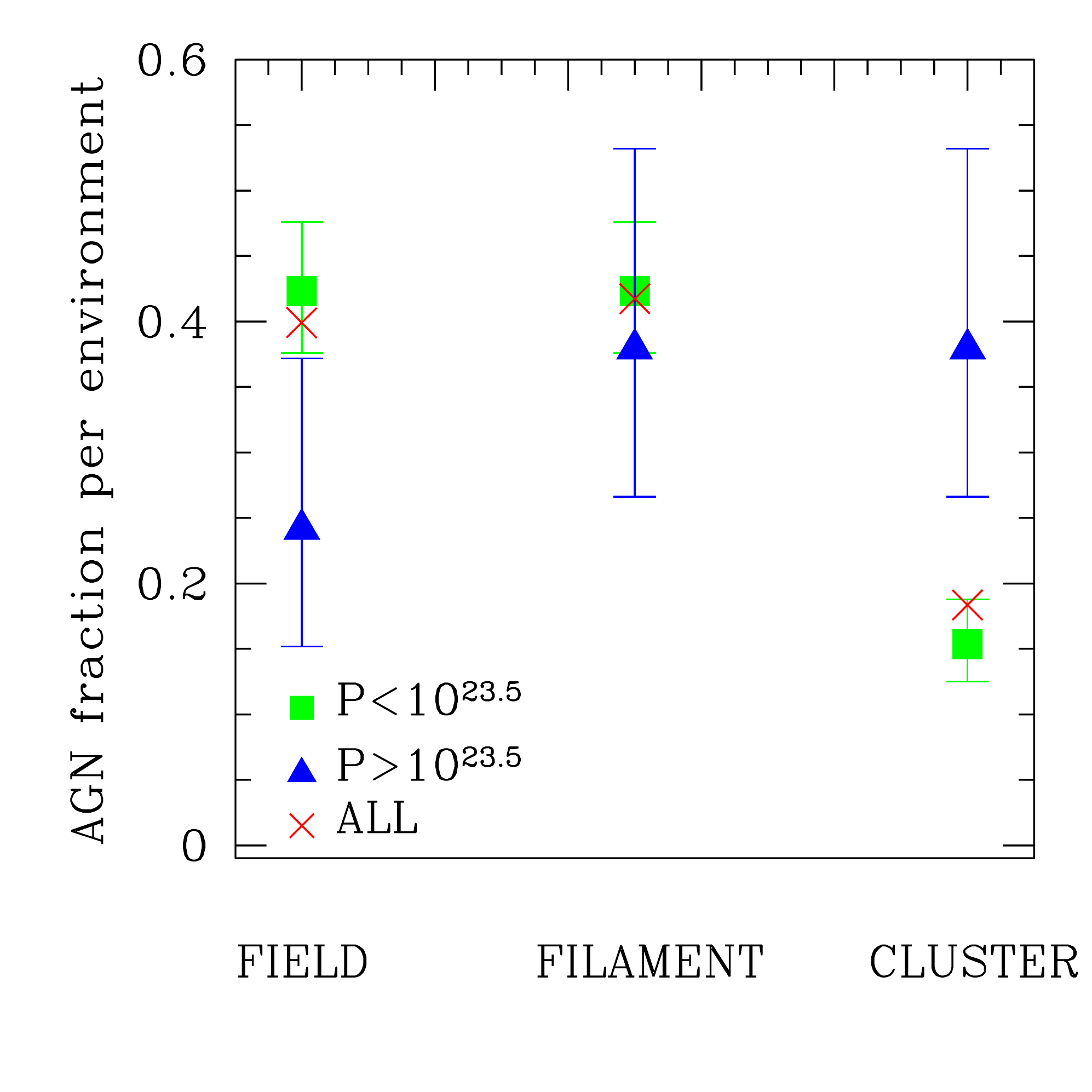}
 \includegraphics[width=0.49\textwidth]{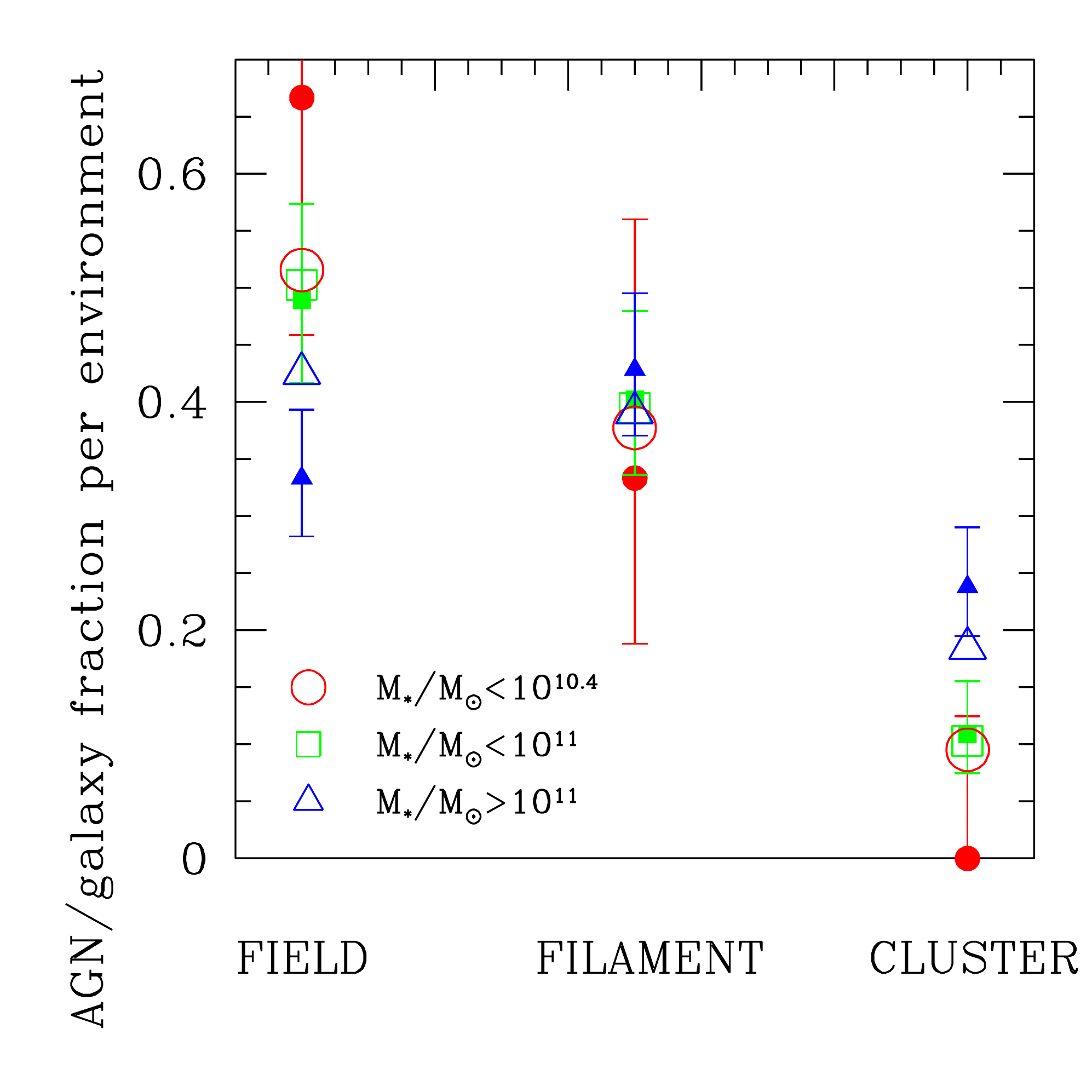}
\caption{Left-hand panel: Distribution of radio-selected AGN within different environments for different values of the AGN radio luminosity (expressed in $\rm W \ Hz^{-1}\ sr^{-1}$).  Right-hand panel: Fraction of radio-selected AGN (small filled symbols) and galaxies from \citet{darvish} (large open symbols without error bars) within different environments for various values of the stellar mass of the hosts. The circles indicate the results obtained for sources with stellar masses $M_*/M_\odot\le 10^{10.4}$, squares for sources with $M_*/M_\odot\le 10^{11}$, and triangles for $M_*/M_\odot> 10^{11}$. In all cases error-bars show $1\sigma$ Poisson uncertainties. Those corresponding to the galaxies from \citet{darvish} are smaller than the symbol sizes. Figures from \citet{maglio17}.}
\label{manu_env}       
\end{figure*} 

\citet*{miraghaei1} also investigates radio and optical AGN activity as a function of environment spanning from voids to clusters of galaxies. Data were taken from the value added spectroscopic catalogue of \citet*{best6}. No dependence of radio-AGN recurrence on environment is found in either blue (i.e., with ongoing star-formation) or green (i.e., in the process of being quenched) galaxies, while it is reported as significant in the case of radio-AGN hosted by red galaxies, whereby radio activity is observed to increase in dense environments.  These findings lead the author to conclude that the efficiency of gas accretion from small- or large-scale environments onto super massive black holes is low in the presence of cold gas within the galaxies while, in its absence, the hot gas from the intergalactic medium filling a dense environment efficiently triggers radio-AGN activity.

Lastly, it is worth mentioning the first attempts to investigate the environments of radio-AGN by pinpointing these sources within known structures at redshifts as high as $z\sim 4$ (\citealt{uchi}).
In this case, no preference for high-redshift radio galaxies to reside within overdense structures is observed.  However, \citet{uchi} report an excess of $g$-dropout galaxies around faint ($L_{1.4 \rm GHz}\sim 10^{26.0-26.5}$ W Hz$^{-1}$) radio-AGN, while no difference is found between the densities of galaxies surrounding brighter radio-AGN and of those belonging to the field. Perhaps most interestingly, these authors observe that the galaxies surrounding their sources tend to distribute along the major axes of the radio jets. If confirmed, this would imply the onset of filamentary structures around radio-AGN at an epoch as early as $z\sim 4$. Masses for the haloes hosting these sources are estimated to be $\sim 10^{13.1}\,M_\odot$, result which extends the observed independence of radio-AGN environmental properties from look-back time (cf.\ Sect.~\ref{sec:3.1}) up to $z\sim 4$.

\subsubsection{The FRI-FRII and HERG-LERG distinction} 
So far we have mainly considered radio-AGN as a single population of sources. However (cf.\ Sects.~\ref{sec:2.1} and \ref{sec:2.2}), they indeed come in a number of flavours, mostly as FRI vs FRII galaxies and/or as LERGs vs HERGs. It is therefore of interest investigating the environmental properties of these sub-classes of sources separately. A small number of early works (e.g., \citealt{prestage,  Allington, zirbel2, best2}, see also later in this Section) already tackled this issue, coming to the general conclusion that in the local universe FRIs and LERGs occupy denser structures than FRIIs and HERGs, even if there were hints for a positive cosmological evolution of FRII environments, at least up to $z\sim 0.5$.  \citet*{wing} extended these studies by considering $\sim 700$ FRI and $\sim 200$ FRII galaxies from the FIRST survey with a counterpart in the SDSS DR7. They observed a preference for extended radio-AGN of any kind (i.e., with straight or bent morphologies) to reside within overdense environments as opposed to point-like sources (up to $\sim 80$\% in the case of bent-lobed AGN vs $\sim 10-30$\% for single objects), and also found more FRI galaxies within $z\simlt 0.5$ to inhabit rich structures (such as groups or clusters or galaxies) than FRIIs (e.g., in the case of straight extended radio emission, $\sim 60$\% vs $\sim 40$\% in group-like structures and $\sim 37$\% vs $\sim 21$\% in cluster-like structures). \citet{gendre} applied the same method of \citet*{wing} to investigate the richness of the structures surrounding 88 LERGs and 70 HERGs set at $z < 0.3$, further divided into FRIs and FRIIs. These authors found that HERGs were more likely associated with poor over-densities regardless of their further division into FRI and FRII morphologies, while LERGs of any type could be found within both relatively poor and rich environments. 
 
\citet{ineson1} and \citet{ineson2} present the results of the Environment of Radio-loud AGN (ERA) program characterizing the cluster environment of radio-AGN. In their first work they concentrate on 26 radio-AGN at $z\sim 0.5$ drawn from the \citet{mclure3} sample in order to beat effects due to cosmological evolution and conclude that there is some weak correlation between radio luminosity and host cluster X-ray luminosity, correlation driven by the sub-population of LERGs, with HERGs  showing no significant correlation. \citet{ineson2} confirmed the above results on 55 lower-redshift ($z<0.2$) AGN, finding strong correlations between radio luminosity, cluster richness and central density in the case of LERGs and no correlation between radio luminosity and cluster richness, or between radio luminosity and central density for HERGs. HERG environments appear to be poorer at lower redshifts with respect to $z\sim 0.5$. No cosmological evolution of structure density is instead observed for LERGs.
 
\citet*{miraghaei} also investigate the environmental properties of radio galaxies selected according to their radio morphology (i.e., FRII, FRI, compact and with unusual morphologies) as well as their accretion properties (LERGs vs HERGs). The AGN sample adopted for their analysis is the one presented in \citet*{best6} which includes low-redshift ($z<0.4$) radio sources from the NVSS and the FIRST  surveys cross-matched with SDSS DR7, while that for the environments was taken from \citet*{sabater}. These authors conclude that the environments of LERGs display higher densities compared to those of HERGs, supporting the hypothesis that the source of AGN fuelling is the main origin of the HERG/LERG dichotomy. Furthermore, by comparing FRI LERGs with FRII LERGs at fixed stellar mass and radio luminosity, they show that FRI galaxies typically reside in richer environments and are hosted by smaller galaxies with higher mass-surface density (cf.\ Sect.~\ref{sec:2.2}), in agreement with a scenario invoking extrinsic effects of jet disruption driving the FR dichotomy.  No environmental difference is instead found between extended and compact LERGs. \citet{croston2} also observe that FRI radio galaxies inhabit systematically richer environments than FRIIs, and that the probability for an FRI to be found within an over-density strongly increases with its radio luminosity, at variance with bright FRIIs that tend to avoid rich structures.

\citet{massaro, massaro1} however argue that the above results on the environmental differences between FRI and FRII galaxies as well as between LERGs and HERGs mostly originate from selection biases and present a new analysis which could minimize them. The adopted catalogues are those of visually-inspected FRIs and FRIIs from \citet{capetti} and \citet{capetti1} (cf.\ Sect.~\ref{sec:2.2}), with a redshift range limited to $z< 0.15$. The environments of the 195 FRI +115 (14 of which classified as HERGs, all the others being LERGs) FRII galaxies are then investigated by making use of various techniques which include both cross-matching the radio sources with known catalogues of clusters and groups of galaxies as well as direct search for over-densities around the selected radio-AGN using them as beacons. By combining these methods together, they find that in the local universe FRIs and FRIIs as well as HERGs (all FRIIs in their catalogue) and LERGs all live in overdense environments having the same richness, independent of the redshift range considered, their radio luminosity or the absolute magnitude of their hosts, even though the probability of finding an FRI in such structures seems to be enhanced with respect to the population of FRIIs. 

More recently, \citet{vardoulaki} brought this analysis to higher redshifts and considered the recurrence of 3GHz-selected radio-AGN down to $\mu$Jy level subdivided into FRI (39 sources), FRII (59 sources) and compact/jetless radio emitters (1818 sources) within 247, $0.08\le z\le 1.53$, X-ray detected groups of galaxies in the COSMOS field (\citealt{gozaliasl}). At variance with most of the results presented so far in the literature,  they find no preference for radio-AGN of any morphology to reside within over-densities. They ascribe their results as possibly due to the lack of high sensitivity and resolution in previous radio observations that could in principle have prevented the detection of fainter and/or smaller radio sources.  These conclusions are however in disagreement with the findings obtained by the same authors when they compare their radio samples with the density field reconstructed in COSMOS by \citet{darvish}. Indeed, in this latter case they observe that point-like radio sources preferentially reside within the field (40\% of the sub-sample), while FRII galaxies are more often hosted by cluster-like environments (46\%), although the small number of objects involved in the analysis does not allow to draw strong conclusions.

\subsection{Methods for the detection of environment 3: search for over-densities around known sources}
\label{sec:3.3}
A third method to investigate the environmental properties of radio galaxies consists in adopting an opposite approach to that discussed in Sect.~\ref{sec:3.2} and observe the  structures that surround pre-selected radio sources. In this case the analysis can be divided into i) targeted searches around radio-AGN and ii) cross-matches of radio and optical (or IR) catalogues 

\subsubsection {Targeted searches around radio-AGN} 
Targeted searches for clusters and protoclusters of galaxies embedding radio-active AGN have been proven throughout the years to be very successful. 
\citet*{hill}  were amongst the first ones to investigate the environment around powerful radio-AGN by estimating the net excess number of galaxies within 0.5 Mpc from the radio galaxy and within a magnitude range $m_1$ and $m_1+3$, where $m_1$ was the magnitude of the radio galaxy itself. They observed 45 radio galaxies and 6 radio-active quasars of radio luminosities spanning 4 dex within the narrow redshift range $0.35< z <0.55$.  By doing so, they found that radio-AGN on average resided in rich environments. This held for both powerful (i.e., above the break of the radio-luminosity function) and less powerful sources, at variance with results from the more local universe obtained by previous studies (i.e., \citealt*{prestage}, cf.\ Sect.~\ref{sec:3.1}) which showed that only FRI galaxies resided in overdense structures. 

Another seminal work was that of \citet{Allington} who applied the same method of \citet*{hill}  to a catalogue of 98, $z<0.5$, radio galaxies selected within a narrow interval in radio luminosity ($L_{408 \rm MHz}$ between $10^{26}$ and $10^{28}$ W Hz$^{-1}$) and subdivided into two redshift ranges, $z< 0.25$ and $0.25<z<0.5$. They found that these sources mostly resided in rather poor clusters (or groups of galaxies), with three to ten members and a richness that roughly doubled from the low-$z$ to the higher-$z$ sample. Furthermore, as also re-stated by \citet{zirbel2}, they managed to reconcile the disagreement between the results of \citet*{hill}  and those of \citet*{prestage} as it was observed that, while at low redshifts FRII galaxies avoided overdense regions,  they indeed existed in rich groups at redshifts $z\sim 0.5$ (cf.\ Fig.~\ref{zibel}). From the same \citet{Allington} sample, \citet{zirbel2} also showed that the high- and low-redshift groups surrounding FRI and FRII galaxies belonged to separate subsets, and that FRI groups were dynamically more evolved than FRII groups.

\begin{figure}
\centering
\includegraphics[scale=0.6]{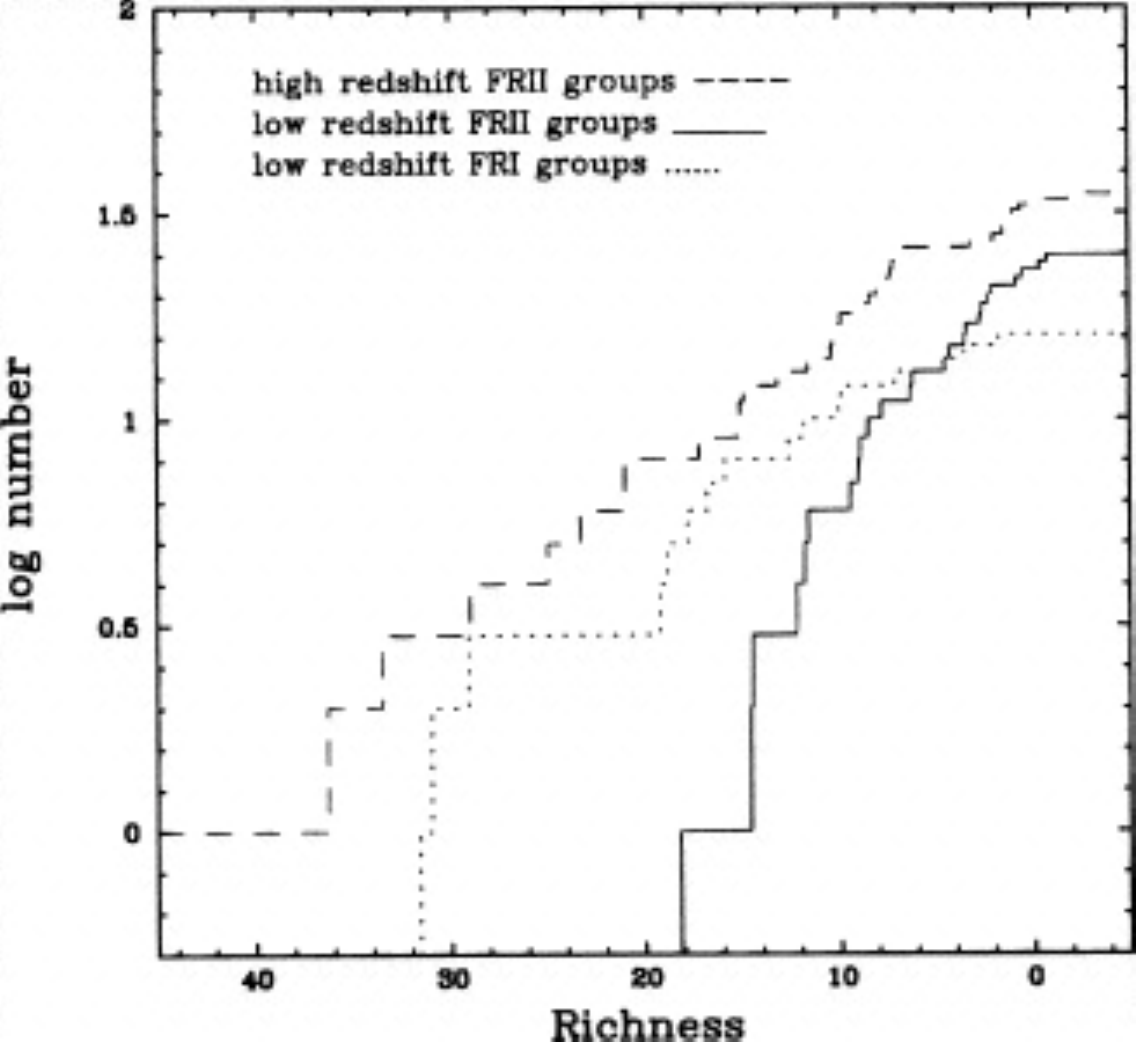}
\caption{Cumulative richness distributions of low-redshift groups surrounding FRI galaxies (dotted line), low-redshift groups surrounding FRII galaxies (solid line), and $z\sim 0.5$ groups surrounding FRII galaxies (dashed line). Figure from \citet{zirbel2}.}
\label{zibel}       
\end{figure}

Thanks to all the work and efforts put into analysing the environmental properties of radio-AGN in the previous years (cf.\ also Sects.~\ref{sec:3.1} and \ref{sec:3.2}), by the end of last century it was then clear that radio sources of AGN origin preferentially inhabited overdense structures. This led to a further step in this kind of targeted studies around radio-AGN: use them as beacons to look for over-densities of galaxies at high (i.e., $z\sim1-2$ and above) redshifts, beating some of the limitations which affected the-then-state-of-the-art methods for the search of high-redshift clusters of galaxies such as the partial absence of evolved galaxies and therefore of a red sequence (e.g., \citealt{mcdonald}), or the difficulty of detecting hot cluster gas by means of X-ray techniques in the more distant universe.
 
\citet{pentericci2} used VLT spectroscopic observations to provide one of the first pieces of evidence for the existence of an overdense region (marked by 14 Ly$\alpha$ emitters and one quasar) surrounding a clumpy radio galaxy set at redshift $z=2.16$. \citet{Venemans1} report the detection of a structure made of 20 Ly$\alpha$ emitters (out of 23 observed galaxies, with a success rate of 87\%) around the luminous radio galaxy TN J1338-1942 set at $z=4.1$. Such a structure extends for at least 2.7 Mpc $\times$ 1.8 Mpc in projection and constitutes an over-density of about 15 when compared with the field distribution of Ly$\alpha$ emitters at the same redshift. The inferred mass of the structure is about $10^{15}\,M_\odot$ and, interestingly enough, the radio galaxy does not constitute the centre of the protocluster. These observations were subsequently complemented with the discovery within the same structure of a large population of Lyman-break galaxies, a factor $\sim 2.5$ higher than the average number observed in random fields (\citealt{miley}, see \citealt*{miley1} for a review on high-redshift radio galaxies and their environments). 

\citet{stern} made use of Keck observations to report the discovery of a galaxy over-density associated with the $z=1.11$ radio galaxy MG1 J04426+0202. At variance with \citet{pentericci2} and \citet{Venemans1}, and likely due to the lower redshift of the source, in this case members of the over-density were identified as passive elliptical galaxies formed at high redshift. \citet{Venemans}  extended the work of \citet{pentericci2} and \citet{Venemans1} by searching for forming clusters of galaxies near 8 powerful ($L_{\rm 2.7 GHz}  > 10^{33}$  erg s$^{-1}$ Hz$^{-1}$ sr$^{-1}$) radio galaxies at $2.0< z< 5.2$. At least six of their eight fields were found to be overdense in Ly$\alpha$ emitters by a factor 3-5 when compared to the field density at similar redshifts. Furthermore, the emitters showed significant clustering in velocity space. From their results, \citet{Venemans} estimated that roughly 75\% of powerful  high redshift radio galaxies reside in a protocluster. \citet{hatch} imaged the fields surrounding six, $2.2<z<2.6$, radio-AGN, finding over-densities of physically associated galaxies for three of them which extended to $\sim 4$ Mpc and contained approximately $\sim 2-7 \cdot 10^{14}\,M_\odot$ mass. 
\citet{doherty} spectroscopically confirmed the existence of two protoclusters  surrounding two high redshift radio galaxies set at $z=2.16$ and $z=2.93$. These authors were also able to identify two massive red galaxies, one of which turned out to be the first red and mainly passively evolving galaxy confirmed to belong to a $z>2$ protocluster.

More recently, \citet{mayo} used counts-in-cells analyses (which consist in counting sources within cells of different sizes) to detect over-densities of 24$\mu$m,  \textit{Spitzer}-selected galaxies around 63 High Redshift ($1\le z \le 5.2$) Radio Galaxies (HzRGs). They concluded that over 95\% of the targeted HzRGs lie in higher than average density fields. 
\citet{Galametz1} also used a counts-in-cells analysis to identify over-densities of \textit{Spitzer}/IRAC-selected galaxies in the fields of 48, $1.2<z<3$, radio sources from the Spitzer High-Redshift Radio Galaxy program (SHzRG -- \citealt{seymour, debreuck1}).  Using relatively shallow IRAC data, these authors showed that radio galaxies preferentially reside within medium-to-dense regions, with 73\% of the targeted fields denser than average, in agreement with the results of \citet{Venemans}. One of these structures was also spectroscopically confirmed (\citealt{Galametz2}).

The CARLA (Clusters Around Radio-Loud AGN) program (\citealt{Wyl}) expanded the work of \citet{Galametz1} and investigated the environment of a larger sample of both obscured (e.g., type 2) and unobscured (e.g., type 1), luminous (rest-frame 500 MHz radio luminosities $\ge 10^{27.5}$ W Hz$^{-1}$) radio-active AGN between redshift $z=1.2$ and $z=3.2$, again by making use of \textit{Spitzer}-IRAC observations. Data were obtained for 387 fields, 187 around radio-active quasars and 200 around radio galaxies, reaching a 95\% completeness of 22.6 mag at 3.6$\mu$m and of 22.9 mag at 4.5$\mu$m. These authors found that 92\% of the selected radio-active AGN resided in environments richer than average (cf.\ left-hand panel of Fig.~\ref{CARLA}), and concluded that these sources constitute ideal beacons for finding high-redshifts clusters and protoclusters.  \citet{Wyl} also investigated how environment depends on different AGN physical properties (i.e., obscured vs unobscured sources) but found no correlation with either AGN type or radio luminosity. 16 out of 20 structures from the original CARLA selection were successively confirmed by using slitless grim spectroscopy on board of the \textit{Hubble Space Telescope} \citep{noirot}. 

\begin{figure*}
  \centering
\includegraphics[width=0.49\textwidth]{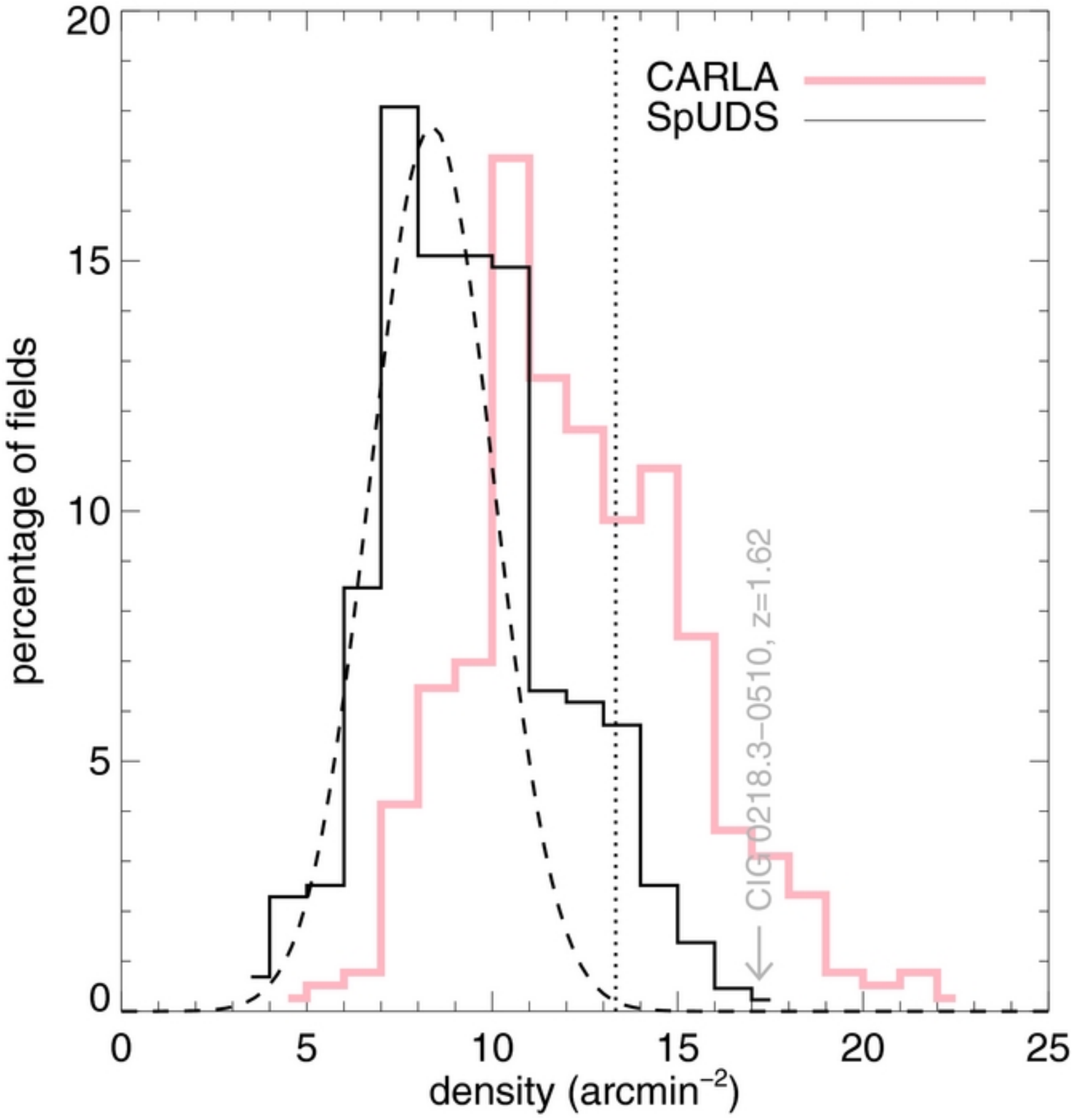}
\includegraphics[width=0.5\textwidth]{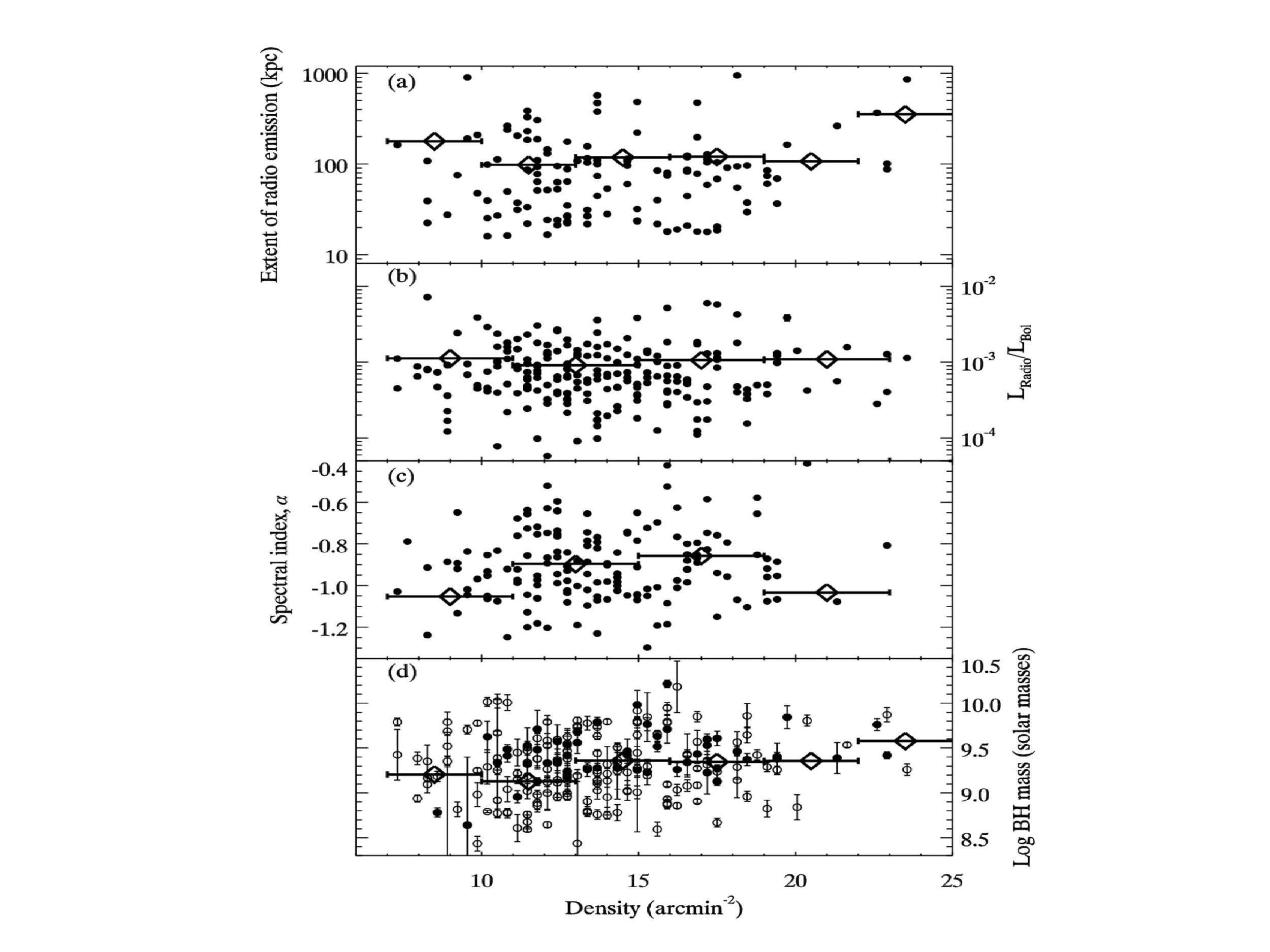}
\caption{Left-hand panel: histogram of the densities of IRAC-selected sources in the CARLA fields and in the control UKIDSS Ultra Deep Survey (SpUDS -- \citealt{Galametz3}) fields. The Gaussian fit to the low-density half of the SpUDS density distribution is shown by the dashed black curve and presents a peak at $\Sigma_{\rm SpUDS} = 8.3 \pm 1.6$ arcmin$^{-2}$. The vertical dotted line corresponds to $\Sigma_{\rm SpUDS} +3\sigma$, with $\sigma$ standard deviation. 92.0\% of the CARLA fields are denser than the SpUDS peak density. 18.7\% of the SpUDS fields are denser than $\Sigma_{\rm SpUDS} +  2\sigma$ in contrast to 55.3\% of the CARLA fields. Only 0.7\% of the SpUDS fields are denser than $\Sigma_{\rm SpUDS} + 5\sigma$ as opposed to 9.6\% of the CARLA fields. Figure from \citet{Wyl}. Right-hand panel: relationships between the large-scale environment of the CARLA radio-AGN and properties of their radio emission and powering black hole (BH). From top to bottom: (a) the extent of the radio emission for resolved CARLA sources, (b) ratio of radio to bolometric AGN power (only for quasars), (c) radio spectral index $\alpha$ (where $S_\nu\propto\nu^\alpha$) and (d) black hole mass for the 211 CARLA AGN matched with the \citet{shen} SDSS catalogue. Filled circles are black hole masses derived from fitting the MgII line in the optical spectra, whilst open circles are obtained from the CIV line. Open diamonds are the mean in each density bin. Figure from \citet{hatch1}.}
\label{CARLA}       
\end{figure*}

\citet{hatch1} compared the environments of powerful radio-AGN at $1.3 < z < 3.2$ from the CARLA survey to those of a sample of inactive galaxies drawn from the UKIDSS Ultra Deep Survey (SpUDS -- \citealt{Galametz3}) matched in mass and redshift, finding that --   at all scales $<2-4$ Mpc -- the structures that host radio-AGN were significantly denser than those hosting radio-quiet galaxies. Based on their results, \citet{hatch1} suggest that the dense Mpc-scale environment may foster the formation of a radio jet from an AGN and further speculate that virtually all the the massive, $M_*>10^{14}\,M_\odot$, clusters and protoclusters go through a radio-active AGN phase during $1.3 < z < 3.2$. This corresponds to a life-time for the radio-active AGN phase of at least 60 Myr, lower than the values from \citet{lin1} and \citet{maglio16}, which were however obtained in different redshift ranges, for different radio luminosity levels (CARLA sources are much brighter than those considered by \citeauthor{lin1} and \citeauthor{maglio16}) and by using entirely different methods. \citet{hatch1} also expanded the work of \citet{Wyl} and reinvestigated the relationships between the large-scale environment of the CARLA AGN and their physical properties. Once again, no correlation was found, except for a low-significance ($2.3\sigma$) positive dependence on black hole mass (cf.\ right-hand panel of Fig.~\ref{CARLA}). 

\citet{rigby} extended the search for over-densities around high-redshift radio-AGN to galaxies detected in the FIR by the SPIRE instrument onboard of the \textit{Herschel} Space Observatory. By imaging regions around 19 extremely powerful radio-AGN  ($L_{500\, \rm MHz} > 10^{28.5}$ W Hz$^{-1}$ at $2.0<z<4.1$), they found that most of these fields were denser than average, with two of them clearly indicating the presence of (previously unknown) protoclusters made of galaxies undergoing intense star-formation activity (SFR $\simgt 500$--$1000\,M_\odot$ yr$^{-1}$), extending up to $\sim$ 2--3 Mpc. On the other hand, \citet{cooke} report of a mature cluster, mostly formed by quiescent galaxies, surrounding a radio-AGN already at $z=1.58$.

\begin{figure}
\centering
\includegraphics[scale=0.4]{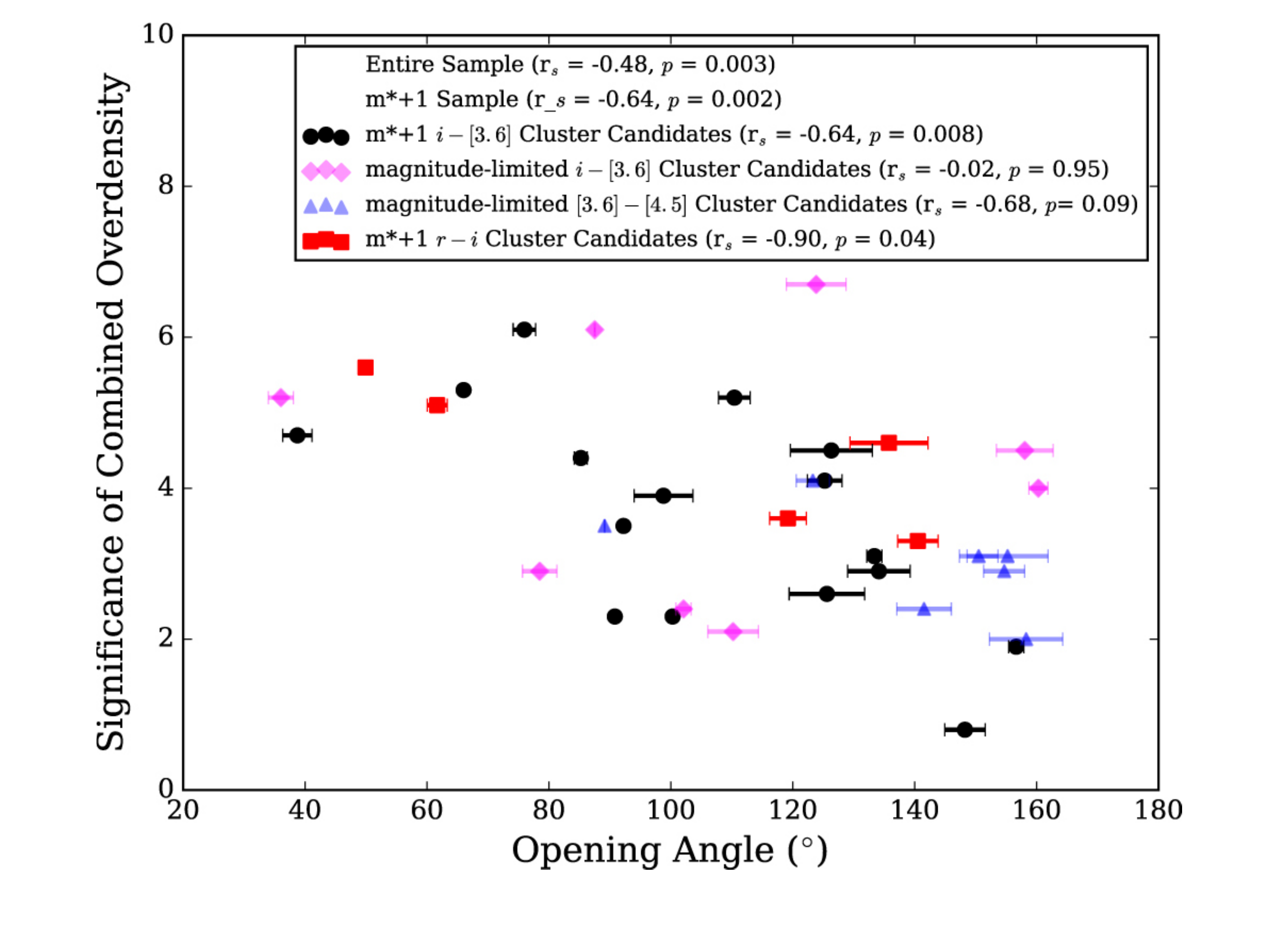}
\caption{Significance of the combined over-density  as a function of the opening angle of the bent radio source. All significances reported here are from \citet{golden-marx}. 
Figure from \citet{golden-marx1}.}
\label{golden}       
\end{figure}

In a similar fashion to what done by the CARLA survey, the COBRA (Clusters Occupied by Bent Radio AGN) program (\citealt{paterno}) searches for overdense regions around radio-AGN with double-lobed structures which are not aligned with each other, but bent by forming angles $<180^\circ$. The rationale behind this search is that the radio lobes of these AGNs are most likely bent because of the ram pressure that occurs due to the relative motion of the AGN host galaxy and the ICM (e.g., \citealt{feretti, blanton, giacintucci, wing}), which makes these sources good tracers for finding galaxy clusters. Indeed, out of 646 bent radio-AGN, 530 (corresponding to $\sim 82$\% of the original sample) are associated with over-densities -- mostly at high, $z=1-3$, redshifts -- in the \textit{Spitzer}/IRAC maps, and 190 are associated with galaxy cluster candidates.\\
By following up on the previous work, \citet{golden-marx1} also show for a subsample of 36 high-$z$ ($0.35<z<2.2$) cluster candidates that radio-AGN with narrower ($\simlt 80^\circ$) opening angles reside in richer clusters (cf.\ Fig.~\ref{golden}), clearly indicating that the cluster environment impacts radio morphology.

More insights on the inner structure of protoclusters surrounding bright radio galaxies and on the physical processes in action are provided in the recent work by \citet{Gilli} (see also \citealt{damato}). These authors make use of deep \textit{Chandra} observations centred on an over-density which develops along a compton-thick FRII galaxy set at redshift $z=1.7$. The region presents significant diffuse X-ray emission. In particular, X-ray emission extending for $\sim 240$ Kpc is found around the eastern lobe of the FRII and most of the star-forming galaxies in the over-density are distributed in an arc-like shape along the lobe edges. \citet{Gilli} propose that the diffuse X-rays originate from an expanding bubble of gas that is shock-heated by the FRII jet, and that star formation within galaxies in the proximity of the bubble is promoted by the compression of the cold interstellar medium. This result provides one of the first pieces of evidence for positive AGN feedback on cosmological (i.e., beyond the extension of the galaxy host of the radio-AGN) scales.

\subsubsection {Cross-matches of catalogues}
With a somehow different approach from that presented in the first part of this Section, a number of studies have also investigated the environmental properties of mainly radio-AGN but also radio-emitting star-forming galaxies by matching catalogues of sources observed in the radio band (mostly at 1.4 GHz) with those obtained in the optical/NIR. The first pioneer works based on cross-correlation analyses appeared in the late 1970s  (e.g., \citealt{seldner, longair, prestage} -- cf.\ Sect.~\ref{sec:3.1}), but a great leap forward in this field was only possible starting from the beginning of the century when technology developments made large galaxy surveys possible. 

\begin{figure}
\centering
\includegraphics[scale=0.4]{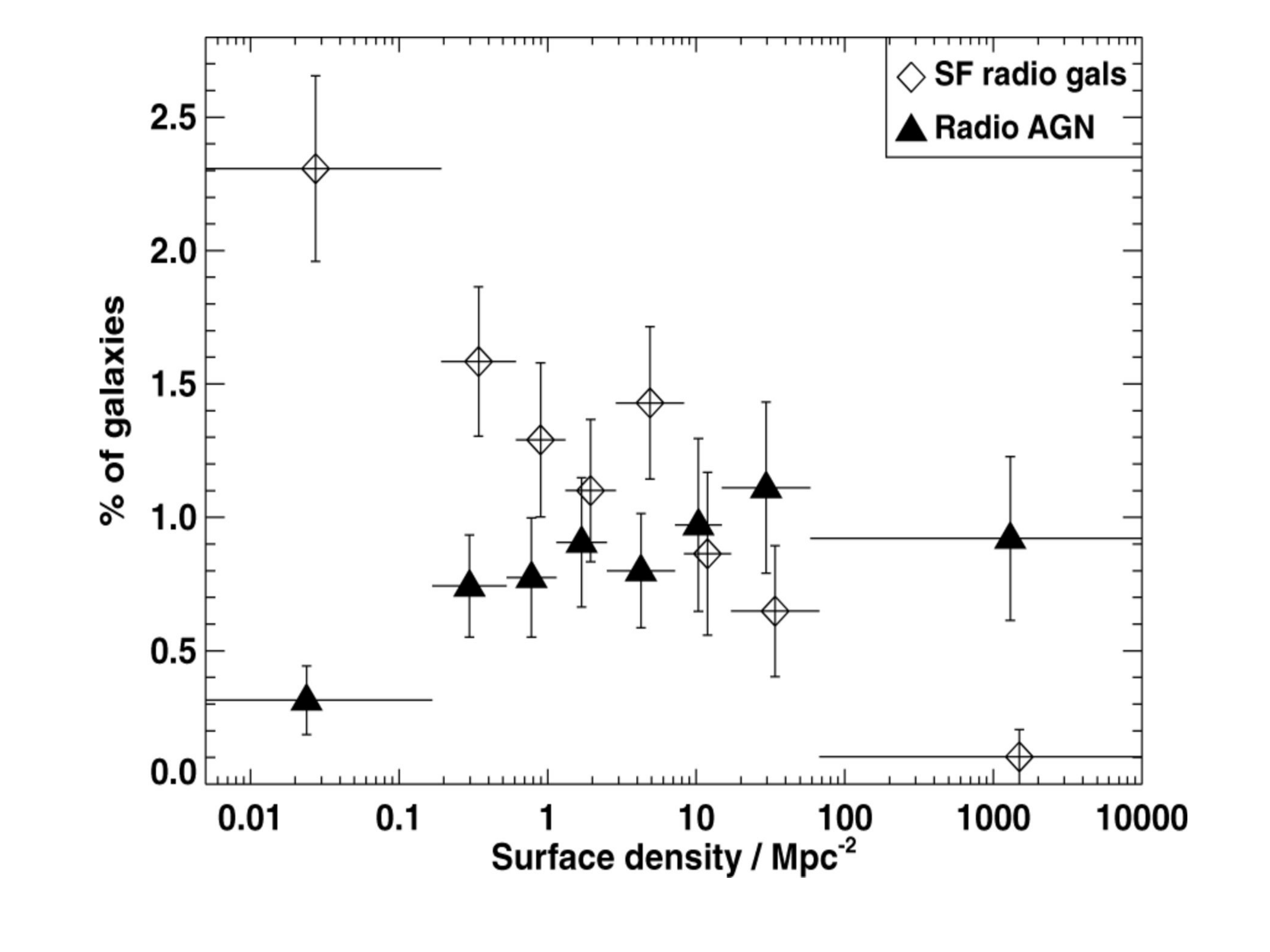}
\caption{Fraction of galaxies within the 2dFGRS catalogue that are associated with radio sources securely classified as either star-forming or AGN, as a function of the local projected galaxy surface density. Error-bars correspond to Poissonian uncertainties. The frequency of star-forming radio sources is greatly suppressed in dense environments, while AGN activity is roughly independent of environment, except in the most underdense regions. Figure from \citet{best2}.}
\label{Best}       
\end{figure}

\citet{best2} combined the optical 2dFGRS with NVSS and used a Friends-of-Friends algorithm to investigate the environment of 91 local ($z<0.1$) radio-AGN mostly endowed with luminosities $L_{1.4 \rm GHz}\ge 10^{22.8}$ W Hz$^{-1}$. It was found that, while AGN activity of radio origin showed little dependence on local galaxy surface density except at the very lowest surface densities where such an activity is suppressed (cf.\ Fig.~\ref{Best}), the larger-scale environment was instead more important, since -- in agreement with earlier works (e.g., \citealt{Allington, zirbel2}) -- radio-AGN were preferentially found in moderate-richness groups and poor clusters of galaxies. It was also found that the ratio of absorption-line to emission-line radio-AGN changed dramatically with environment, with essentially all radio-AGN in rich environments showing no emission lines in their optical spectra, result that led to conclude that such a difference could be due to the lack of cool gas in cluster galaxies.
\citet{best2} also repeated the investigation for a sample of 154 radio-emitting star-forming galaxies taken from the same NVSS parent catalogue. At variance with radio-AGN, these exhibited a strong correlation with environment, with star formation activity suppressed in high-density structures (cf.\ Fig.~\ref{Best}), a result which was well-known at optical wavelengths (e.g., \citealt{dressler1}), but not yet seen in radio-selected objects. 

\begin{figure*}
\centering
\includegraphics[scale=0.5]{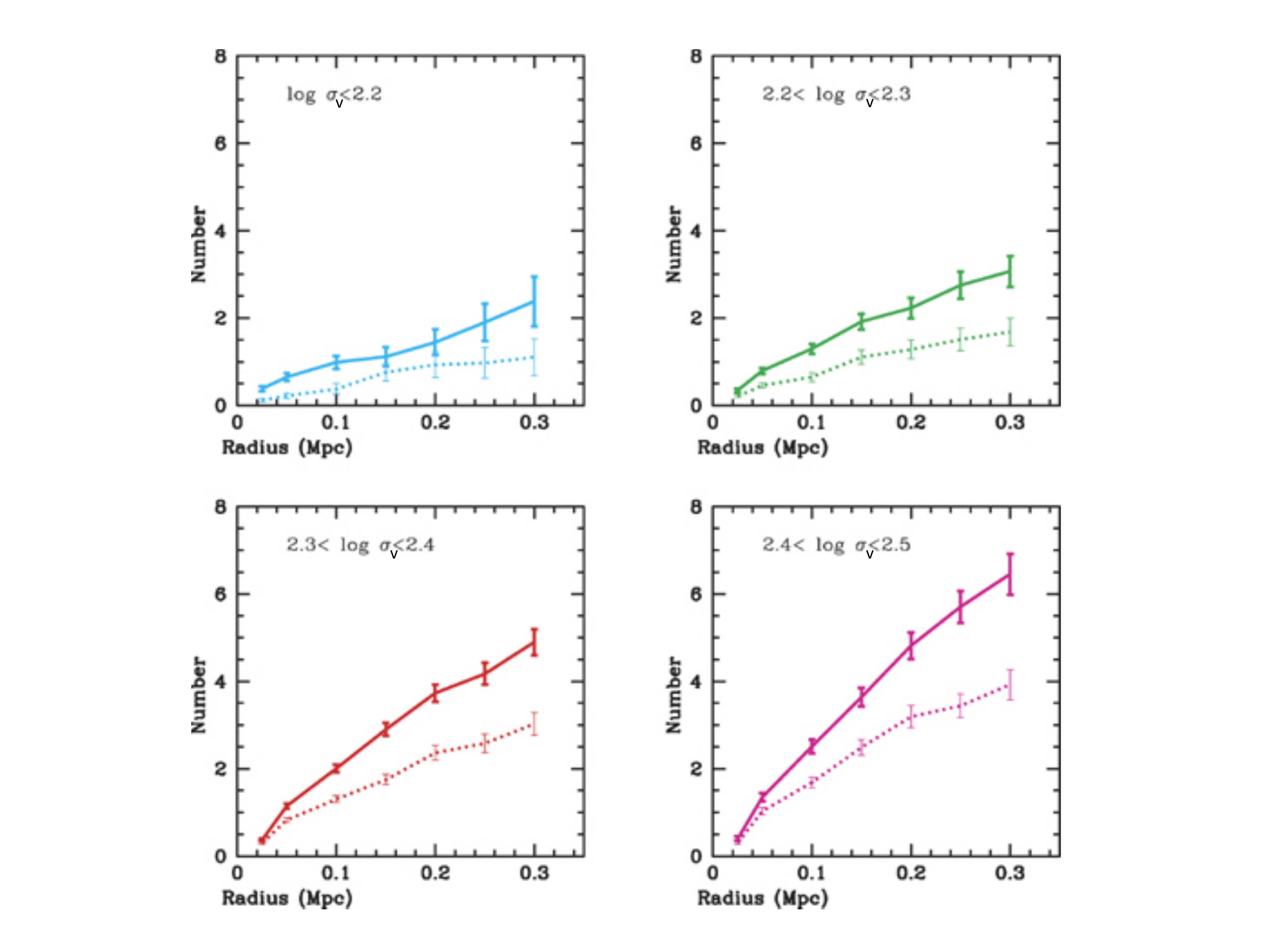}
\caption{Galaxy counts within projected radius R for emission-line radio-active AGN  (solid) and matched samples of radio-quiet AGN (dotted). Results are shown for four different ranges in the velocity dispersion $\sigma_{\rm v}$ of the hosts. 
Figure from \citet*{Kauffmann}.}
\label{Kauffmann_2008}       
\end{figure*}

\citet*{Kauffmann} instead considered a particular sample of $\sim 1600$, local radio-active AGN with strong emission lines in their optical spectra, obtained by combining together NVSS and FIRST with spectroscopic information from the Sloan Digital Sky Survey DR4.  AGN were selected from the population of radio emitters on the basis of a correlation between radio and $H_\alpha$ (proxy for FIR) luminosity, corrected for extinction. These sources where then compared to a sample of radio-quiet AGN matched in redshift, stellar mass and velocity dispersion $\sigma_{\rm v}$ (proxy for the central black hole mass) of the host. While these authors found a remarkable agreement between the properties of radio-active and inactive AGN in terms of host galaxy structure, stellar populations and emission line properties in their optical spectra, they also observed a factor $\sim 2-3$ enhancement in the local density surrounding radio-active AGN (cf.\ Fig.~\ref{Kauffmann_2008}):  in other words, radio-active AGN reside in denser environments with respect to radio-quiet ones, even though no dependence on radio luminosity was found.

\begin{figure}
\centering
\includegraphics[scale=0.4]{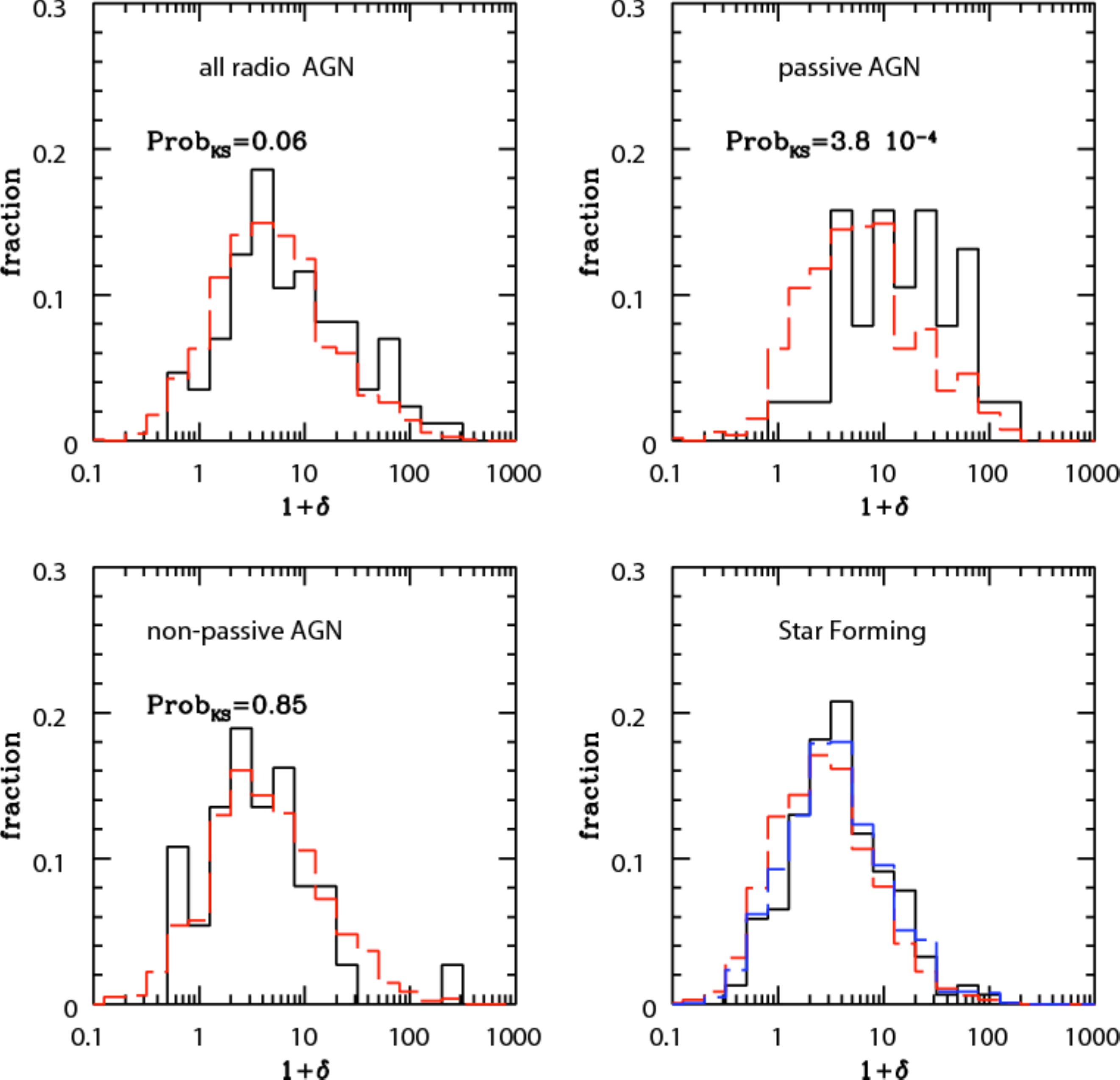}
\caption{Black histograms: over-density distribution $1+\delta$ of the radio sources in the \citet{bardelli} sample. Red dashed histograms show the same quantity but for the control samples. Upper left panel: total AGN sample. Upper right panel: passive AGN. Lower left panel: non-passive AGN. Lower right panel: star-forming galaxies. The blue histogram corresponds to the control sample corrected for stellar mass distribution. Figure from \citet{bardelli}.}
\label{bardelli}       
\end{figure}

Few years later, \citet{bardelli} moved the analysis to higher redshifts by investigating a sample of 315, $0.1<z<0.9$,  radio sources obtained from the VLA-COSMOS survey (\citealt{schinnerer, bondi}) via cross-correlation with galaxies endowed with a spectroscopic redshift from the z-COSMOS bright catalogue (\citealt{lilly3}). Radio galaxies were divided into passive AGN (i.e., very little star-forming activity), non-passive AGN (ongoing star-forming activity) and star-forming according to a combined method that first used the excess of their radio-derived SFRs with respect to those obtained from their optical SEDs to select AGN (cf.\ Sect.~\ref{sec:1}), and then a criterion based on [8.0--4.5] $\mu$m colours vs specific star-formation rates (as derived from the SEDs) to distinguish between passive and non-passive AGN.
The authors conclude that the only class which is preferentially hosted within overdense structures is that of passive AGN (fraction increasing from  $\sim 2$\% in underdense regions to  $\sim 15$\% in overdense ones, cf.\ Fig.~\ref{bardelli}, where $1+\delta$ is the density contrast), while both non-passive AGN and radio-emitting star-forming galaxies are embedded in structures which do not differ from those hosting galaxies drawn from control samples. \citet{bardelli} further observe a dependence of the radio luminosity on the stellar mass of the hosts for the sub-class of passive radio-AGN, dependency which however disappears for $L_{1.4\rm GHz}\simgt 10^{23.5}$ W Hz$^{-1}$, luminosity beyond which also almost all passive radio-AGN are found to reside within overdense regions. 

\citet{malavasi} present an analysis similar to that of \citet{bardelli} on the same COSMOS field, but based on photometric redshifts  (\citealt{Ilbert}), which on one hand allow to expand the statistical significance and redshift range of the sample (1427 objects up to $z\sim 2$) but at the price of larger uncertainties in the redshift determination. By using control samples with the same stellar mass and sSFR distributions of the original radio dataset, \citet{malavasi} conclude that the environments surrounding radio-AGN are significantly denser than those  occupied by inactive galaxies, especially for low-power -- $24 \le \log(L_{1.4 GHz}/[\rm{W\, Hz}^{-1}] )< 24.5$ -- radio-AGN, while the trend disappears for AGN of higher radio luminosities. 
Roughly at the same time, \citet{castignani} focus on a sample of 32, $1<z<2$, FRI galaxies in the COSMOS field extracted from the catalogue of \citet{chiaberge2}. These authors investigate the environments surrounding such sources subdivided into 21 low-luminosity radio galaxies (LLRGs) and 11 high-luminosity radio galaxies (HLRGs) -- where LLRGs were selected to have 1.4 GHz rest frame luminosities lower than the fiducial FRI/FRII divide -- and find over-densities extending up to $\sim 2$ Mpc around $\sim$70\% (22 out of 32) of the sources, independent of their radio luminosity. This is in agreement with e.g., the conclusions of \citet{Wyl} but -- puzzling enough -- at variance with those of \citet{malavasi} obtained on the same COSMOS field and for a comparable redshift range and radio luminosity divide. A similar result on the independence of AGN radio luminosity from environment was also recently reported in \citet{Kolwa} for 2716, $z\simlt 0.8$ radio sources with luminosities $10^{23} < L_{\rm 1.4 GHz}/[{\rm W \; Hz^{-1}}] < 10^{26}$ observed in the SDSS Stripe 82 field. 
By comparing their objects to a sample of control galaxies matched in redshift, colour and stellar mass of the hosts, \citet{Kolwa} also observe that the environments of their radio sources are similar to those of the control sample, except for the fact that radio-AGN are more prevalent within the most overdense regions such as galaxy clusters. These results lead the authors to conclude that environment is expected to play some (relatively minor) role on the production of radio activity from an AGN, but it is not the crucial ingredient. However, given the luminosity range probed by their sources,  it is likely that the \citet{Kolwa} results are at least partially affected by the presence in their radio-AGN sample of a non-negligible fraction of radio-emitting star-forming galaxies (cf.\ Sect.~\ref{sec:1}) which generally inhabit structures indistinguishable from those occupied by their radio-quiet counterparts.


\section{Discussion}
\label{sec:4}
Given the established evidence for a strong connection between extragalactic radio source activity and the formation and evolution of galaxies and larger structures such as groups and clusters, this review has concentrated on the Kpc-to-Mpc scale properties of radio-AGN and star-forming galaxies by providing an overview on their hosts and environments. The conclusions based on the most recent works are summarized below.

\subsection{Hosts}
\label{sec:4.1}
\paragraph{\underline {AGN}} -- 
It is widely agreed that radio-AGN are mainly hosted by the most massive galaxies both in the local (e.g., \citealt{matthews, ekers, Auriemma, jenkins, best4, mauch, sabater1, capetti2})
and in the higher-redshift (e.g., \citealt{lilly1, eales, jarvis, debreuck, willott, seymour, smolcic1, debreuck1, gurkan, drouart1, maglio14})
universe. As a matter of fact, there are large variations in the level and relevance of radio activity of AGN origin within different extragalactic populations, with the probability for a source to host a radio-AGN observed to vary from as little as $\sim 2-5$\% in moderate-to-high ($z\sim 0.5-2$) redshift AGN (e.g., \citealt{hickox, maglio18}) up to virtually 100\% in the most massive local ellipticals (e.g., \citealt{brown, capetti2, grossova}). 

Many works had already investigated in the local universe the connection between the fraction of radio-AGN within a chosen population of galaxies and their optical/NIR luminosity (e.g., \citealt{colla, Auriemma, hummel}). These have been implemented in the more recent years with studies connecting such a fraction with the stellar mass of the hosts (e.g., \citealt{best4, mauch, sabater1, maglio18}). In all cases it has been found that the probability for a galaxy to host a radio-active AGN is a strong function of both their optical/NIR luminosity (proxy for the stellar mass in absence of this information) and mass, with a functional form of the kind $\propto M_*^{2.5}$ (e.g., \citealt{best4}). This dependence also appears to be stronger than that estimated for the mass of the black hole powering the radio-AGN ($\propto M_*^{1.6}$ -- e.g., \citealt{best4, sabater1}), evidence that would hint to a stronger link between radio-AGN activity and gas fuelling with respect to nuclear properties. However, in spite of this relation, no connection is observed between level of radio activity and host mass (e.g., \citealt{mauch}). No connection has either been found between probability for radio-AGN emission and  nuclear accretion properties, cosmological evolution or star-formation activity of the hosts (e.g., \citealt{maglio18}). 
It therefore appears that the host mass is the main factor driving the formation and evolution of a radio-AGN.  

These findings have also been confirmed up to $z\sim 2$ (e.g., \citealt{smolcic1, williams, lin3, mo2}), together with the result of a global increase as a function of look-back time of the fraction of radio-active AGN within galaxies, observed to happen at all stellar masses. Such an increment is however differential, since relatively low-mass, $M_*\simlt 10^{10.75}\,M_\odot$, galaxies are found to be up to 10 times more likely to host a radio-AGN at higher redshifts with respect to local results, while galaxies with a larger stellar mass content show little evolution. This shift of radio-AGN activity towards lower-mass, mainly star-forming galaxies has been interpreted as evidence for a change in the nuclear accretion properties of radio-emitting AGN from mostly hot/inefficient in the local universe, to cold/efficient as we approach cosmic noon at $z\sim 2$. Indeed, already in the local universe, radio-AGN are observed to show different accretion modes: cold/efficient in high-excitation radio galaxies or HERGs and hot/inefficient in low-excitation radio galaxies or LERGs (e.g., \citealt{hardcastle}), although with a numerical prevalence of LERGs over HERGs. However, since at least up to $z\sim 1$ HERGs display a stronger cosmological evolution than LERGs (e.g., \citealt{best7, pracy, butler2}), this would explain the larger fraction of radio-AGN within HERG hosts (locally found to be less massive, bluer and more star-forming than LERGs -- e.g., \citealt{best6, jannsen, hardcastle2, sadler}) observed at higher redshifts. 


Things however do not seem to be too crystal clear beyond $z\sim 1$. The most recent studies indeed show a much less marked distinctions between HERG and LERG host properties with respect to what observed locally: their mass distributions tend to overlap (e.g., \citealt{fernandes, delvecchio}), except for a more pronounced tail observed below $M_*\simlt 10^{10.5}\,M_\odot$ in the case of HERGs (e.g., \citealt{williams2, butler}). Also the star-forming properties of HERG and LERG hosts are found to be more similar than what found in the local universe. These results likely stem from two main factors: 1) the cosmological evolution of the general galaxy population, which becomes more star-forming as cosmic noon is approached, with a global star-forming activity mainly shifting to larger-mass galaxies (cosmic downsizing), and 2) the different methods (spectroscopic vs photometric) applied to select HERGs and LERGs at low and high redshifts which produce more contaminated samples beyond $z\sim 0.5-1$.

Within this scenario, a special place is occupied by the morphological differences exhibited by radio-active AGN,  also in connection with their accretion properties. 
Early works based on bright radio sources attributed different hosts to the two populations of FRI and FRII galaxies, with FRIs being almost ubiquitously observed within massive and passively evolving red galaxies (but see \citealt{ledlow2, ledlow3} for few examples of FRIs within spirals), and FRIIs mainly associated to bluer, fainter and smaller galaxies (e.g., \citealt{lilly, owen1, baum, zirbel1}). \citet{ledlow1} found in the optical properties of the host galaxy a crucial ingredient to account for the FRI-FRII dichotomy, which then had to be added to the already known ingredient of radio luminosity (\citealt*{fanaroff}). Indeed, it was shown that \emph{all} the radio-AGN above the locus $L_{\rm radio}\propto L_{\rm optical}^{1.8}$ were FRIIs, while those below it were FRIs. This meant that powerful FRI galaxies with radio luminosities comparable to those of FRIIs could exist, just as long as their host galaxies were also about two magnitudes brighter. However, despite its enormous success and the fact that it was further converted into a relationship between AGN power and black hole mass (\citealt*{ghisellini}), it was subsequently found that the \citet{ledlow1} result suffered from a number of biases, most importantly luminosity issues since FRI and FRII galaxies had been sampled in different redshift ranges. 

With the advent of deeper and wider surveys which started to provide more homogeneous large samples of fainter radio galaxies, it has become clear that, although there are indeed very few (although not zero) FRIs above the \citet{ledlow1} divide, there is a large overlap between FRI and FRII galaxies below it (e.g., \citealt{best5, lin2, wing, mingo1}). To make things more complicated, the almost one-to-one correspondence between FRIs and LERGs and FRIIs and HERGs which was thought to hold at least to a zero-th order approximation, has in the recent years proven not to be true. Indeed, despite the fact that few FRIs are observed in association with a HERG, LERGs can instead assume both FRI and FRII morphologies (e.g., \citealt{laing1, tadhunter, chiaberge, chiaberge1, hardcastle5, buttiglione, capetti, capetti1, miraghaei, mingo2}). However, while there is nowadays a general consensus on this result, disagreement still remains on the possible mechanisms causing the FRI-FRII dichotomy, also in connection with their HERG and LERG distinction.

In particular, two main scenarios  have emerged: the first one attributes differences in radio morphology to the large-scale (i.e., host galaxy and environment, see also below in this Section) properties of radio-AGN which influence the interaction between the radio jet and the external medium (e.g., \citealt*{kaiseraltro, wing}; \citealt{miraghaei, mingo1, mingo2}), with those between HERGs and LERGs being instead dictated by different fuelling mechanisms as discussed earlier.  According to this framework, HERGs are powered by accretion of cold gas, provided by e.g., a recent merger with a gas-rich galaxy, while LERGs accrete hot intergalactic gas from dense environments at a low rate (e.g., \citealt*{best6}), with fuelling from major mergers strongly disfavoured by recent observations (e.g., \citealt{ellison}). On the other hand, other works do not observe any difference in the host and/or environmental properties of FRI and FRII galaxies, except in the rare cases of FRII HERGs (e.g., \citealt{lin2, capetti1, jimenez, massaro, massaro1, vardoulaki}) or even between those of HERGs and LERGs (e.g., \citealt{fernandes}), so that an alternative mechanism for their large-scale radio behaviour has to be invoked.  In this case, ageing processes can be thought as the main driver for the observed morphological differences, with an evolutionary pattern that proceeds from FRII HERGs that switch from efficient to inefficient accretion due to gas starvation and transform themselves into FRII LERGs, sources that still maintain their large radio structures thanks to the past nuclear activity at high efficiency (e.g., \citealt*{ghisellini, tadhunter1}; \citealt{macconi, grandi}). The switch off/change in accretion mode will eventually show in the radio morphology with the delay needed to reach Kpc-to-Mpc distances, and the source will ultimately turn into an FRI galaxy.

A further important point to stress is that, while as already discussed, local -- $z\simlt 1$ -- observations indicate that radio-active AGN mostly reside within passive ellipticals, hosts of very little ongoing star-forming activity (e.g., \citealt{siebenmorgen, dicken1, gurkan1}), the situation becomes dramatically different at higher redshifts (e.g., \citealt{dey, pentericci, pentericci1, williams, lin3}). Indeed, the advent of facilities such as SCUBA, \textit{Spitzer} and \textit{Herschel} probing the IR and sub-millimeter regimes to high sensitivity levels, has opened a new window on our knowledge of the physical status of radio-AGN during cosmic noon at $z\sim 1.5-2$ and beyond. All the works based on these observations (e.g., \citealt{Archibald, reuland, seymour, seymour1, rawlings, drouart, podigachoski, maglio13, maglio14, maglio15}) have clearly shown that in the $z\sim 1-3$ range and beyond, radio-AGN hosts produce stars at very high rates, up to $\sim 10^3\,M_\odot$ yr$^{-1}$. Such an activity is observed to increase with look-back time, up to the highest redshifts probed by the observations, and in most cases also to be in excess with respect to those exhibited at the same epochs by inactive galaxies or by the hosts of radio-quiet AGN (e.g., \citealt{drouart, kal, kal1}). It is then clear that the \emph{negative} radio-mode feedback (e.g., \citealt{croton, fanidakis, weinberger}), which invokes the presence of a radio jet to warm up or even expel the cold gas contained within massive galaxies (therefore halting their star-formation processes so to reproduce the locally observed physical properties and number density) can only be valid in the nearby universe. In the high-redshift universe instead, radio-AGN activity and build-up of the stellar mass at vigorous rates co-exist within the same galaxy. This would imply the presence of \emph{positive} feedback (\citealt{silk}), whereby radio jets increase the star-formation activity by compressing the intergalactic medium. Alternatively, as some observational results seem to indicate (e.g., \citealt{seymour1, rawlings, drouart, maglio13, maglio14, maglio15, podigachoski}) it is possible that radio-AGN and star-formation activities proceed parallel to each other and only result coeval thanks to the availability of large reservoirs of gas within massive galaxies at $z\sim 2$. More work in this direction is needed to assess the validity of one of the two scenarios and also to understand the drastic change in feedback behaviour when moving from $z\sim 1.5$ down to the local universe.

\paragraph{\underline {Star-Forming Galaxies}} -- 
At all the (relatively low, $z\simlt 1$) redshifts probed by present facilities, radio-emitting star-forming galaxies appear in all respect identical to star-forming galaxies selected at any other wavelength, with spectral types ranging from spirals to irregulars and dwarfs. Although only $\simlt 10^{-4}$ of the bolometric luminosity produced by a non-AGN powered galaxy is radiated at radio wavelengths (\citealt{condon}), the importance of radio selection for star-forming galaxies relies on the fact that it can provide estimates of the star-formation activity of such sources in a way which is unaffected by the presence of dust up to the earliest epochs. In order to do that, a solid calibration of the relation between radio luminosity and infrared luminosity (or alternatively SFR) is mandatory. This is why a lot of effort has been put in the recent years to provide calibrations as precise as possible at all redshifts and for all galaxy types. 
From an observational point of view, it has been found at least in the local universe that radio and infrared luminosities show an incredibly tight, approximately linear, correlation over more than 5 orders of magnitude in infrared luminosity (e.g., \citealt{helou, Dejong, condon, yun}), irrespective of galaxy type and level of star-forming activity, as long as the galaxy hosts episodes of star-formation. This relation, originally found at 1.4 GHz, has also been observed to hold at different frequencies (e.g., \citealt{delhaize, gurkan2}). The theoretical reasons behind it are still not fully understood, as many factors involved in the process of shaping a galaxy should intervene to produce at least some modifications which are not (yet) clearly visible in the available data.  

One of the most important open questions associated to the IR-radio correlation is its eventual evolution with redshift, as erroneous estimates of the $q_{\rm IR}\propto L_{\rm IR}/L_{\rm radio}$ value at high $z$'s would hamper the huge potential of present and future radio surveys such as SKA to investigate the history of formation and evolution of galaxies since the cosmic dawn. Early studies which exploited IR data from the \textit{Spitzer} and \textit{Herschel} satellites found a general independence of the IR-radio correlation from redshift (e.g., \citealt{Appleton, vlahakis, Ibar, garn, jarvis1, chapman1, sargent, mao}). Then, more recent works based on different source selection and stacking analyses, have started to infer a slight but statistically solid evidence for the 
ratio between IR and radio luminosity to decrease with look-back time (\citealt{ivison, ivison1, bourne, magnelli, calistro, delhaize}). The reason for such discrepant results is to be found in the different selection criteria, as in principle while e.g., the radio selection favours higher $L_{\rm radio}$/SFR ratios, the opposite is true for the FIR and sub-millimetre selections which favour high SFRs. 

More insights on the cosmological evolution of the IR-radio correlation have been recently provided by the results of \citet{molnar} and \citet{delvecchio1}. Indeed, while the first work shows that a different evolution is observed for star-forming galaxies of different morphological types, hinting to an increasing presence and contribution of radio-emitting AGN within bulk-dominated star-forming galaxies at higher redshifts, the second study finds in the stellar mass the main driver of such a redshift dependence, with galaxies of higher stellar masses displaying lower values of the $q_{\rm IR}$ parameter.  However, despite all the progress made in the very recent years, a convergence on the IR-radio relation for what concerns both its deviation from linearity and eventual redshift evolution is still far from being reached (e.g.,  \citealt{bonato, molnar1, smith, tisanic}).

\subsection{Environment}
\label{sec:4.2}
\paragraph{\underline {AGN}} -- 
While --  starting from the very early works (e.g., \citealt{seldner, longair, hill, peacock, Allington, zirbel2}) --  virtually all the studies presented in the literature converge at indicating that radio-AGN preferentially reside within overdense structures, the three (or rather five if one also includes cross-correlation studies) methods presented in Sect.~\ref{sec:3} provide different information on the large-scale structure behaviour of these objects. To summarize them in a few words, we might say that the method based on clustering returns more information on the very large/cosmological (i.e., at Mpc level and beyond) scales traced by radio sources and also on the dark matter content of the regions that host them. On the other hand, very different results are obtained if one searches for structures around known radio-AGN or if one pinpoints radio-AGN within known structures. Indeed, in the first case one finds that virtually all radio-AGN are surrounded by over-densities (e.g., \citealt{Venemans, mayo, Galametz1, castignani, Wyl, rigby}), while the second method shows that only about 20--30\% of them inhabit rich (group- and cluster-like) structures (e.g., \citealt{best8, maglio7, lin1, croft, maglio17, croston2}). 
The reason for this discrepancy is not known, but it is likely related to the different redshift ranges probed by the two methods (much more local sources are considered in the second one), and/or -- under the assumption of a strong correlation between radio luminosity and environmental density (e.g., \citealt{bardelli, maglio17, croston2, mo2}, but see further in this Section for different points of view) -- to the fact that generally the first method images much brighter radio sources than the second one. In any case, we note that these findings also have implications for the life-time of the radio-AGN phenomenon, and it is therefore of no surprise if works based on the different methods illustrated above find different values, ranging from $\sim 60$ Myr up to a few Gyr (e.g., \citealt{lin1, smolcic2, hatch1, maglio16}). We also note that recent studies based on direct LOFAR observations which -- we remind -- sample lower frequencies and therefore older emission,  would tend to better agree with the high-end values provided above for the radio-active phase of an AGN ($\tau > 200$ Myr -- \citealt{heesen}).

The fact that -- at variance with what observed for e.g., optical AGN (e.g., \citealt{ellingson, porciani, Kauffmann, mandelbaum, donoso1, sabater, retana}) -- most of the radio-AGN reside in overdense/cluster-like structures, has important implications not only for the investigation of the radio-AGN phenomenon \textit{per se'}, but also for the study of the interaction between radio-AGN activity and their surrounding structures, both in terms of companion galaxies and ICM.  First of all, regardless of the adopted method, all studies report estimates for the masses of the dark matter hosts of radio-AGN to be of the order of $10^{13.5}$--$10^{14}\,M_\odot$ (e.g., \citealt{maglio6, hickox, mandelbaum, Lindsay1, hatch1, maglio16, retana, hale, damato, uchi}), with a value which remains unchanged at all probed epochs, from as early as $z\sim 4$ (\citealt{uchi}) down to the local universe (cf.\ \citealt{maglio16} and references therein). This leads us to draw the fundamental conclusion that \emph{the environmental properties of radio-AGN remain unchanged throughout all cosmic ages}. 

Furthermore, it has been found that the BCG of a cluster or group of galaxies is much more likely to host a radio-AGN than any other galaxy belonging to the over-density, and that this likelihood is a strongly increasing function of its stellar mass and of the AGN radio luminosity (e.g., \citealt{best8, lin1, croft, smolcic2, mo1, maglio17, croston2}). The above findings clearly push towards a strong connection between radio activity and over-density/ICM properties. 
Although a whole review would be needed to tackle this fundamental aspect of galaxy evolution known under the name of `feedback', here we suffice saying that indeed plenty of evidence has been reported for the above interaction, both in the case of extended radio sources presenting clear jetted structures (e.g., \citealt{maglio7, paterno, Gilli, damato, golden-marx1}) and for the whole population of radio-AGN regardless of their morphological division (e.g., \citealt{croston1, lin1, ineson1, ineson2}), although no dependence is inferred between environmental richness and extension of the radio-AGN, at least in the case of Giant Radio Galaxies (e.g., \citealt{lan}). A strong connection has been also reported between radio-AGN activity, environmental properties and star-forming processes within the host galaxy (e.g., \citealt{bardelli, miraghaei1}), whereby mostly radio-AGN associated to passive galaxies are found to reside within over-densities. 

However, despite all the progress made in the field, the details of the interaction between radio-AGN and ICM are far from being fully understood. For instance, there is still no consensus on which (large) scales are affected by the presence of a radio-AGN, or how radio feedback precisely works. This lack of knowledge is somehow related to the uncertainties we still have on the environmental properties of radio-AGN of different types, differences that likely reflect -- amongst others -- different accreting modes. So, while for instance there seems to be a general consensus on the fact that locally FRI and/or radiatively inefficient galaxies tend to reside in denser environments than FRII and/or radiatively efficient ones (e.g., \citealt{prestage, zirbel2, best2, wing, ineson1, ineson2, miraghaei, croston2}, but see \citealt{massaro, massaro1} for a discording view), just as radio-AGN hosted by early-type/red galaxies do (e.g., \citealt{bardelli, miraghaei1}), it is still not clear whether FRII galaxies are embedded in lower density environments only at low redshifts (e.g., \citealt{Allington, zirbel2, ineson2}) or at all cosmological epochs. To our knowledge, except for very few cases that present somehow discording conclusions (e.g., \citealt{hale, vardoulaki}), this kind of systematic investigation has only been performed up to $z\sim 0.5$. An extension of the analysis to higher redshifts is therefore mandatory in order to provide a definite answer. Just as an example, the mere existence of a protocluster serendipitously found surrounding a FRII galaxy at $z=1.7$ (\citealt{Gilli}), and the fact that most high-redshift radio galaxies are indeed FRIIs (\citealt*{miley1}), show that at high redshifts these sources can indeed reside in overdense structures. And if this is the case, does it imply a different cosmological evolution for FRII and FRI galaxies also in terms of their environmental properties? Also, do the above results unequivocally point towards different fuelling mechanisms for the different sub-populations of radio-AGN mentioned above as studies of their host galaxies seem to indicate (cf.\ Sect.~\ref{sec:2})?

Even more intricate is the case of AGN of different radio luminosities. Indeed, no consensus has been reached for what concerns their environmental properties despite all the effort put in by the scientific community. The number of works that claim no dependence (e.g., \citealt*{hill};  \citealt{maglio6, best8, Kauffmann, Fine, Wyl, castignani, maglio16, Kolwa}) almost equals that of those that do find a difference, with brighter radio-AGN belonging to denser structures (e.g., \citealt{longair, yates, overzier, bardelli, hale, maglio17, croston2, mo2}). The reason for such a striking disagreement is unknown, although some works advocate the different luminosities and/or redshift intervals considered in the various analyses. It is also not clear whether the claimed dependency holds for the whole radio-AGN population or only for some sub-class of objects (e.g., \citealt{ineson1, ineson2}). Furthermore, a few studies mainly sampling the higher-redshift universe (e.g., \citealt{donoso1, malavasi, uchi}) refer of the opposite effect, i.e., fainter radio-AGN being surrounded by denser environments. 

There are other open issues connected with the physics of the radio-AGN phenomenon and of its accreting black hole. Amongst a few, whether it is true that AGN that emit both in the radio and in the MIR have a preference to reside within underdense structures (at variance with radio-AGN which are also X-ray emitters), as observed by \citet{tasse} and \citet{maglio17} but with a somehow low significance level, finding that if confirmed would imply an intimate connection between nuclear activity and environmental properties even for AGN that are already radio emitters.  Also, if there is indeed a dependency of the radio-AGN environmental properties on their black-hole mass, or at least on some threshold value of $\sim 10^9\,M_\odot$, as the works by \citet{hatch1} but mostly by \citet{retana} seem to suggest. And lastly, it is still has to be further confirmed whether indeed radio-active galaxies and radio-active quasars cluster differently (i.e., are hosted by different structures), as suggested by the work of \citet{donoso1}.

All these answers require wider radio surveys which would be able to return large samples of sources divided per morphological type, redshift interval, radio luminosity, accretion mode, etc, deep enough to probe the whole population of radio-active AGN up to the early phases of our Universe. Luckily, all this is within reach with the advent of SKA in the next decade. 

\paragraph{\underline {Star-Forming Galaxies}} -- 
Radio-active star-forming galaxies are a much more straightforward case than AGN: they exhibit no environmental differences when compared to their radio-quiet counterparts (e.g., \citealt{bardelli, maglio16, maglio17, hale, chakraborty}), and are mainly found in average-to-under-dense structures. Dense environments are shown to inhibit the presence of radio-active star-forming galaxies (\citealt{best2, sabater}), precisely in the same way it is observed for their radio-quiet counterparts. This holds at least up to $z\sim 1.5$. At earlier epochs, evidence collected at all wavelengths but radio (due to lack of sensitivity of present facilities) indicates that intense, $SFR > 20-30\,M_\odot$ yr$^{-1}$, star-forming activity shifts to galaxies embedded within denser/protocluster-like structures (e.g., \citealt{maglio12}). Once again, the advent of SKA will unequivocally confirm or refuse whether this conclusion is also valid at radio wavelengths.


%

\begin{acknowledgements}
I wish to thank my friends and colleagues from IAPS, A. Traficante, S. Molinari, S. Pezzuto, A. Di Giorgio, M. Benedettini, E. Schisano and also my kids Tommaso D. Metcalf and Eva G. Metcalf for putting up with me during these long months of writing further affected by the pandemic, F. De Angelis for being the safest safe for the various versions of the manuscript and J. Noceti for unconditional support (both in presence and in absence) towards the completion of the task. I would also like to thank S.J. Maddox, G. de Zotti, L. Danese, A. Celotti and J.V. Wall for everything they taught me on extragalactic radio astronomy and surveys in general. Last but not least, a warm thank you is due to the editor L. Feretti who offered me the opportunity to write this review and also gave me the freedom to lay it out the way it was closest to my chords. 
\end{acknowledgements}

%
%

\phantomsection
\addcontentsline{toc}{section}{References}
\bibliographystyle{spbasic-FS}      
\bibliography{refs}   

\end{document}